\documentclass[fleqn,usenatbib]{mnras}
\usepackage{newtxtext,newtxmath}
\usepackage[T1]{fontenc}
\usepackage{threeparttable}
\DeclareRobustCommand{\VAN}[3]{#2}
\let\VANthebibliography\thebibliography
\def\thebibliography{\DeclareRobustCommand{\VAN}[3]{##3}\VANthebibliography}

\usepackage{graphicx}	
\usepackage{amsmath}	

\title[From shell burning to shock revival]{Three-dimensional Core-Collapse Supernova Simulations: From shell burning to shock revival}

\author[H. Andresen et al.]{
Haakon Andresen$^{1,*}$, 
Evan P. O'Connor$^{1}$, 
C.~E.~Fields$^{2}$, 
Sean M. Couch$^{3}$ \\
{$^{1}$The Oskar Klein Centre, Department of Astronomy, Stockholm University, AlbaNova, SE-106 91 Stockholm, Sweden}\\
{$^{2}$Department of Astronomy and Steward Observatory, University of Arizona, 933 North Cherry Avenue, Tucson, AZ 85721-0065, USA}\\
{$^3$Department of Physics and Astronomy, Michigan State University, East Lansing, MI 48824, USA}
}

\date{Accepted XXX. Received YYY; in original form ZZZ}

\pubyear{\the\year{}}

\begin{document}
\label{firstpage}
\pagerange{\pageref{firstpage}--\pageref{lastpage}}
\maketitle

\begin{abstract}
The outcome of core-collapse supernova simulations depends sensitively on
the multi-dimensional structure of the progenitor star at the onset of
collapse. We perform
three-dimensional simulations of the final $\sim$10--15 minutes of stellar
evolution for five non-rotating solar-metallicity progenitors with zero-age main-sequence
masses of 20, 21.5, 24.5, 26, and 29 solar masses, mapped from one-dimensional
\textsc{MESA} models into the \textsc{FLASH} hydrodynamics code. Convection
develops in the oxygen-rich layers of all five models, with convective
velocities reaching several hundred km\,s$^{-1}$, and in some models strong
convection also develops in the inner silicon- and oxygen-burning shells.
For the 24.5 solar mass progenitor, we carry out three core-collapse
simulations: one initialised from the fully three-dimensional model, one
from its angle-averaged counterpart, and one from the original
one-dimensional \textsc{MESA} progenitor. We find that the multi-dimensional
progenitor leads to 35 to 50\% higher non-radial kinetic energy in the
post-shock region and an average shock radius 5 to 10\% larger than in the
angle-averaged model, and shows the earliest shock revival of the three.
The gravitational-wave emission is similar in all three models and
strengthens after shock revival, driven by a change in the downflows
reaching the protoneutron star rather than by progenitor asymmetries. 
\end{abstract}




\section{Introduction}
The core-collapse supernova problem consists of following the final evolution of a massive star 
from the onset of iron core collapse to the successful disruption of the stellar envelope. 
To simulate core-collapse supernovae, it is necessary to track the non-linear interactions between hydrodynamics, 
neutrino transport, nuclear physics, and gravity (see \citealt{Burrows_21,Yamada_24,Janka_25,Mezzacappa_2026} for recent reviews). 
Over the past decades, progress has been made in modelling core-collapse supernovae, 
with modern simulations \citep{Hanke_12,Hanke_13,mosta_14,Melson_15,Melson_15a,lentz_15,Muller_17,Just_18,Summa_18,oconnor_18b,Kuroda_18,Muller_19,
Vartanyan_19a,Powell_19,Burrows_19,Melson_20,Burrows_20,obergaulinger_20,
Powell_20,Kuroda_20,wang_22,Powell_21,Obergaulinger_21,Eggenberger-Andersen_21,Vartanyan_22,obergaulinger_22,Kuroda_22,Matsumoto_22,Mori_22,bugli_23,Powell_23,Mezzacappa_23,Bruenn_23,Jakobus_23,Shibagaki_24,Nakamura_25,Jakobus_25,Takata_25,Mori_25,Vartanyan_25,Chen_26} incorporating increasingly sophisticated treatments of neutrino 
interactions, nuclear equations of state, and general relativistic effects either approximately (typically using effective potential prescriptions \citealt{Marek_06,fromm_26}) or fully 
\citep{Kuroda_20,Shibagaki_24}. However, despite these advances, numerical simulations still do not capture the full physics of core-collapse supernovae;
see \cite{Mezzacappa_2026} for an overview of current problems and challenges for the future. 
While the modelling community continuously improves their simulation frameworks,
simulations of core-collapse supernovae are constrained not only by the fidelity of the input physics and numerical methods, but also by the quality of the 
initial conditions. In practice, simulations can only be as realistic as the progenitor models 
from which they are initialised.

It is now accepted that neutrino heating plays a central role in powering most core-collapse supernovae; 
energy deposited by neutrinos behind the stalled shock drive 
its revival \citep{Bethe_85,Bethe_90,Herant_94,Janka_95,Janka_95a,Burrows_95,Janka_96}. 
Multi-dimensional effects are important since they both provide pressure support
to the shock and increase the neutrino heating \citep{Murphy_08,Couch_15,Melson_15a,Muller_17,Couch_19}, 
while spherically symmetric simulations generally fail to explode 
for most progenitors \citep{Rampp_00,Liebendorfer_01,Kitaura_06,Couch_14}.
A key source of such multi-dimensional structure is the progenitor star itself \citep{Couch_13a,yoshida_19,Yadav_20,Fields_20,yoshida_21,Fields_21,Varma_21,Griffiths_26a,Griffiths_26b,Varma_26}. In the final 
stages of stellar evolution, vigorous convection can develop in the burning shells. 
These convective motions are 
inherently three-dimensional (3D) and can imprint significant 
asymmetries on the collapsing core. When accreted through the shock, these 
perturbations seed turbulence in the post-shock region, thereby influencing the explosion 
dynamics \citep{Couch_13a,Muller_17,OConnor_18,Bollig_21,Vartanyan_22,Chen_26}. 

Early simulations incorporated progenitor asymmetries by imposing physically
motivated velocity perturbations onto one-dimensional (1D) stellar progenitors.
\cite{Couch_13a} found that progenitor perturbations turned a failed supernova
into a successful explosion. In contrast, \cite{Couch_15a}, who performed core-collapse simulations
in octant symmetry with and without imposed perturbations, reported
that while the perturbations increased turbulence behind the shock, leading to
enhanced neutrino heating and larger shock radii, they did not qualitatively
change the outcome of their models. Similarly, \cite{OConnor_18}, using
full 3D simulations, found that progenitor perturbations led to
increased shock radii and higher neutrino heating, but their models
ultimately did not explode. 
\cite{Muller_17} performed three core-collapse simulations, two of which
were based on 3D progenitors and one on a spherically symmetric progenitor.
They found that large-scale structures
in the oxygen shell were crucial for explosion: the models based on
3D progenitors exploded, whereas the model based on the
1D progenitor did not. \cite{Muller_17} traced the explosions to advantageous
large-scale forcing and found that
pre-collapse perturbations reduce the critical neutrino luminosity
required for shock revival by approximately 20\%.
\cite{Bollig_21} collapsed the 18.88 solar masses ($M_\odot$)
progenitor evolved in three dimensions by \cite{Yadav_20}
self-consistently through bounce, obtaining a successful explosion with an energy of
$\sim 10^{51}\,$erg. \citet{Vartanyan_22} carried out direct
comparisons between 1D and 3D progenitors, using the 12.5\,$M_\odot$ progenitor
of \cite{Muller_19} and the 15\,$M_\odot$ progenitor of
\cite{Fields_20} with the \textsc{Fornax} code. They 
found that models initialised from 3D progenitors were
more prone to explosion than their 1D
counterparts.

Recently, based on axisymmetric simulations, \cite{Chen_26} found that
progenitor perturbations do not significantly affect the outcome of their
models. Instead, they reported that nuclear burning during collapse generates
small-scale perturbations that are amplified in the collapsing core, seeding
convection in a manner similar to imposed progenitor perturbations. Furthermore, they
reported that the standing accretion shock instability (SASI)
\citep{Blondin_03} and large-scale shock deformations are the primary drivers
of turbulence, independent of the progenitor structure.
It is not clear that the results of \cite{Chen_26} can be directly extrapolated from axisymmetric simulations to full 3D models.
Turbulence is fundamentally different in two and three dimensions. Turbulent energy cascades
towards smaller scales in three dimensions, but this energy cascade is reversed in two dimensions (2D).
The inverse cascade leads to the development of large-scale fluid patterns in axisymmetric simulations \citep{Hanke_12,Takiwaki_14}. At least partly due to the different nature of turbulence in two
and three dimensions, axisymmetric simulations are known to be more ``explodable'' than their 3D counterparts \citep{Takiwaki_14}.

In this work, we follow \citet{Fields_20,Fields_21} and perform
3D stellar burning simulations using \textsc{FLASH}.
We carry out five simulations, mapped from 1D stellar evolution models approximately
$10$--$15$ minutes prior to core collapse. The stellar evolution models were computed with Modules for Experiments in Stellar Astrophysics (\textsc{MESA}) 
\citep{Paxton2011}.
We compare the stellar evolution in \textsc{FLASH} to the corresponding
\textsc{MESA} models.

From the five stellar models, we select one progenitor and perform three
fully 3D core-collapse simulations. We self-consistently follow the evolution
of the fully 3D progenitor through collapse, bounce,
and shock revival. In addition, we evolve a model initialised from
the angle-averaged version of the 3D progenitor, as
well as a model based directly on the corresponding
\textsc{MESA} simulation. We investigate how differences in the stellar
structure impact the subsequent core-collapse evolution in the three models.

This paper is structured as follows. In Section~\ref{sec:methods}, we describe the numerical 
methods, including the stellar evolution models, the mapping procedure, and the 3D 
hydrodynamic simulations. In Section~\ref{sec:shell}, we present the properties of the 
convective shell-burning phase and characterise the resulting multi-dimensional structure of the 
progenitors. In Section~\ref{sec:cc}, we analyse the core-collapse simulations and compare 
the evolution of models initialised from 1D and 3D progenitors. Finally, we summarise our findings and give our conclusions in Section~\ref{sec:conclusions}.

\section{Computational Methods} \label{sec:methods}
We evolve our models in three steps. First, stellar models are evolved
through the main sequence and subsequent shell-burning phases using
\textsc{MESA}. The models are followed until the onset of iron core
collapse. We then select a snapshot $\sim10$--$15$ minutes prior to
collapse, map the corresponding \textsc{MESA} model into \textsc{FLASH}
and follow the final minutes of evolution in 3D.
Finally, for one progenitor we perform three core-collapse supernova simulations: one starting from the original \textsc{MESA} model and two based on the 3D \textsc{FLASH} model. These simulations are evolved through core collapse
and shock revival using \textsc{FLASH}.

\subsection{Stellar evolution models}
The progenitor models are evolved using version 24.08.01 of the stellar evolution code
\textsc{MESA}
\citep{Paxton2011, Paxton2013, Paxton2015, Paxton2018, Paxton2019, Jermyn2023, Itoh1996, Cyburt2010, Angulo1999, Chugunov2007, Fuller1985, Oda1994, Langanke2000, Iglesias1993, Iglesias1996, Poutanen2017, Cassisi2007, Blouin2020, Ferguson2005, Irwin2004, Timmes2000, Saumon1995, Potekhin2010, Rogers2002, Jermyn2021, Eggleton1983, Ritter1988, Dewitt1973, Salpeter1954, Alastuey1978, Itoh1979}. 
We compute non-rotating stellar
models with zero-age main-sequence (ZAMS) masses of $20$, $21.5$, $24.5$, $26$, and
$29\,$$M_{\odot}$ at solar metallicity ($Z = 0.0142$) using the Asplund
abundances \citep{asplund_09}.
Mass loss is included using the Dutch wind scheme with the
low-temperature prescription of \cite{dejager_88} and a scaling
factor of $0.8$. Nuclear burning is followed using the
\texttt{approx21\_cr56} (hereafter referred to as \texttt{approx21}) reaction network. Convective energy transport is
treated using mixing-length theory (MLT) with $\alpha=1.5$ and the
time-dependent convection formulation of \cite{kuhfuss_86}. See
\cite{Jermyn2023} for more information regarding the \textsc{MESA} implementation.
The models are evolved from the ZAMS until the onset of
iron core collapse.

\subsection{Shell Burning in \textsc{FLASH}}
The 3D simulations were carried out with a modified
version~\citep{Dubey_09,Couch_13,Couch_14,Fields_20,Fields_21} of
the \textsc{FLASH4} simulation framework \citep{Fryxell_2000}. The
simulations use a 3D Cartesian grid with adaptive mesh
refinement. The computational domain extends to
$\pm 10^{10}\,$cm from the origin in each dimension. Each model
uses up to $9$ levels of refinement with a finest grid spacing of
$\sim24.4\,$km. The $24.5\,M_\odot$ model employs a slightly
different grid configuration, achieving a finest spacing of
$\sim19.5\,$km. We enforce a maximum effective angular resolution of $1.1^\circ$, which
limits the maximum allowed refinement at larger radii. Refinement is
triggered by gradients in density, pressure, and velocity. The hydrodynamics is evolved with Spark~\citep{couch_21}, a second-order-in-time Runge--Kutta based method of lines solver.  We use the nominally 5th-order WENO5-Z method \citep{borges_08} for the reconstruction and the HLLC Riemann solver~\citep{Einfeldt_88}.
Self-gravity is computed using a multipole expansion
assuming a spherically symmetric (monopole, $\ell = 0$)
gravitational potential \citep{Couch_13}.
We employ the Helmholtz equation of state
\citep{Timmes2000} and the same \texttt{approx21} nuclear reaction
network used in the \textsc{MESA} calculations. While we use the same nuclear network
in the \textsc{FLASH} and \textsc{MESA} simulations, the exact implementation of the network differs
between the two codes. Our simulation setup closely
follows previous stellar evolution simulations carried out with
\textsc{FLASH} \citep{Fields_20,Fields_21} and we refer the reader to these works for further details regarding the exact numerical implementation.

The general structure of the stellar models considered in this work consists of an iron core surrounded by nuclear burning shells.
The nuclear burning regions above the iron core can contain both
stable layers and convectively unstable shells. Typically, the iron core is followed by a Si-burning shell, then an O-burning shell, which is then surrounded by an extended O-rich layer. 
The exact extent of
these regions is model-dependent, and convectively unstable Si-burning and O-burning shells
are not generic features of all the progenitors considered in this work.

To break spherical symmetry and seed convection, we impose
divergence-free velocity perturbations,
$\nabla \cdot (\rho\, \mathbf{v}) = 0$ \citep{Muller_15,OConnor_18,Fields_20,Fields_21}. Here $\rho$ is the fluid density and $\mathbf{v}$ the fluid velocity. We define three perturbation
layers going outward from the iron core. For each model we determine
four radii ($r_1$, $r_2$, $r_3$, and $r_4$) from the \textsc{MESA}
progenitor at the time of mapping, which define the radial extent of
these layers:
\begin{enumerate}
    \item \textbf{Si-burning shell} ($r_1$ to $r_2$): the region where
    silicon is actively burned into iron-group elements.
    \item \textbf{O-burning shell} ($r_2$ to $r_3$): the region where
    oxygen is burned into silicon and sulphur.
    \item \textbf{O-rich layer} ($r_3$ to $r_4$): the extended region dominated by $^{16}$O, which is convectively unstable in our models.
\end{enumerate}
The boundary $r_2$ between the Si- and O-burning shells is defined as
the radius where $X(^{28}\mathrm{Si})$ drops below
$0.2$ moving outward. The boundary $r_3$ between the O-burning shell
and the O-rich layer is defined as the radius where the $^{16}$O
mass fraction exceeds $X(^{16}\mathrm{O}) = 0.6$. The inner and outer
bounds, $r_1$ and $r_4$, are determined by inspection of the stellar
structure profiles for each progenitor. The boundaries for each progenitor are given in Table~\ref{tab:layers}.

For each layer, we write the velocity perturbations as
\begin{equation} \label{eq:dv}
\delta \mathbf{v} = \frac{C}{\rho} \nabla \times \boldsymbol{\psi},
\end{equation}
where $C$ is a normalisation constant that sets the amplitude
of the perturbations and
\begin{equation} \label{eq:psi}
\boldsymbol{\psi} = \mathbf{e}_\phi \frac{\sqrt{\sin\theta}}{r}\sin\left(
n\pi \frac{r-r_{\min}}{r_{\max}-r_{\min}}\right)\Re{\Big(Y_{\ell,m}(\theta,\phi)}\Big).
\end{equation}
Here $r$ is the radial coordinate, $r_{\max}$ the upper boundary of
the layer, $r_{\min}$ the lower layer boundary, $n$ denotes the radial order, and
\begin{equation} \label{eq:sph}
Y_{\ell,m}(\theta,\phi) =
\sqrt{\frac{2\ell+1}{4\pi}
\frac{(\ell-m)!}{(\ell+m)!}}
P_\ell^m(\cos\theta)\,e^{im\phi}
\end{equation}
is the spherical harmonic of degree $\ell$ and order $m$. 

\begin{figure}
    \centering
    \includegraphics[width=\columnwidth]{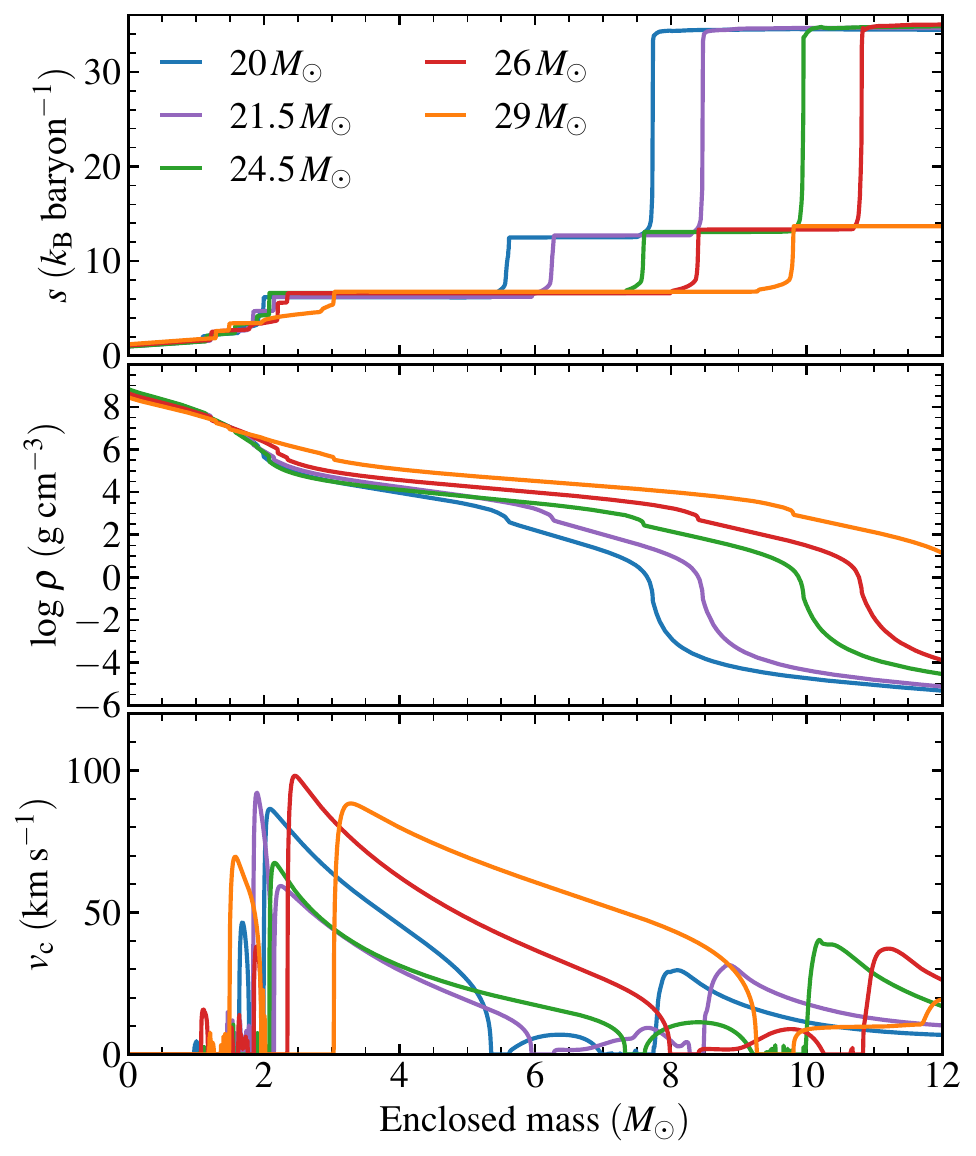}
    \caption{Profiles of the specific entropy
    (top), density (middle), and convective velocity (bottom) as a
    function of enclosed mass for the five \textsc{MESA} progenitor
    models at the time of mapping into \textsc{FLASH}.}
    \label{fig:mesa_structure}
\end{figure}

From Eqs.~\eqref{eq:dv} and~\eqref{eq:psi}, we see that
\begin{equation}
    \delta v \sim \frac{C}{\rho r}.
\end{equation}
We determine $C$ such that the resulting velocity perturbations
correspond to a specified fraction of the convective velocity
($v_c$) in the \textsc{MESA} model at the time of mapping,
\begin{equation} \label{eq:dvscale}
    \delta v \sim \frac{C}{\rho r} = f_i\, v_c,
\end{equation}
where $f_i$ specifies the chosen fraction of the convective velocity
in the $i$-th layer. The convective velocities of the \textsc{MESA} models, at time of mapping, are shown in Fig.~\ref{fig:mesa_structure}. Since Eq.~\eqref{eq:dvscale} depends on radius, we
compute volume-weighted averages of $v_c$, $\rho$, and $r$ from the
\textsc{MESA} model over each radial interval,
\begin{equation}
     \bar{q}_i =
    \frac{\int_{r_i}^{r_{i+1}} q(r)\, r^2 \, \mathrm{d}r}
         {\int_{r_i}^{r_{i+1}} r^2 \, \mathrm{d}r},
\end{equation}
where $i$ denotes the perturbation layer. The normalisation
constant is then
\begin{equation}
    C_i = f_i\,\bar{\rho}_i\,\bar{r}_i\,
    \bar{v}_{c,i}.
\end{equation}
Following \citet{Fields_20}, we adopt $f_i = 0.01$ for the Si-burning
and O-burning shells and $f_i = 0.05$ for the O-rich layer. We set
$n=1$ in all layers. For the Si-burning and O-burning layers we set $(\ell,m)=(9,5)$ and $(\ell,m)=(7,5)$ for the O-rich region.
The layer boundaries and normalisation constants for each
progenitor are listed in Table~\ref{tab:layers}.
Seeding convection using this approach, rather than allowing the finite grid resolution to provide the seeds, avoids unwanted Cartesian grid artefacts during the early onset of convection \citep{Fields_20}.

\begin{table*} 
\centering
\caption{Radial boundaries and perturbation amplitudes used in the
multi-dimensional simulations. Radii are given in km, the perturbation amplitudes are given in $\mathrm{g\,cm^{-1}\,s^{-1}}$}.
\label{tab:layers}
\begin{tabular}{c cccc ccc}
\hline
Model & $r_1$  & $r_2$  & $r_3$ & $r_4$  & $C_{\mathrm{Si\text{-}burn}}$ & $C_{\mathrm{O\text{-}burn}}$ & $C_{\mathrm{O\text{-}rich}}$  \\
\hline
20$\,M_{\odot}$   & 2200 & 2727 & 4009 & 68000 & $3.98\times10^{26}$ & $4.80\times10^{27}$ & $1.70\times10^{28}$ \\
21.5$\,M_{\odot}$ & 2000 & 3197 & 6105 & 68000 & $2.05\times10^{27}$ & $1.40\times10^{28}$ & $9.64\times10^{27}$ \\
24.5$\,M_{\odot}$ & 2000 & 2521 & 4222 & 92000 & $3.83\times10^{26}$ & $6.17\times10^{25}$ & $1.09\times10^{28}$ \\
26$\,M_{\odot}$   & 2000 & 3206 & 5015 & 76000 & $1.86\times10^{27}$ & $2.28\times10^{27}$ & $2.19\times10^{28}$ \\
29$\,M_{\odot}$   & 2000 & 3874 & 7603 & 65000 & $3.18\times10^{28}$ & $5.87\times10^{25}$ & $3.96\times10^{28}$ \\
\hline
\end{tabular}
\end{table*}

\subsection{Core-collapse}
The core-collapse simulations were carried out with the same version of 
\textsc{FLASH} used for the late-stage stellar evolution, but the nuclear burning network was switched off and neutrino transport switched on. In \textsc{FLASH},
neutrino transport is handled with an energy-dependent M1 scheme
\citep{Cardall_12,Shibata_11,oconnor_18b} that evolves three neutrino species: electron neutrinos, electron antineutrinos, and a third species representing all heavy-lepton neutrinos and their antineutrinos. Neutrino opacities are calculated with the \textsc{NuLib} library \citep{OConnor_15}, using the
standard set of opacities outlined in \cite{OConnor_18} with the addition of mean-field and virial corrections \citep{horowitz_17}, and inelastic scattering on electrons \citep{bruenn_85}. We solve the Newtonian hydrodynamic equations, but with a modified general-relativistic effective potential (case~A of \citealt{Marek_06}). We use adaptive mesh refinement with 12 refinement levels and a finest grid resolution of $\sim610\,$m, and we keep the same $1.1^{\circ}$ effective angular resolution used in the shell-burning simulations. We use SFHo for the high-density equation of state (EOS) \citep{steiner_13}.
We refer the reader to \cite{Dubey_09,Couch_13,Couch_14,OConnor_18,oconnor_18b} for additional details regarding the code.

\section{Pre-collapse Shell Burning} \label{sec:shell}
The structure for each of the five \textsc{MESA} models at the time of mapping is shown in Fig.~\ref{fig:mesa_structure}.
The density outside $\sim2\,M_{\odot}$ is ordered by the progenitor ZAMS mass (and also by compactness; see Table~\ref{tab:shell_masses}). However, the density ordering does not hold for the innermost regions of the stars. The central density of the models peaks at
$\sim10^9\mathrm{g}\,\mathrm{cm}^{-3}$. The entropy per baryon sits between 6 and 8 $k_b\,\mathrm{baryon}^{-1}$ ($k_b$ denotes the Boltzmann constant) in the oxygen-rich layer and decreases towards the centre.

\begin{figure*}
    \centering
    \includegraphics[width=1\linewidth]{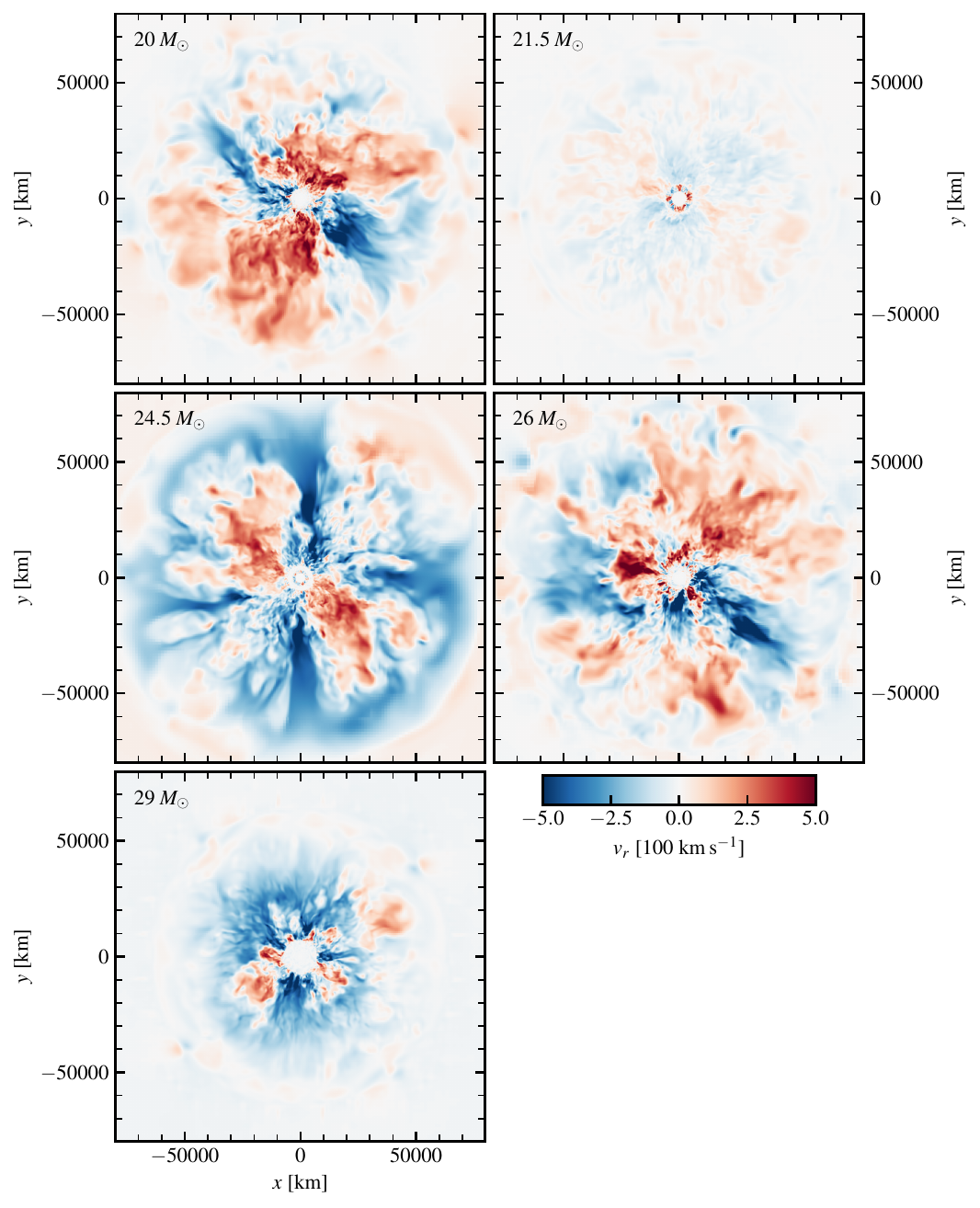}
    \caption{Slices of the radial velocity through the centre of the
    computational domain for all five progenitor models
    $\sim10\,$s prior to core collapse. The spatial scale is
    chosen to show the full extent of the O-burning shell. Red
    indicates outward-directed flow and blue indicates inward-directed
    flow.}
    \label{fig:vr_outer}
\end{figure*}

\begin{figure*}
    \centering
    \includegraphics[width=1\linewidth]{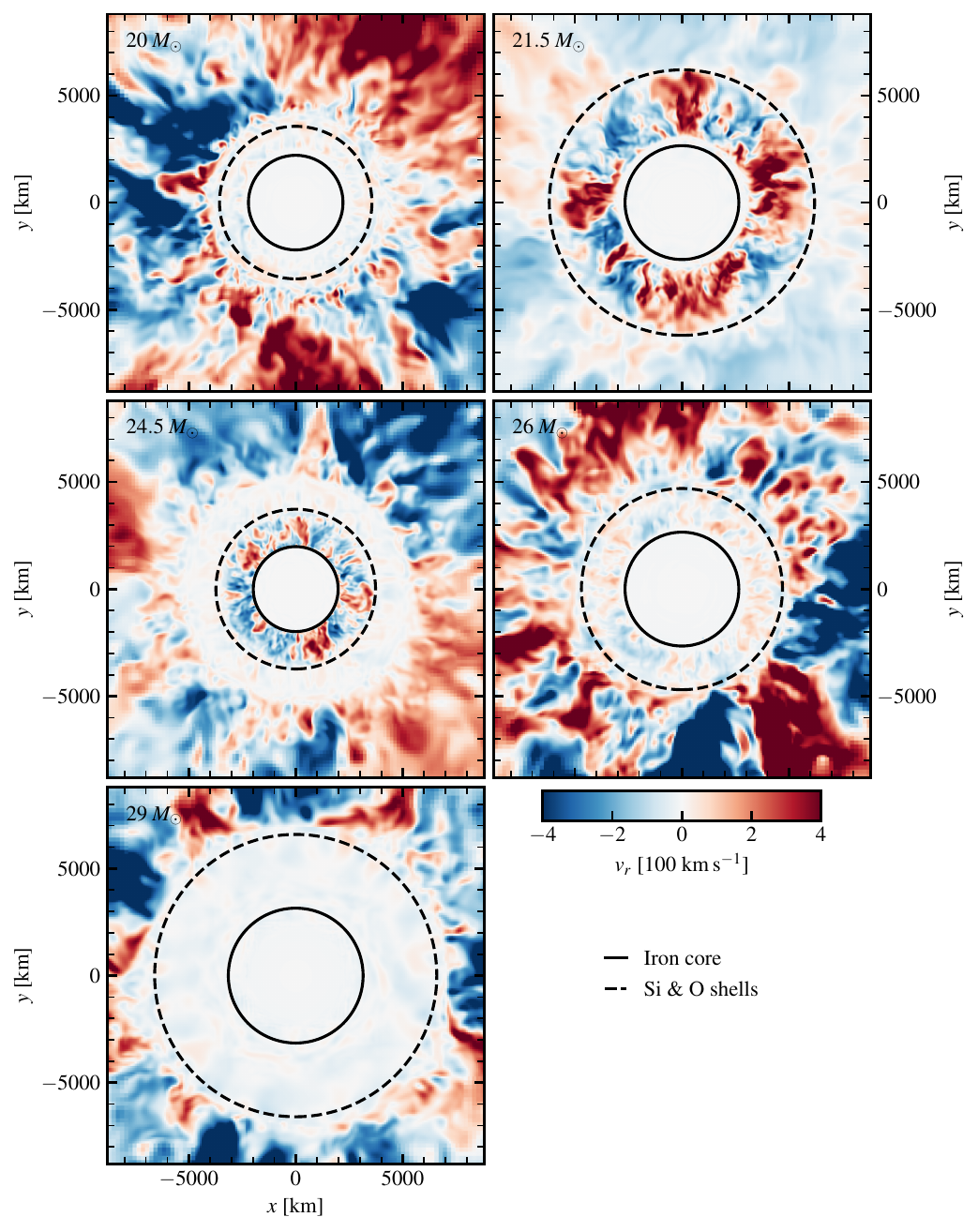}
    \caption{Same as Fig.~\ref{fig:vr_outer} but zoomed in on the
    inner core and burning shells. The solid black circles indicate
    the edge of the iron core, defined as the radius where the
    iron-group mass fraction drops below $0.75$. The dashed black
    circles enclose the Si- and O-burning shells (the start of the O-rich layer). The boundary is defined as the radius where $X(^{16}\mathrm{O}) > 0.6$. The
    enclosed masses at these boundaries are listed in
    Table~\ref{tab:shell_masses}. }
    \label{fig:vr_inner}
\end{figure*}

The velocity perturbations imposed at the time of mapping break spherical
symmetry and lead to the development of convection, after an initial transient
phase of $\sim$150\,s. For reference, the simulations were evolved for
15.4, 15.4, 11.7, 14.1, and 12.0 minutes for the 
20, 21.5, 24.5, 26, and 29 $M_{\odot}$ models, respectively.
Figs.~\ref{fig:vr_outer} and~\ref{fig:vr_inner} show the
radial velocity in the $xy$-plane for all five models $\sim10\,$s prior to core
collapse. We define the
onset of collapse as the time at which the central density begins increasing
by more than a quarter of a decade per second
($\mathrm{d}\log_{10}\rho_{\max}/\mathrm{d}t > 0.25\,$s$^{-1}$).
At the moment the criterion is satisfied, the maximum infall velocities of the stellar cores are $\sim25$--$60\,\mathrm{km\,s^{-1}}$. The low infall velocities indicate that our
criterion captures the early onset of collapse\footnote{A common criterion for terminating 1D stellar
evolution calculations is an infall velocity of 
$1000\,\mathrm{km\,s^{-1}}$ in the
iron core \citep{woosley_07}. While the core collapse is underway and
accelerating for all our models, our 3D simulations do not
reach this threshold. Given the available computational resources, we prioritised evolving
all models well into the onset of collapse rather than following a
subset to higher infall velocities.
At the end of the simulations, the maximum infall
velocities in the inner $2000\,$km range from 
$\sim400\,\mathrm{km\,s^{-1}}$ to $\sim800\,\mathrm{km\,s^{-1}}$. The central densities at the end of
the simulations lie in the range $(1.5$--$4.5) \times 10^{9}\,$g$\,$cm$^{-3}$. We
expect that all models would reach infall velocities of $1000\,\mathrm{km\,s^{-1}}$
within a fraction of a second from the end of the simulations. }.

Fig.~\ref{fig:vr_inner} shows the inner $9000\,$km of the star, while
Fig.~\ref{fig:vr_outer} shows nearly the full simulation domain to highlight
the oxygen-rich layer. All models develop convection in the outer oxygen-rich
layer, with convective plumes reaching several hundred $\mathrm{km\,s^{-1}}$ in four of
the five models ($20$, $24.5$, $26$, and $29\,M_\odot$). The $21.5\,M_\odot$
model is an outlier, with much weaker O-layer convection than 
in our other models. 
The \textsc{MESA} progenitors show the same qualitative behaviour: the
$21.5\,M_\odot$ model has the lowest convective velocity in the oxygen-rich
layer of the five models, both at the time of mapping and at collapse. The
relative difference is, however, considerably smaller than in three dimensions.
In \textsc{MESA} the volume-averaged convective velocity in the oxygen-rich layer of model $21.5\,M_\odot$ is at most a factor of $\sim2$ below the other models.
Models $20$, $21.5$, $24.5$, and $26\,M_\odot$ additionally develop convection in the inner burning shells.

\begin{figure}
    \centering
    \includegraphics[width=\columnwidth]{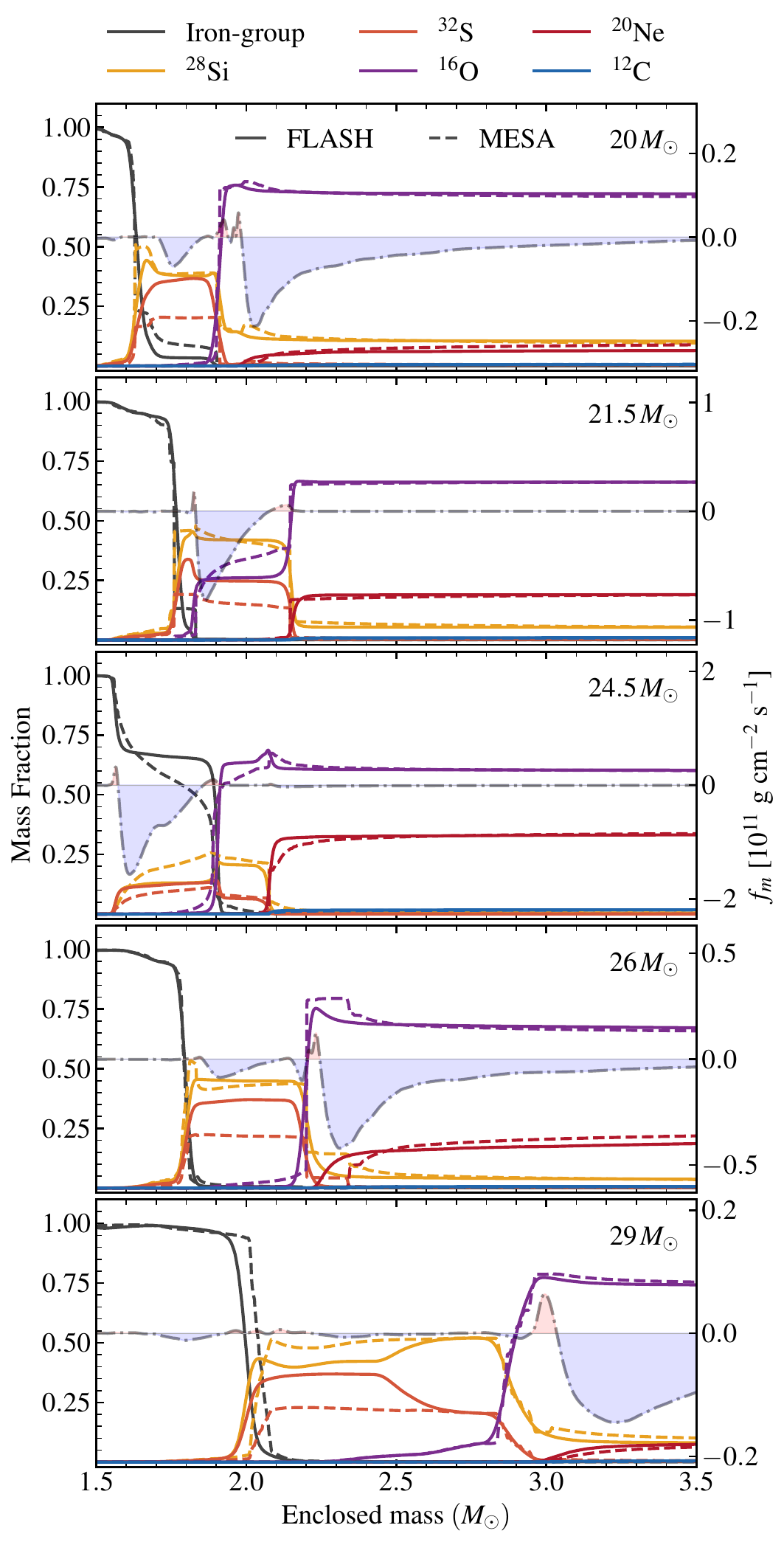}
    \caption{Mass fractions $10\,$s before the onset of collapse, as a function of enclosed mass for the
    five progenitor models and zoomed in on the inner burning shells surrounding the iron
    core. The shaded regions and thin black line show the turbulent
    mass flux $f_m$ (right axis), with red and blue indicating
    positive and negative values, respectively.}
    \label{fig:composition_inner}
\end{figure}

To identify which inner burning layers become convective, we overplot the
turbulent mass-flux on the composition profiles in
Fig.~\ref{fig:composition_inner}. 
The turbulent mass-flux is defined as 
\begin{equation}\label{eq:turb_mass_flux}
    f_m(r) = \langle \rho' v_r' \rangle
           = \langle \rho\,v_r \rangle
             - \langle \rho \rangle \langle v_r \rangle,
\end{equation}
where $\langle \cdot \rangle$ denotes the spherical average and primes denote
deviations from that average \citep{Reynolds_95} (see, for example, \citealt{Nordlund_09,Viallet_13,Andresen_17}).
A negative turbulent mass flux signifies a convectively active
region: underdense fluid elements ($\rho' < 0$) rise
($v_r' > 0$) while overdense elements ($\rho' > 0$) sink
($v_r' < 0$), so that $\rho' v_r' < 0$ in both cases. Positive
values of $f_m$ instead track overshooting regions. Above the
upper boundary of a convective zone, rising plumes are denser
than their stably stratified surroundings ($\rho' > 0$,
$v_r' > 0$), while below the lower boundary sinking plumes are
less dense than the surrounding medium ($\rho' < 0$,
$v_r' < 0$).
Model $24.5\,M_\odot$ is the only one to
develop unambiguous convection in the Si-burning shell, evident as a strong
negative turbulent mass-flux in the region dominated by $^{28}$Si and
$^{32}$S, where iron-group elements are also being produced. The $20$ and
$26\,M_\odot$ models show a weaker turbulent mass-flux in the Si-shell,
indicating far less vigorous Si-shell convection than in the $24.5\,M_\odot$
model.
The $21.5\,M_\odot$ model develops convection in the O-burning shell.
Its $^{16}$O profile shows an initial drop followed by a flattening, and this
flattened region coincides with a negative turbulent mass-flux extending down
to the Si/O interface. No iron-group elements are produced there, consistent
with O-shell rather than Si-shell convection. The $26\,M_\odot$ model may also
exhibit weak O-burning convection, indicated by a small dip in the turbulent
mass flux near the $^{16}$O composition drop (fourth panel of
Fig.~\ref{fig:composition_inner}). The $29\,M_\odot$ model shows no evidence of
convective activity interior to the O-rich layer.

Using a simulation setup very close to ours, \cite{Fields_21} found that the convective velocities were $\sim2$--$5$ times larger in their 
3D \textsc{FLASH} simulations than in the corresponding \textsc{MESA} models, particularly in the oxygen-rich layer. Similar results have been reported in several other studies \citep{muller_16,Jones_17,Georgy_24,Griffiths_26a}.
Following \cite{Fields_21}, we calculate the
convective velocity from the tangential velocity
\begin{equation} \label{eq:vt}
v_c\sim\sqrt{\langle{v}_{t}^2\rangle} = \sqrt{\langle |\mathbf{v}|^2 - v_r^2\rangle}.
\end{equation}
We also estimate the convective velocity as
\begin{equation} \label{eq:vrtrb}
    v_c = \sqrt{\langle  v'^2_r\rangle} = \sqrt{\langle  (v_r-\langle v_r\rangle)^2\rangle},
\end{equation}
which accounts for turbulent motions in the radial direction.
Lastly, we estimate the convective velocity predicted by standard MLT for the hydrodynamic structure of our \textsc{FLASH}
simulations. Following the formulation of \cite{muller_16}, 
the Brunt--V\"ais\"al\"a frequency, during Ledoux convection, is
\begin{equation}
\omega_\mathrm{BV}^2 = g\bigg(\frac{1}{\langle\rho\rangle}\frac{\partial \langle\rho\rangle}{\partial r}
-\frac{1}{\langle\rho\rangle \langle c_s\rangle^2}\frac{\partial \langle P\rangle}{\partial r}\bigg),
\label{eq:wbv}
\end{equation}
where $\langle\rho\rangle$, $\langle P\rangle$, $\langle c_s\rangle$, and $g$ are the angle-averaged density,
pressure, sound speed, and local gravitational acceleration.
$\omega_\mathrm{BV}^2 > 0$ marks convectively unstable regions. 
The mixing length is set to one pressure scale height under the assumption of hydrostatic equilibrium,
\begin{equation}
\Lambda_\mathrm{mix} = H_P = -\langle P\rangle\bigg(\frac{\partial \langle P\rangle}{\partial r}\bigg)^{-1}= \frac{\langle P\rangle}{\langle\rho\rangle g}.
\label{eq:hpbig}
\end{equation}
Balancing the buoyancy work over
one mixing length against the kinetic energy of the parcel gives the
MLT estimate of the convective velocity,
\begin{equation}
v_c = \alpha\omega_\mathrm{BV}\Lambda_\mathrm{mix}
                 = \alpha\bigg(g\Lambda_\mathrm{mix}\frac{\delta\rho}{\rho}\bigg)^{1/2},
\label{eq:vconv}
\end{equation}
where $\alpha$ is a dimensionless mixing-length parameter of order unity. We
set $\alpha = 1.5$ (matching the \textsc{MESA} setup), and ${\delta\rho}/{\rho} = \Lambda_\mathrm{mix}\omega_\mathrm{BV}^2/g$ denotes the relative density contrast between convective plumes and the spherically averaged background. Note that 
Eq.~\eqref{eq:vconv} is only valid where
$\omega_\mathrm{BV}^2 > 0$; for plotting purposes we show $\alpha\,|\omega_\mathrm{BV}|\,\Lambda_\mathrm{mix}$.
While our \textsc{MESA} simulations handled convective energy transport with a
time-dependent convection formulation \citep{kuhfuss_86,Jermyn2023} in addition to
MLT with $\alpha_{\rm MLT}=1.5$, it is interesting to see how well the convective velocities we observe in the simulations agree with predictions from standard MLT. 
The convective velocities in the burning shells are time and model dependent in both the \textsc{MESA} and \textsc{FLASH} simulation sets: in some models they grow monotonically towards collapse, while in others they peak and then decline by as much as 
$50\,\%$ from their maximum. The oxygen-rich layer behaves more uniformly, with the
convective velocity rising gradually up to collapse in all five models and in
both codes.

\begin{figure}
    \centering
    \includegraphics[width=1\linewidth]{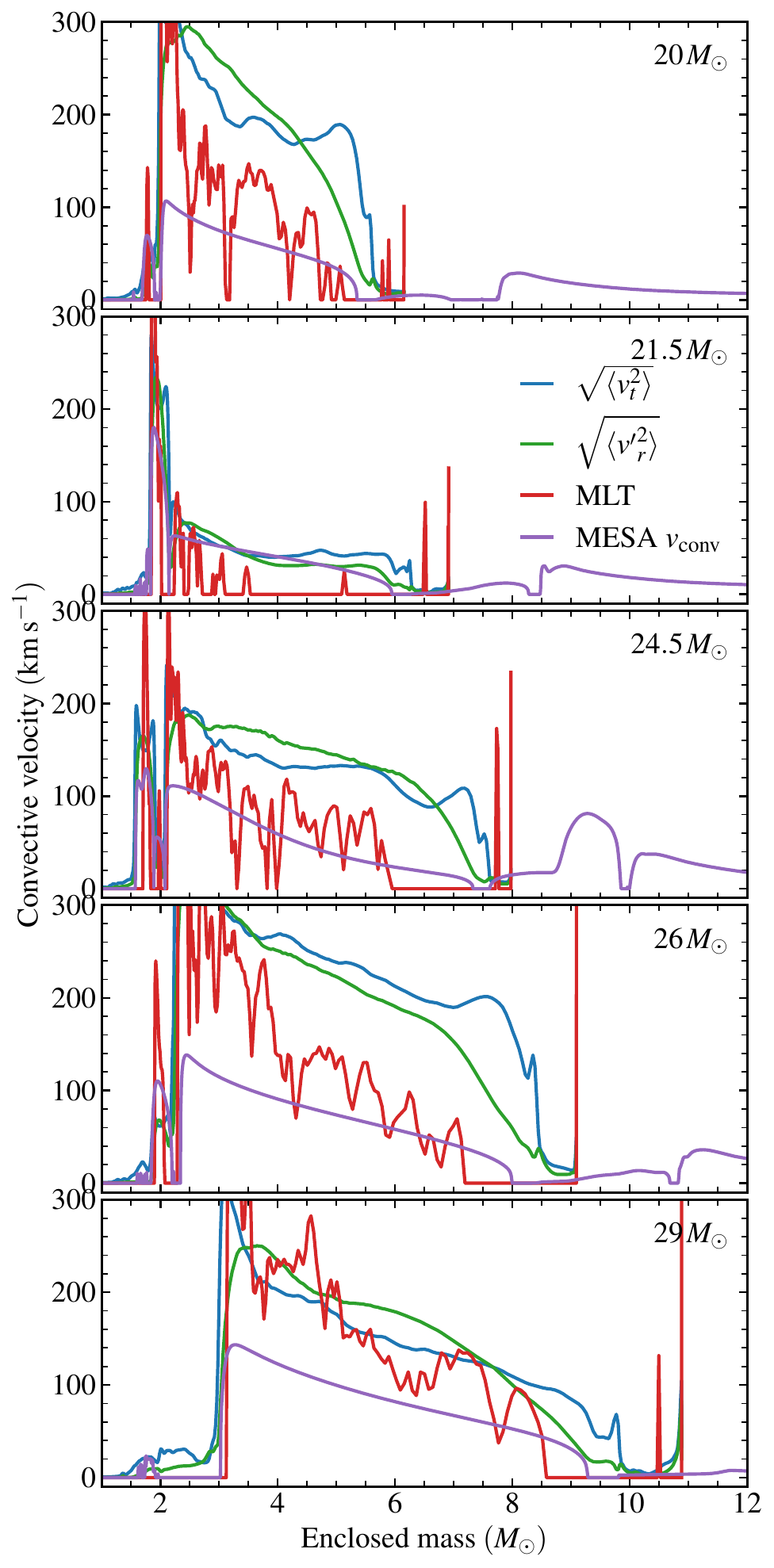}
    \caption{Convective velocity estimates for our \textsc{FLASH} simulations and the convective velocity
    from the \textsc{MESA} models. The blue lines represent Eq.~\eqref{eq:vt}, green lines
    show the estimate according to Eq.~\eqref{eq:vrtrb}, red lines show the MLT result calculated from Eq.~\eqref{eq:vconv},
    and the purple lines show convective velocities from the \textsc{MESA} models. Each panel represents one model, with the model name
    indicated in the top right-hand corner. The figure shows profiles $10\,$s prior to collapse.}
    \label{fig:conv}
\end{figure}

Fig.~\ref{fig:conv} shows the three estimates of the convective velocity in our
\textsc{FLASH} simulations, and the convective
velocity from the \textsc{MESA} models. We show the profiles $10\,$s before the onset of collapse.
The \textsc{FLASH} models show systematically larger (a factor of $\sim 1.5$--$4$) convective velocities in the O-rich layer than
the \textsc{MESA} models, with the exception of the $21.5\,M_{\odot}$ progenitor
(for which the two codes agree to within $\sim50\%$ across the layer). The predictions from MLT tend to lie between the \textsc{MESA} and \textsc{FLASH} results, with
the exception of model $29\,M_{\odot}$ where the MLT prediction agrees well with the estimates from Eq.~\eqref{eq:vt} and Eq.~\eqref{eq:vrtrb}.
Again, model $21.5\,M_{\odot}$ is an outlier: here MLT predicts convective velocities close to
zero throughout large parts of the oxygen-rich layer, while both \textsc{FLASH} and \textsc{MESA}
give convective velocities of a few tens of km per second.

The agreement between the convective velocity
in the \textsc{MESA} and \textsc{FLASH} simulations is much better in the inner burning shells where we find 
that the estimates from Eq.~\eqref{eq:vrtrb} agree with the \textsc{MESA} models to within 
$\sim\,10\%$. The 
two exceptions are model $26\,M_{\odot}$, where \textsc{FLASH} and \textsc{MESA} agree 
within a factor of $2$, and model $29\,M_{\odot}$, which
does not develop strong convection in the inner shells.

The relative poor agreement between the two codes in the O-rich layer is likely a consequence of the duration of our
3D simulations.
To truly capture the properties of convection in the O-rich layer one would need significantly longer simulations. The Si-burning and O-burning shells, for the models where convection develops, have
convective turnover times of $\tau_{\rm conv}\sim 15$--$30\,$s and reach a quasi-steady state within a few hundred seconds, completing
tens of convective turnovers over the course of the simulations. The 
O-rich layer has significantly longer turnover times, 
$\tau_{\rm conv} \sim 100$--$200\,$s, and completes at most a few turnovers by the end of the simulations. 
It has likely not reached a fully developed   
convective steady state. We have calculated the convective turnover times
as $\tau_{\rm conv}\sim 2 h/v_c$, where $h$ is the height of the convective layer estimated 
from Fig.~\ref{fig:vr_outer} and with $v_c$ from Eq.~\eqref{eq:vt}.

\begin{table*}
\centering
\caption{Iron core and burning shell properties at the end of the
\textsc{FLASH} simulations and the corresponding results from the
\textsc{MESA} simulations. The iron
core edge is defined where the iron-group mass fraction drops below
$0.75$, and the O-rich boundary where $X(^{16}\mathrm{O}) > 0.6$.
The boundaries between layers are not sharp interfaces and the extent of the iron core depends on the definition we chose. In particular, model $24.5\,M_{\odot}$ has an extended region outside the iron core where the mass fraction of iron-group elements remains high.
The Si+O shell mass is the difference between the enclosed masses at the
two boundaries. $\xi_{2.5}$ is the compactness
parameter evaluated at a baryonic mass coordinate of $2.5\,M_\odot$.}
\label{tab:shell_masses}
\begin{tabular}{c c cc cc cc cc cc}
\hline
 & $\rho_c$ [$10^9\,$g\,cm$^{-3}$] & \multicolumn{2}{c}{$r_{\rm IC}$ [km]} & \multicolumn{2}{c}{$M_{\rm IC}$ [$M_\odot$]} & \multicolumn{2}{c}{$r_{\rm O}$ [km]} & \multicolumn{2}{c}{$M_{\rm Si+O}$ [$M_\odot$]} & \multicolumn{2}{c}{$\xi_{2.5}$} \\
$M_{\rm ZAMS}$ [$M_\odot$] & \textsc{FLASH} & \textsc{FLASH} & \textsc{MESA} & \textsc{FLASH} & \textsc{MESA} & \textsc{FLASH} & \textsc{MESA} & \textsc{FLASH} & \textsc{MESA} & \textsc{FLASH} & \textsc{MESA} \\
\hline
20   & 2.99 & 1721 & 1708 & 1.62 & 1.63 & 3065 & 2880 & 0.30 & 0.28 & 0.21 & 0.23 \\
21.5 & 4.31 & 2102 & 2047 & 1.76 & 1.75 & 5987 & 5711 & 0.40 & 0.40 & 0.24 & 0.24 \\
24.5 & 2.38 & 1873 & 2116 & 1.58$^\dagger$ & 1.64$^\dagger$ & 3603 & 3599 & 0.34 & 0.27 & 0.21 & 0.21 \\
26   & 1.86 & 2253 & 2203 & 1.78 & 1.78 & 4367 & 4033 & 0.43 & 0.42 & 0.31 & 0.35 \\
29   & 1.53 & 2648 & 2736 & 1.98 & 2.02 & 6222 & 5813 & 0.94 & 0.86 & 0.57 & 0.59 \\
\hline
\end{tabular}
\begin{tablenotes}
\item{\tiny {$^\dagger$The shape of the iron-group composition profile makes the exact definition of the 
iron core ambiguous for model $24.5\,M_\odot$; see the third panel of Fig.~\ref{fig:composition_inner}.}}
\end{tablenotes}
\end{table*}

Table~\ref{tab:shell_masses} lists the iron core and burning shell properties at the end of each \textsc{FLASH} simulation and the corresponding \textsc{MESA} snapshots, which we determine by matching the central density of the \textsc{MESA} models to that of our \textsc{FLASH} runs. The table shows the iron core radius ($r_{\rm IC}$) and enclosed mass ($M_{\rm IC}$), the radius of the O-rich shell ($r_{\rm O}$), the mass of the Si+O shells ($M_{\rm Si+O}$), and the compactness parameter ($\xi_{2.5}$) evaluated at $M = 2.5\,M_\odot$. 
The compactness parameter \citep{OConnor_11} at a mass coordinate of $M = 2.5\,M_{\odot}$ is given by
\begin{equation}
\xi_{2.5} = \frac{2.5}{R(M_{\mathrm{bary}} = 2.5\,M_\odot) / 1000~\mathrm{km}},
\end{equation}
where $R(M_{\mathrm{bary}} = 2.5\,M_\odot)$ is the radius enclosing a baryonic mass of $2.5\,M_\odot$. We also list the central density ($\rho_c$) in the \textsc{FLASH} simulations.
The iron core edge is defined as the location where the iron-group mass fraction drops below $0.75$, while the O-rich boundary is defined by $X(^{16}\mathrm{O}) > 0.6$. The mass of the Si and the O shells is given by the difference in enclosed mass between these two boundaries.

At the end of our \textsc{FLASH} simulations, we find central densities of $\sim(1.5$--$4.5) \times 10^9\,\mathrm{g\,cm^{-3}}$. 
The iron core radii lie in the range $\sim1700$--$2700\,\mathrm{km}$, with enclosed masses between $1.6$ and $2.0\,M_\odot$. The O-rich shell is located at radii of $\sim3000$--$6200\,\mathrm{km}$, and the Si+O shell masses range from $\sim 0.3$ to $0.94\,M_\odot$.
The compactness parameter $\xi_{2.5}$ spans $\sim 0.2$--$0.6$ and the
\textsc{MESA} simulations tend to be more compact than the \textsc{FLASH} simulations. However, the differences are relatively small, up to $\sim 0.04$ and several models show identical compactnesses across the \textsc{MESA} and \textsc{FLASH} runs.

With only five models, it is difficult to establish trends in terms of differences between the two simulation sets. However, \textsc{MESA} tends to produce slightly more massive iron cores and less massive burning shells when compared to \textsc{FLASH}. The strong convective activity in the 
Si-burning shell in model $24.5\,M_\odot$ leads to a flattening of the iron mass fraction curve and the iron core is $\sim250\,$km smaller in the \textsc{FLASH} model compared to the 
\textsc{MESA} simulation. Similarly, in the two $29\,M_\odot$ models we see that the iron core is $\sim90\,$km larger in the \textsc{MESA} simulation than in the corresponding \textsc{FLASH} model. The differences cannot be explained by strong convection, since this model does not develop vigorous convection in the inner burning shells. It is, therefore, likely that the larger iron cores seen in the \textsc{MESA} simulations are at least partly due to differences in the nuclear burning. We see that \textsc{MESA} burns Si at larger mass coordinates than \textsc{FLASH}, which is likely related to discrepancies between the nuclear reaction rates in the two implementations of \texttt{approx21}.

The spherically averaged composition profiles show systematic   
differences between the \textsc{FLASH} and \textsc{MESA} simulations (see Fig.~\ref{fig:composition_inner}).
We observe significant differences between the two codes in the $^{16}$O and $^{20}$Ne mass fractions in 
the oxygen-rich layer for several models, most prominently for models $26$ and $20\,M_{\odot}$.
Furthermore, the $^{32}$S mass fraction in the inner burning shells is
consistently $20$--$50\,\%$ higher in \textsc{FLASH} simulations than in \textsc{MESA} runs.
By comparing the composition profiles at time-steps where the full composition
is available in our \textsc{FLASH} simulations, we found that 
this excess appears to come predominantly at the expense of $^{40}$Ca and $^{28}$Si. The
differences are present in all five progenitors and do not seem to be
directly linked to multi-dimensional effects. 
Furthermore, the differences are not due to differences in the temperature or density profiles,
which agree to within a few per cent between the two codes. 

By comparing the reaction rates implemented in \textsc{FLASH} with those in \textsc{MESA}, 
we found large discrepancies in certain rates, which are the most likely cause of the composition differences. 
We traced these to an issue with the implementation of certain rates in \textsc{MESA}.
The differences in the rates, the source of the differences, and the impact of correcting the rates are 
described in Appendix~\ref{apx:rates}.

\section{Core Collapse} \label{sec:cc}
We performed three 3D simulations of the core-collapse phase of the $24.5\,M_\odot$ model:
one initialised from our multi-dimensional stellar model (\texttt{m24.5-3D}),
one from a corresponding \textsc{MESA} snapshot (\texttt{m24.5-MESA}),
and one from the angle-averaged 3D progenitor (\texttt{m24.5-$\langle$3D$\rangle$}).
The third model has the same spherically averaged hydrodynamic properties as \texttt{m24.5-3D} but lacks the aspherical perturbations induced by turbulent burning.
As noted above, the \textsc{MESA} snapshots do not perfectly align with the output cadence of the \textsc{FLASH} simulations. 
Model \texttt{m24.5-3D} and model \texttt{m24.5-MESA} were, therefore, initialised at slightly different times.
Because the stellar core is already in a dynamically collapsing state at this stage, this offset leads to different collapse durations. In addition, the structure of the iron core differs between the \textsc{MESA} and \textsc{FLASH} stellar evolution simulations, which will also affect the collapse 
time. \texttt{m24.5-3D} and \texttt{m24.5-$\langle$3D$\rangle$} reach bounce after $0.327\,\mathrm{s}$, compared to $0.240\,\mathrm{s}$ for \texttt{m24.5-MESA}. Consequently, in this section, we normalise the time such that $t=0$ refers to bounce for all three models.

We chose model $24.5\,M_\odot$ because it develops vigorous convection in the Si-burning
shell (see Section~\ref{sec:shell}). Perturbations originating in the
inner burning shells are advected through the shock earlier and with
larger Mach numbers than those originating in the much more extended O-rich
layer, which makes it a useful case for comparing
1D and 3D progenitors.

\subsection{Overall Evolution}
\begin{figure}
    \centering
    \includegraphics[width=\columnwidth]{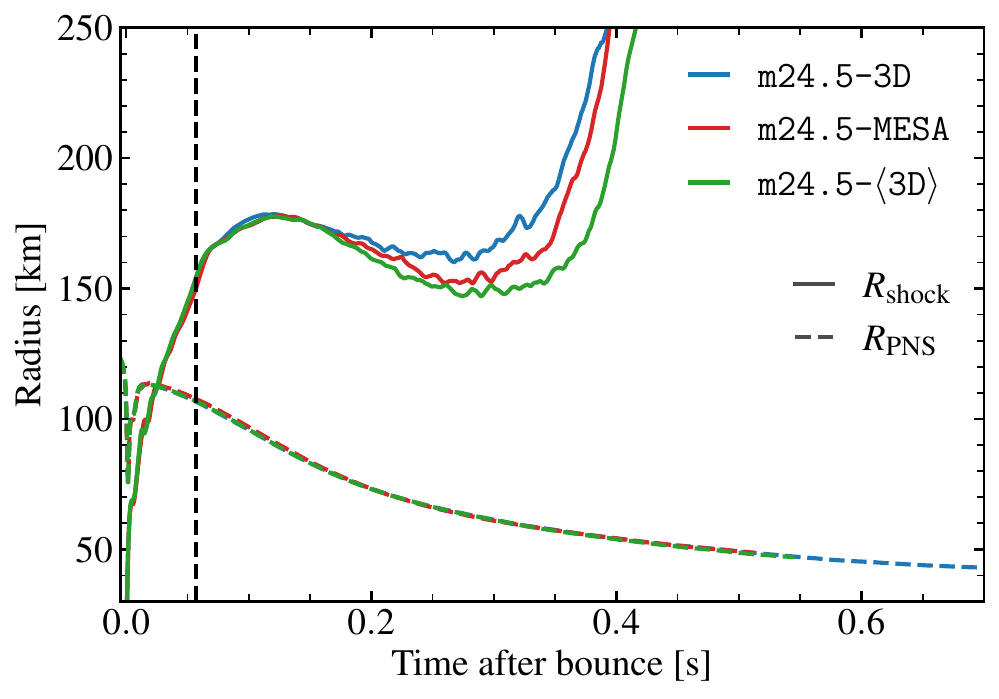}
    \caption{Shock radius (solid lines) and PNS radius (dashed lines)
    as a function of time after bounce for models
    \texttt{m24.5-3D} (blue), \texttt{m24.5-MESA} (red), and
    \texttt{m24.5-$\langle$3D$\rangle$} (green). The vertical
    dashed black line denotes $t=0.057\,$s after bounce, when the
    lower edge of the Si shell falls through the shock.}
    \label{fig:shock_pns}
\end{figure}

Fig.~\ref{fig:shock_pns} shows the evolution of the shock radius and
the protoneutron star (PNS) radius for all three models. Following
bounce, the shock expands to a maximum radius of
$\sim180\,$km before stalling. The shock
evolution of the three models is similar during the first
$\sim0.4\,$s after bounce, with
\texttt{m24.5-3D} maintaining a slightly larger average shock
radius than the other two. All models undergo shock revival by
$\sim0.4\,$s post-bounce, after which the shock expands rapidly.
Shock revival occurs first in model \texttt{m24.5-3D}, followed by 
model \texttt{m24.5-MESA}, and then \texttt{m24.5-$\langle$3D$\rangle$}. 

There are two density jumps in the stellar progenitor that sit at mass coordinates of
$\sim 1.86\,M_{\odot}$ and $\sim 2.09\,M_{\odot}$. The density jump at $\sim 1.86\,M_{\odot}$ is located at the upper
overshooting region of the Si-burning shell and the jump at $\sim 2.09\,M_{\odot}$ sits at the 
interface between the burning shells and the O-rich layer. The shock overtakes the 
first interface around $0.41\,$s after bounce and the second around 
$0.5\,$s, so neither has fallen through by the time shock revival sets in.
It is worth noting here that both
interfaces are smoother in the 3D stellar model, a feature inherited by
\texttt{m24.5-3D} and \texttt{m24.5-$\langle$3D$\rangle$}. However, since the explosion is well underway
when the interfaces fall through the shock, the shape of the interfaces does not play a central role in the dynamics of the explosion.

The PNS radii are nearly indistinguishable between the three
models, contracting steadily from $\sim100\,$km at bounce to
$\sim30\,$km by the end of the simulations. \texttt{m24.5-MESA}
has a marginally larger PNS throughout, most clearly between
$0.05$ and $0.1\,$s after bounce (see Fig.~\ref{fig:shock_pns}).

\subsection{\texttt{m24.5-3D} versus \texttt{m24.5-$\langle$3D$\rangle$}}
We will return to \texttt{m24.5-MESA} below, but for now we turn our attention
to models \texttt{m24.5-3D} and \texttt{m24.5-$\langle$3D$\rangle$} for a clear
comparison of 1D and 3D progenitors. The evolution of the two models is
essentially identical until $0.175\,$s after bounce, at which point the shock radii
of the two models diverge. The shock starts to recede in model \texttt{m24.5-$\langle$3D$\rangle$} and reaches a minimum around $0.3\,$s post-bounce before starting to expand again and eventually reaching shock revival. The shock trajectory of model \texttt{m24.5-3D} is similar to that of model \texttt{m24.5-$\langle$3D$\rangle$}, but the average shock radius is $5$--$10\,$\% larger. Furthermore, the period of shock recession between $0.175\,$s and shock revival is
much less pronounced in model \texttt{m24.5-3D} compared to \texttt{m24.5-$\langle$3D$\rangle$}.
The average shock radius stays relatively flat between $0.2\,$s and the onset of shock
revival in model \texttt{m24.5-3D}.
\begin{figure}
    \centering
    \includegraphics[width=\columnwidth]{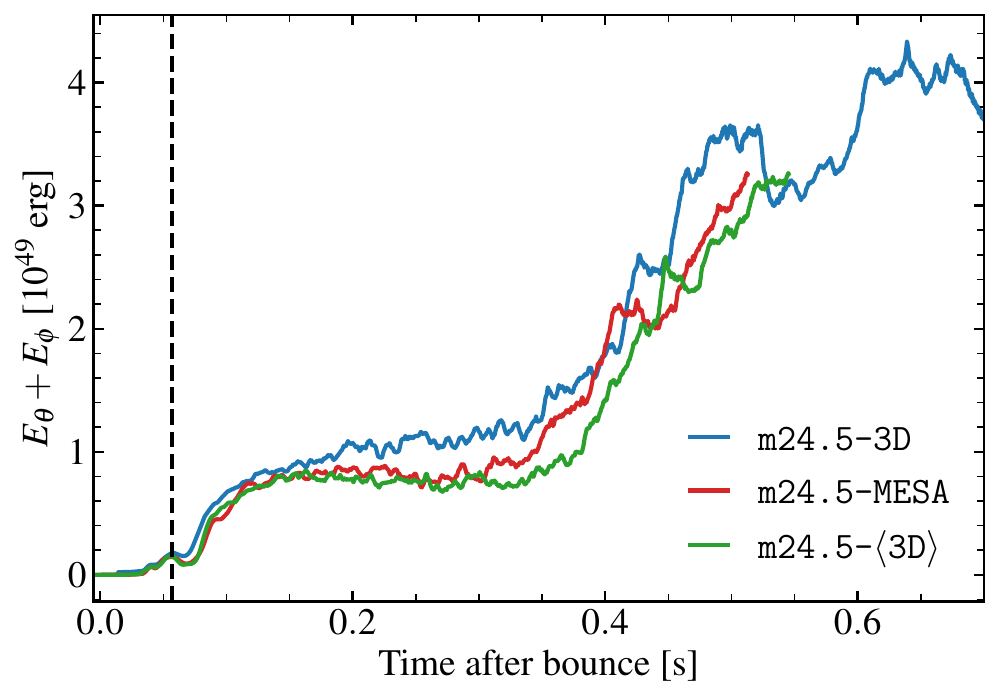}
    \caption{Non-radial kinetic energy ($E_\theta + E_\phi$) as a
    function of time after bounce for all three core-collapse models.
    The vertical dashed black line denotes $t=0.057\,$s after bounce,
    when the lower edge of the Si shell falls through the shock.}
    \label{fig:turb_ke}
\end{figure}

The smaller shock radius of \texttt{m24.5-$\langle$3D$\rangle$} compared to \texttt{m24.5-3D} is accompanied by weaker heating and lower non-radial kinetic energy in the gain layer.
Fig.~\ref{fig:turb_ke} shows the non-radial kinetic energy,
$E_\theta + E_\phi$, as a function of time after bounce. 
The non-radial kinetic energy of the two models starts to diverge $\sim 0.05\,$s 
after bounce, which is when the Si-shell starts to fall through the shock (denoted by the black dashed line). The non-radial kinetic energy of model \texttt{m24.5-$\langle$3D$\rangle$} 
settles at $\sim 0.8\times10^{49}\,$erg around $0.17\,$s post-bounce and remains at this level 
until shock revival. On the other hand, from the point at which the turbulent Si-burning shell starts to fall through the shock, the non-radial kinetic energy grows steadily in model \texttt{m24.5-3D}.
By $0.2\,$s post-bounce the non-radial kinetic energy is
$\sim35\%$ larger in \texttt{m24.5-3D} than in
\texttt{m24.5-$\langle$3D$\rangle$}, and the gap widens to as
much as $50\%$ as the simulations evolve.
\begin{figure}
    \centering
    \includegraphics[width=\columnwidth]{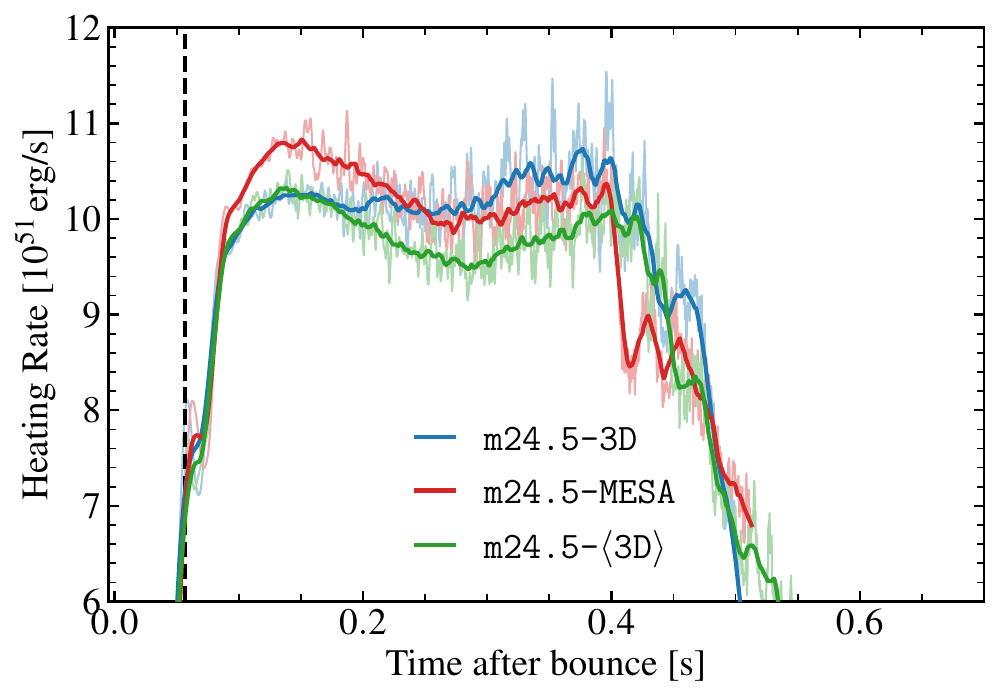}
    \caption{Heating rate in the gain layer as a function of time
    after bounce for all three core-collapse models. The vertical
    dashed black line denotes $t=0.057\,$s after bounce, when the
    lower edge of the Si shell falls through the shock. Before plotting, we apply
    a first-order Savitzky-–Golay filter \citep{scipy} with a $15\,$ms window to smooth the curves. 
    The raw data are shown in the background as low-opacity curves.}
    \label{fig:heating}
\end{figure}
In Fig.~\ref{fig:heating}, we show the total heating rate, which is
defined as the volume integral of the 
change in the internal energy due to neutrino-fluid interactions, 
where the change is positive. The integral is limited to
regions where the density is smaller than $3\times 10^{10}\,$g\,cm$^{-3}$ and the
entropy is greater than $6\,k_b$ per baryon.

As was the case for the shock radius and the non-radial kinetic
energy, the heating rate is initially identical in
\texttt{m24.5-3D} and \texttt{m24.5-$\langle$3D$\rangle$}. The
two curves deviate from $\sim0.175\,$s post-bounce, with
\texttt{m24.5-$\langle$3D$\rangle$} settling at a lower heating
rate, at roughly the same time as the average shock radii begin
to differ.

The progenitor asymmetries seed turbulent motions behind the shock, which in turn provide turbulent stresses that support a larger shock radius. The larger shock radius increases the mass in the gain layer, which, in turn, increases the neutrino heating. Consequently, the shock moves outwards, which again increases the heating. The non-radial kinetic energy continues to grow as more of the progenitor perturbations fall through the shock and this non-linear loop continues and ultimately causes the larger shock radius and earlier explosion we see in model \texttt{m24.5-3D}, compared to \texttt{m24.5-$\langle$3D$\rangle$}. This picture is consistent with previous results in the literature
\citep{Couch_15a,Muller_17,OConnor_18}.

\subsection{\textsc{MESA} versus \textsc{FLASH}}
The 3D shell-burning evolution leaves the \texttt{m24.5-3D} progenitor with an inner-shell structure that differs from its \textsc{MESA} counterpart. As shown in Fig.~\ref{fig:composition_inner}, convection mixes the Si shell and flattens the composition profiles in the 3D simulation. The transition from the iron core
to the Si-burning shell is much more gradual in the \textsc{MESA} simulation than in the full 3D \textsc{FLASH} simulation. This is reflected in Table~\ref{tab:shell_masses}, where
we report a larger and more massive iron core in the \textsc{MESA} model than in the \textsc{FLASH} for model $24.5\,M_{\odot}$. 
However, the exact values of these quantities depend on the threshold we use to define the
iron core and should be interpreted with that in mind.

\begin{figure}
\centering
\includegraphics[width=1\linewidth]{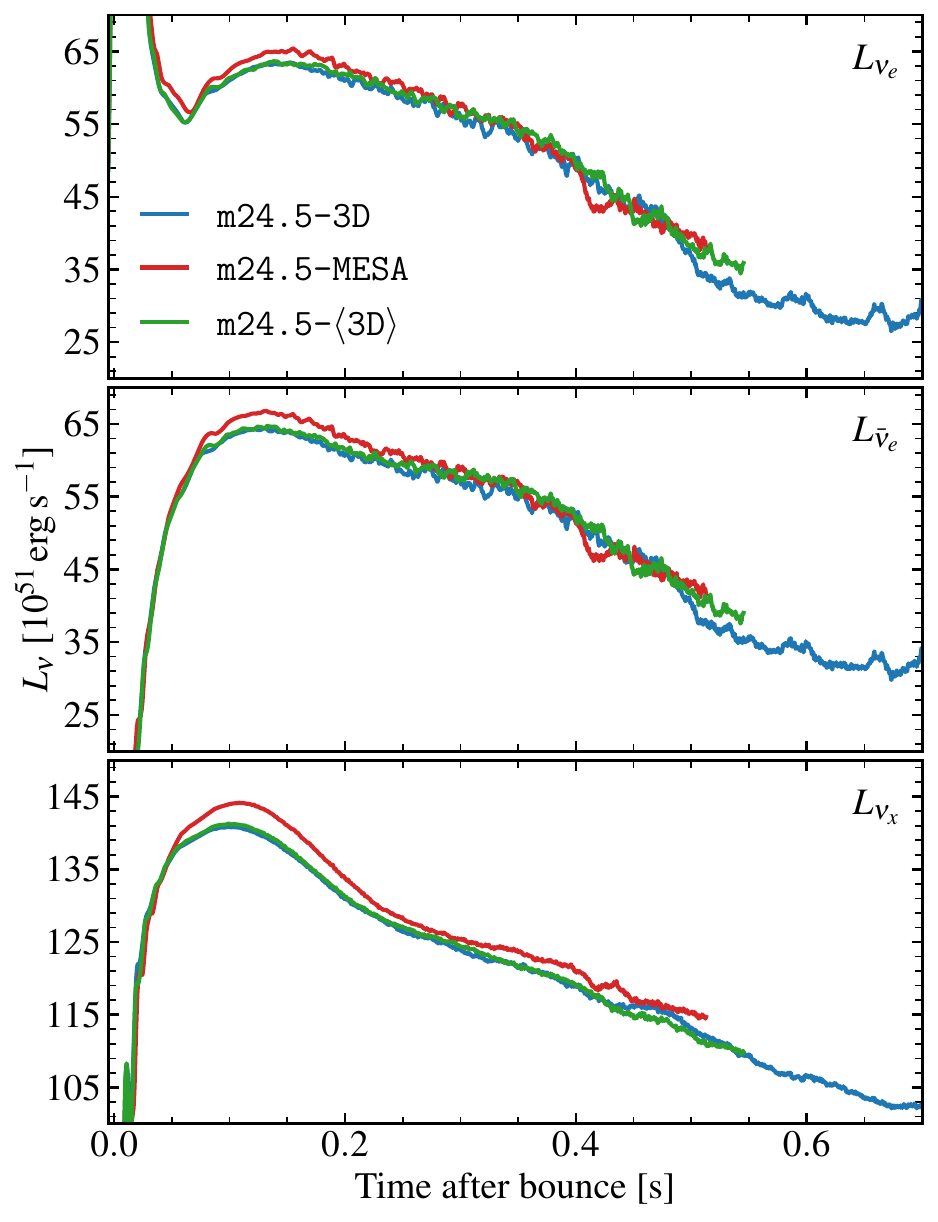}
\caption{Neutrino luminosities as a function of time after bounce for models
\texttt{m24.5-3D} (blue), \texttt{m24.5-MESA} (red), and
\texttt{m24.5-$\langle$3D$\rangle$} (green). The panels show the electron neutrino (top), electron antineutrino (middle), and heavy-lepton neutrino (bottom) luminosities.}
\label{fig:lums}
\end{figure}

Interestingly, model \texttt{m24.5-MESA} consistently lies between the two other models, in terms of the average shock radius, heating rate, and non-radial kinetic energy in the gain layer. This is because the neutrino luminosity of \texttt{m24.5-MESA}
is larger than that of the two models based on the
\textsc{FLASH} progenitor, at least prior to shock revival.
Fig.~\ref{fig:lums} shows the neutrino luminosities for the three simulations. The panels show the electron neutrino luminosity, the electron antineutrino luminosity, and the heavy-lepton neutrino luminosity, from top to bottom.
The higher neutrino luminosities in \texttt{m24.5-MESA}, compared to the 3D models, correspond to the higher heating rates seen in Fig.~\ref{fig:heating}. As the differences in the neutrino luminosities decrease towards $0.3\,\mathrm{s}$ after bounce, the excess heating rate decreases and eventually falls below the heating rate of model \texttt{m24.5-$\langle$3D$\rangle$}. The increased heating prior to $\sim 0.3\,$s is not enough to compensate for the lack of turbulence seeded by progenitor perturbations, and the average shock radius remains smaller than in \texttt{m24.5-3D} until the onset of shock revival. On the other hand, the shock radius is consistently larger in \texttt{m24.5-MESA} than in \texttt{m24.5-$\langle$3D$\rangle$}, which demonstrates the effect of the increased heating in the absence of progenitor perturbations.

The PNS is marginally hotter and denser in model \texttt{m24.5-MESA} than in the simulations based on the 3D progenitor. At $0.1\,\mathrm{s}$ after bounce, the PNS radius is $64.5\,\mathrm{km}$ in \texttt{m24.5-MESA} and $63.8\,\mathrm{km}$ in the other two models. The density is $1$--$8\,$\% higher throughout the PNS in \texttt{m24.5-MESA}, compared to model \texttt{m24.5-3D}, with the difference increasing towards the surface. Similarly, the temperature is higher by $\sim 1$--$2\,$\% throughout the PNS in \texttt{m24.5-MESA} compared to \texttt{m24.5-3D}. 
The relative differences in the $Y_e$ profiles of the two models
depend on radius, but integrated over the PNS, $Y_e$ is on average
$1.1\%$ lower in \texttt{m24.5-MESA} than in \texttt{m24.5-3D} at $0.1\,$s post-bounce.
The deficit is established prior to collapse. Comparing our
final 3D snapshot of the $24.5\,M_\odot$ model with the
corresponding \textsc{MESA} profile, we find $Y_e$ to be on
average $\sim1\%$ lower in \textsc{MESA} over the mass range
$1.1$--$1.4\,M_\odot$. This coincides with the region where the $^{32}$S and $^{28}$Si
mass fractions differ between the two codes, which points to the
reaction-rate discrepancies discussed in Appendix~\ref{apx:rates} as the underlying cause.
Differences in the pre-collapse composition will affect collapse, bounce, and the PNS structure.

The slightly denser and hotter PNS of model \texttt{m24.5-MESA} leads to neutrinospheres that are located $1$--$2\,$\% further out than in \texttt{m24.5-3D} (we do not show the time-dependent neutrinosphere radii here). While these differences are small, they lead to systematically larger neutrino luminosities.
We have verified that the increased neutrinosphere radii and temperatures account for the higher luminosities. From the Stefan-–Boltzmann law, one expects
\begin{equation}
L_{\nu} \sim r_{\rm ns}^2 T_{\rm ns}^4,
\end{equation}
where $L_{\nu}$ is the neutrino luminosity, $r_{\rm ns}$ the neutrinosphere radius, and $T_{\rm ns}$ the fluid temperature at the neutrinosphere. We find that the ratio of $r_{\rm ns}^2 T_{\rm ns}^4$ between the \texttt{m24.5-MESA} and \texttt{m24.5-3D} models agrees well with the corresponding ratio of the neutrino luminosities in the two models for all three species.

\subsection{Gravitational waves}
The gravitational-wave (GW) signals are extracted from the simulations using
the quadrupole formula. In the transverse-traceless (TT) gauge, the GW tensor
can be expressed in terms of two independent components, $h_{+}$ and $h_{\times}$. In the slow-motion limit, at a large distance $D$
from the source, we have
\begin{align}
\label{eq:hp}
h_{+}^{\rm TT} = \frac{G}{c^4 D} & \Big[ \ddot{Q}_{11} (\cos^2{\phi} - \sin^2{\phi} \cos^2{\theta})  \\ \nonumber
& + \ddot{Q}_{22} (\sin^2{\phi} - \cos^2{\phi} \cos^2{\theta}) - \ddot{Q}_{33} \sin^2{\theta} \\ \nonumber
& - \ddot{Q}_{12}\sin2\phi (1 + \cos^2{\theta}) + \ddot{Q}_{13} \sin{\phi} \sin{2\theta} \\ \nonumber
& + \ddot{Q}_{23} \cos{\phi} \sin{2\theta} \Big]
\end{align}
and
\begin{align}
\label{eq:hc}
h_{\times}^{\rm TT} = \frac{G}{c^4 D} & \Big[ (\ddot{Q}_{11} - \ddot{Q}_{22}) \sin{2\phi}\cos{\theta}\\ \nonumber
& +2\ddot{Q}_{12} \cos{\theta} \cos{2\phi} - 2\ddot{Q}_{13} \cos{\phi} \sin{\theta} \\ \nonumber
& +  2\ddot{Q}_{23} \sin{\phi} \sin{\theta} \Big]. 
\end{align}
In Eqs.~\eqref{eq:hp} and~\eqref{eq:hc}, $\ddot{Q}_{ij}$ denotes the second time derivative of the $ij$-component of the quadrupole moment. The two angles $\phi$
and $\theta$ denote the orientation of the observer in the coordinate system of the simulation.
In principle, the second time derivative can be calculated numerically from the standard quadrupole formula. However, this introduces numerical errors.
It is possible to eliminate both time derivatives and compute $\ddot{Q}_{ij}$ directly \citep{oohara_97,finn_89,blanchet_90}. Instead, we compute the first time derivative
\begin{align} \label{eqT:quaddt2}
\dot{Q}_{ij} &=
\int \rho \left(x_i v_j + v_i x_j - \frac{2}{3} \delta_{ij} v_l x_l \right) \mathrm{d}^3x,
\end{align}
during the simulation. Here, $v_i$ and $x_i$ denote the Cartesian velocity components and coordinates, respectively, and $\rho$ is the local fluid density. The second time derivative, $\ddot{Q}_{ij}$, is then obtained in post-processing by numerically differentiating $\dot{Q}_{ij}$ using \texttt{numpy.gradient}~\citep{numpy}.

\begin{figure*}
    \centering
    \includegraphics[width=\linewidth]{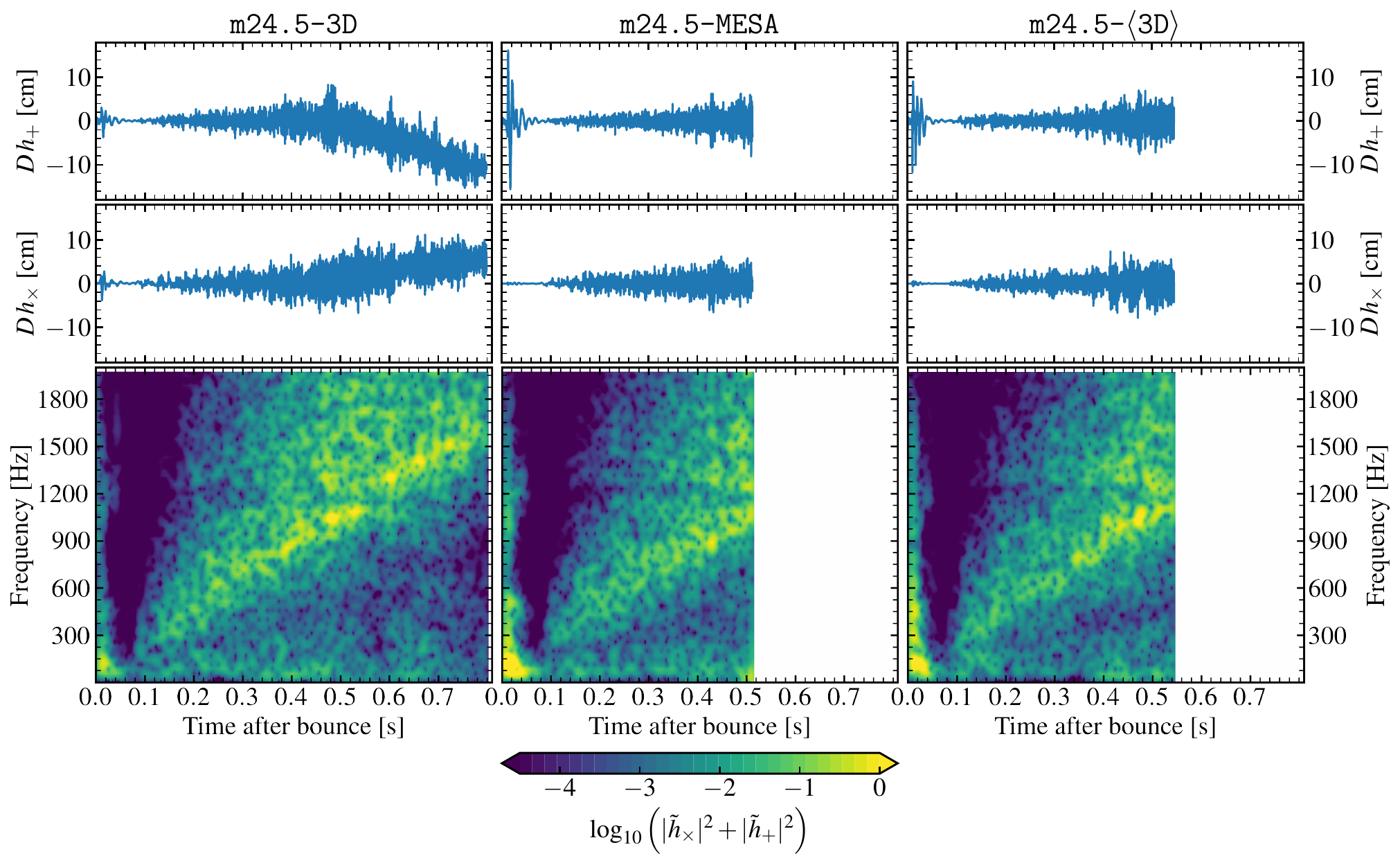}
    \caption{GW signals from the three core-collapse
    simulations, for an observer situated along the $z$-axis in the coordinate system of the
    numerical simulations.
    Each column corresponds to one model:
    \texttt{m24.5-3D} (left), \texttt{m24.5-MESA} (centre), and
    \texttt{m24.5-$\langle$3D$\rangle$} (right). The top two rows
    show the $h_+$ and $h_\times$ polarisation modes of the GW
    strain (scaled by the source distance $D$). The bottom row shows
    the corresponding spectrograms.}
    \label{fig:gw_spectrogram}
\end{figure*}

Fig.~\ref{fig:gw_spectrogram} shows the GW strain for all three
models: $Dh_+$ in the top panels, $Dh_\times$ in the middle
panels, and the corresponding spectrograms in the bottom panels.
We show the GW signals for an observer situated along the
$z$-axis of the simulations.
The spectrograms are computed with short-time Fourier transforms (STFTs).
The STFTs are computed with \texttt{scipy.signal.stft} using a Blackman window~\citep{scipy}, with a $35\,$ms window and $90$\% overlap between windows. We normalise the STFTs such that their logarithmic values lie in
the range $(-\infty, 0]$, using the same normalisation for all
three spectrograms.
We filter the signals using high-pass and low-pass filters, removing any part
of the signals below $25\,$Hz and above $5000\,$Hz before calculating the STFTs.

Right after bounce, models \texttt{m24.5-MESA} and \texttt{m24.5-$\langle$3D$\rangle$} show strong GW emission
associated with prompt convection \citep{Murphy_09}.
{However, we only observe the bounce signal in $h_+$ and it is much weaker for \texttt{m24.5-3D}, which indicates that the strong prompt convection is at least in part due to a grid alignment issue. The asymmetries present in model \texttt{m24.5-3D} break the symmetry that is enforced on the two other models during collapse and bounce, which reduces the strength of the bounce signal for models \texttt{m24.5-MESA} and \texttt{m24.5-$\langle$3D$\rangle$}.}

Beyond the prompt-convection signal, all models exhibit the characteristic GW signal of core-collapse supernovae \citep{Marek_09,Murphy_09,Muller_12,muller_13,Andresen_17,Kotake_17,Morozova_18,Powell_19,Andresen_19,Kawahara_18,OConnor_18,Radice_19,Mezzacappa_20,Andresen_21,Eggenberger-Andersen_21,Nakamura_22,Abdikamalov_22,Jakobus_23,Vartanyan_23,Murphy_25,Jakobus_25}.      
The strain amplitudes reach values of $\sim 5$--$10\,\mathrm{cm}$, and the spectrogram is
characterised by a narrow emission feature, 
with a steadily increasing central
frequency, superimposed on a broader background.
The power gap \citep{Morozova_18,andresen_26} is visible in all models,
but it is difficult to make out against the background signal. 

\begin{figure}
    \centering
    \includegraphics[width=\columnwidth]{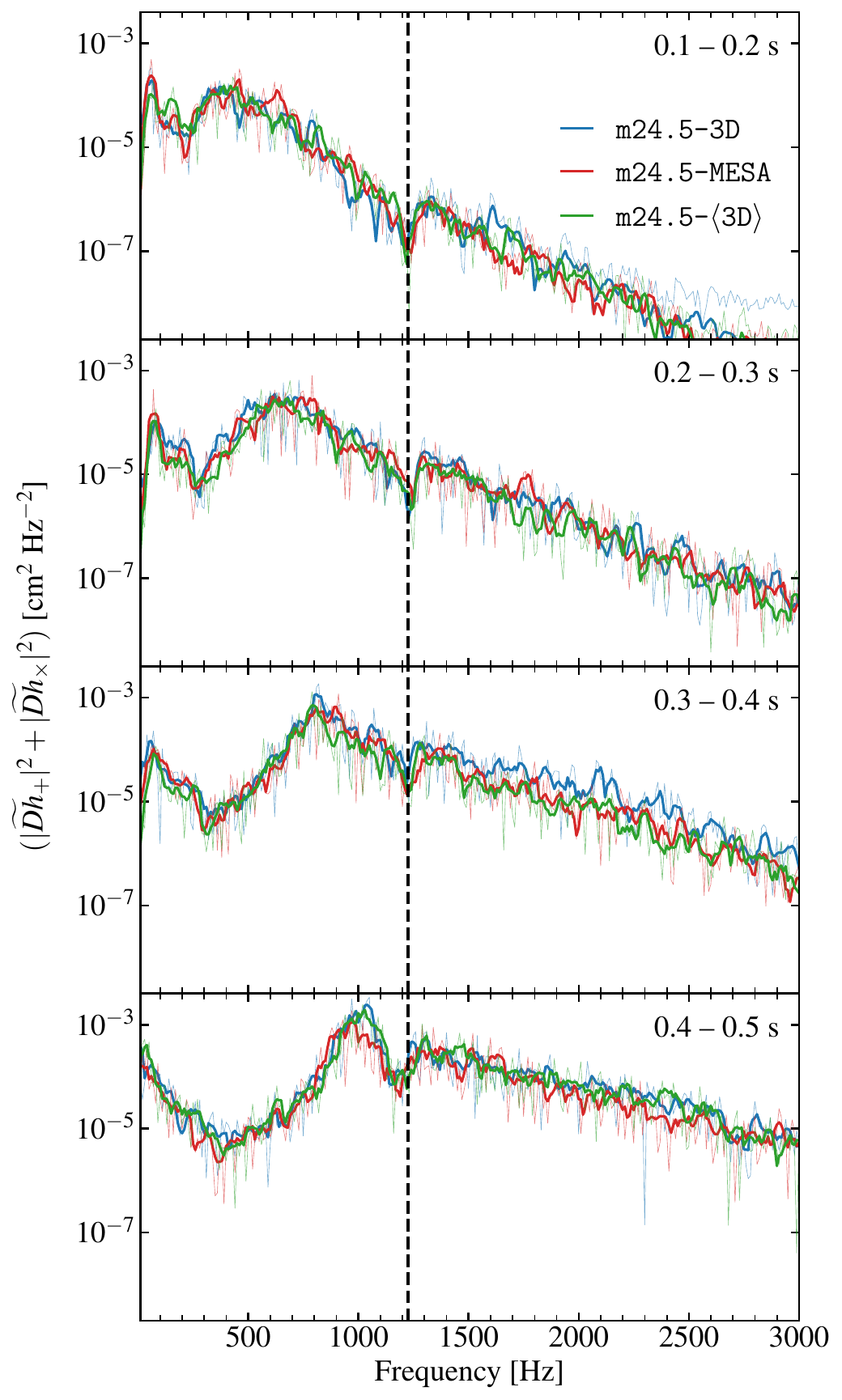}
    \caption{GW power spectral density,
    for models
    \texttt{m24.5-3D} (blue), \texttt{m24.5-MESA} (red), and
    \texttt{m24.5-$\langle$3D$\rangle$} (green), in four post-bounce
    time windows: $0.1$--$0.2\,$s (first row), $0.2$--$0.3\,$s (second row),
    $0.3$--$0.4\,$s (third row), and $0.4$--$0.5\,$s (fourth row). Faint lines show the Fourier transform and solid lines
    show the multitaper estimate for the power spectral density. The vertical dashed lines
    indicate $f=1225\,$Hz. }
    \label{fig:gw_psd}
\end{figure}

The power gap becomes visible in the power spectral density of the GW amplitude\footnote{This is not the spectral energy density of the radiated GWs; the latter has an additional factor of $f^2$.}, which is
shown in Fig.~\ref{fig:gw_psd} for four time windows:
$0.1$--$0.2\,$s, $0.2$--$0.3\,$s, $0.3$--$0.4\,$s, and $0.4$--$0.5\,$s post-bounce.
In each window we
estimate the spectrum with the multitaper method of \cite{thomson_82},
which reduces the variance in each frequency bin and limits spectral
leakage compared with a standard periodogram, making it better suited
to recovering the underlying spectrum of a stochastic time series
\citep{park_87,komm_99}. We compute the tapers using
\texttt{scipy.signal.windows.dpss} \citep{scipy} with $NW = 2.5$ and $K = 4$.
In all four time intervals, clear gaps in power are visible around $1225\,$Hz in Fig.~\ref{fig:gw_psd}. In addition
to the power gap, we see a local emission minimum below $500\,$Hz. The second minimum is a broad valley rather than a narrow gap, with a central frequency that shifts from  $f\sim200\,$Hz in the first window to $f\sim300\,$Hz in the last time window.
Furthermore, we see that both the overall power of the signals and the spectral shape of the three signals agree
very well between $0.1$ and $0.3\,$s after bounce. However, the signal from \texttt{m24.5-MESA} is slightly weaker than the signals from the two other models, at frequencies above 1500\,Hz during the first $0.2\,$s post-bounce.
We see an increase in the strength of the signal from \texttt{m24.5-3D} starting between $0.3$ and $0.4\,$s post-bounce 
(see the third panel of Fig.~\ref{fig:gw_psd}).
During this time, the emission above $1200\,$Hz in model \texttt{m24.5-3D} is up to $\sim 50$\% larger than the corresponding emission in the two other models. For \texttt{m24.5-3D}, the increase in emission above $1200\,$Hz is also visible in the spectrograms. In fact, the two other models also exhibit an increase in their signal strengths around $0.4\,$s post-bounce (see Fig.~\ref{fig:gw_spectrogram}). In the bottom panel of
Fig.~\ref{fig:gw_psd}, we see that the strengths of the three signals are similar in the $0.4$--$0.5\,$s post-bounce window, with a slightly weaker signal for model \texttt{m24.5-MESA}.

We expect that the energy emitted in GWs ($E_{\rm GW}$) scales
with the turbulent energy accreted by the PNS \citep{Radice_16,andresen_26}.
The total accreted turbulent energy is
\begin{equation} \label{eq:eturb}
    E_{\rm turb}(r) = 4\pi\int \big(F_K'(r) + F_H'(r) \big) \mathrm{d}t,
\end{equation}
where $F_K'(r)$ and $F_H'(r)$  are the angle-averaged rates of turbulent energy transfer:
\begin{equation}
    F_K' = \langle r^2 \left(\tfrac{1}{2}\rho v_i v_i\right) v_r' \rangle,
\end{equation}
and
\begin{equation}
    F_H' = \langle r^2 (\rho \epsilon + P) v_r' \rangle.
\end{equation}
$F_K'(r)$ and $F_H'(r)$ are evaluated at a given radius $r$.
$F_H'$ corresponds to thermal energy and $F_K'$ to kinetic energy.
\begin{figure}
    \centering
    \includegraphics[width=1\linewidth]{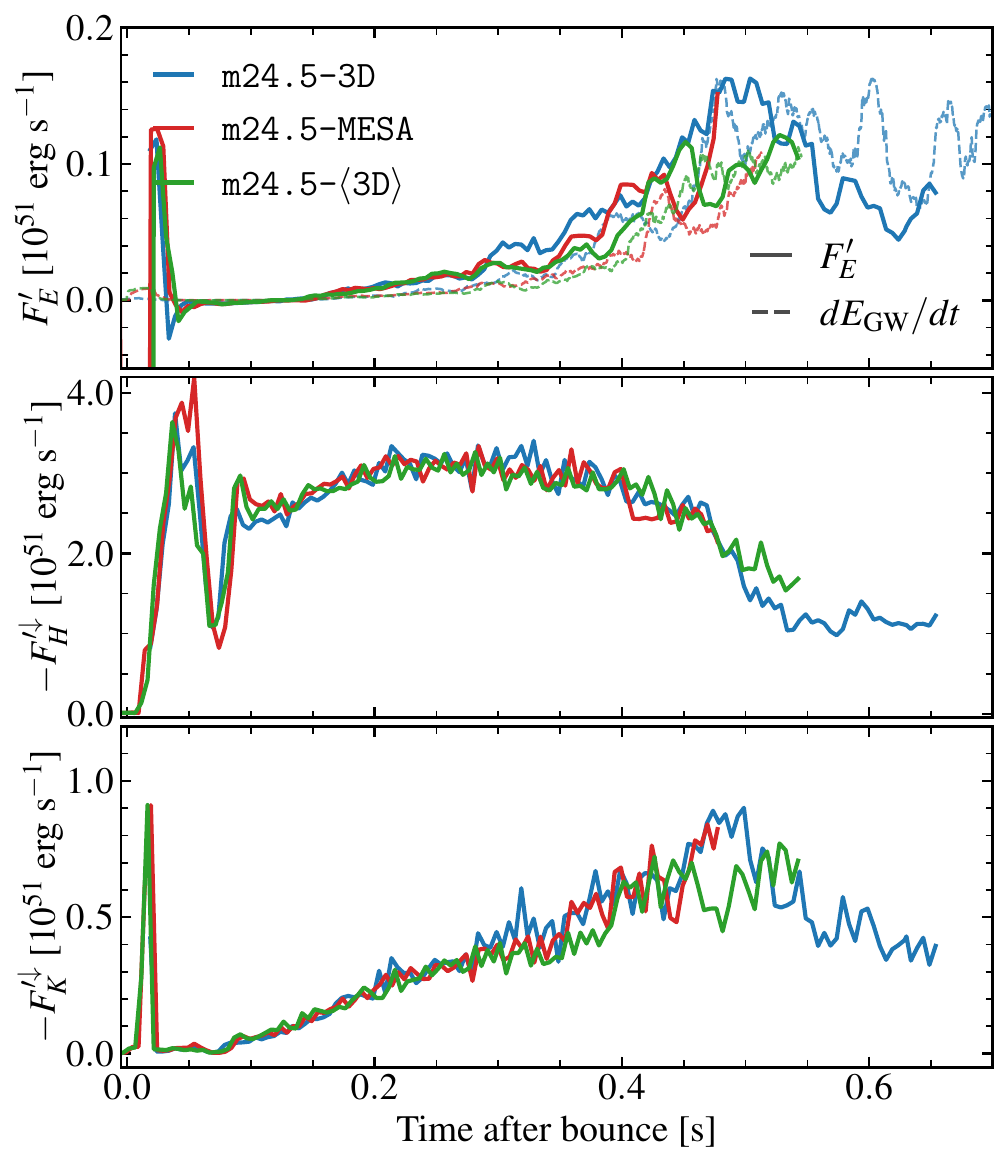}
\caption{Top: $F'_E = F'_K + F'_H$ at the PNS surface ($\rho = 10^{11}\,\mathrm{g}\,\mathrm{cm}^{-3}$) as a function of time after bounce (solid lines). Emitted GW energy per unit time, rescaled by a constant factor to lie in the same range as $F'_E$ (dashed lines). Middle and bottom: $-F^{\prime\,\downarrow}_H$ and $-F^{\prime\,\downarrow}_K$ for material flowing inward, sampled $10\,$km above the PNS surface. The symbol $\downarrow$ denotes downflows ($v_r < 0$).}
    \label{fig:trubgw}
\end{figure}
In the top panel of Fig.~\ref{fig:trubgw}, we plot $F^{\rm turb}_{E} =F_K'(t) + F_H'(t)$ for our three models. 
We also show the emitted energy in GWs per unit time ($\mathrm{d}E_{\rm GW}/\mathrm{d}t$) 
normalised to lie in the same range
as $F^{\rm turb}_{E}$. We see a clear correlation between the turbulent energy accreted
by the PNS and the strength of the GW signal. The increase in
the signals around $0.4\,$s post-bounce is directly linked to an
increase in the amount of turbulent energy accreted by the PNS.

An increase in the turbulent energy accreted by the PNS has to be linked to either an increase in the accretion rate or in the specific turbulent energy of the infalling material. The net accretion rate decreases steadily during the simulations and we do not see any evidence for an episode of increased accretion around $0.4\,$s post-bounce.
It is possible that the PNS undergoes a phase of asymmetric accretion. The net accretion rate could decrease during an episode of increased inward accretion if it is accompanied by a corresponding increase in material flowing out from the regions around the PNS, which could be the case for an asymmetric explosion like what we observe in model \texttt{m24.5-3D}. However, we
did not find evidence to support this scenario.
We instead find an increase in the specific turbulent energy of the material impinging on the PNS surface. This increase is gradual from $0.2\,$s post-bounce, but the growth is more rapid after $0.4\,$s and reaches a maximum around $0.5\,$s post-bounce. We observe the same trend in all three models, which means that this behaviour is not connected to asymmetries present in the progenitor.
We further characterise the structure of the flow above the PNS by decomposing
\begin{equation}
    v_r'(r,\theta,\phi) = \sum_{\ell=1}^{\ell_\mathrm{max}}\sum_{m=-\ell}^{\ell}
        a_{\ell m}(r)\,Y_{\ell, m}(\theta,\phi),
\end{equation}
with coefficients
\begin{equation}
    a_{\ell m} = \int v_r'(r,\theta,\phi)\,
        \big(Y_{\ell,m}(\theta,\phi)\big)^{*}\,
        \sin\theta\,\mathrm{d}\theta\,\mathrm{d}\phi.
\end{equation}
Here the asterisk denotes the complex conjugate.

We then define the power in each $\ell$-number as
\begin{equation}\label{eq:sh_power}
    P_\ell = \sum_{m=-\ell}^{\ell} \left|a_{\ell m}\right|^2,
\end{equation}
but when plotting we show $P_\ell/P_{\ell=1}$. We normalise by
$P_{\ell=1}$ to show the shift towards larger scales and because $P_{\ell=0} = 0$ by
construction.
\begin{figure}
    \centering
    \includegraphics[width=1\linewidth]{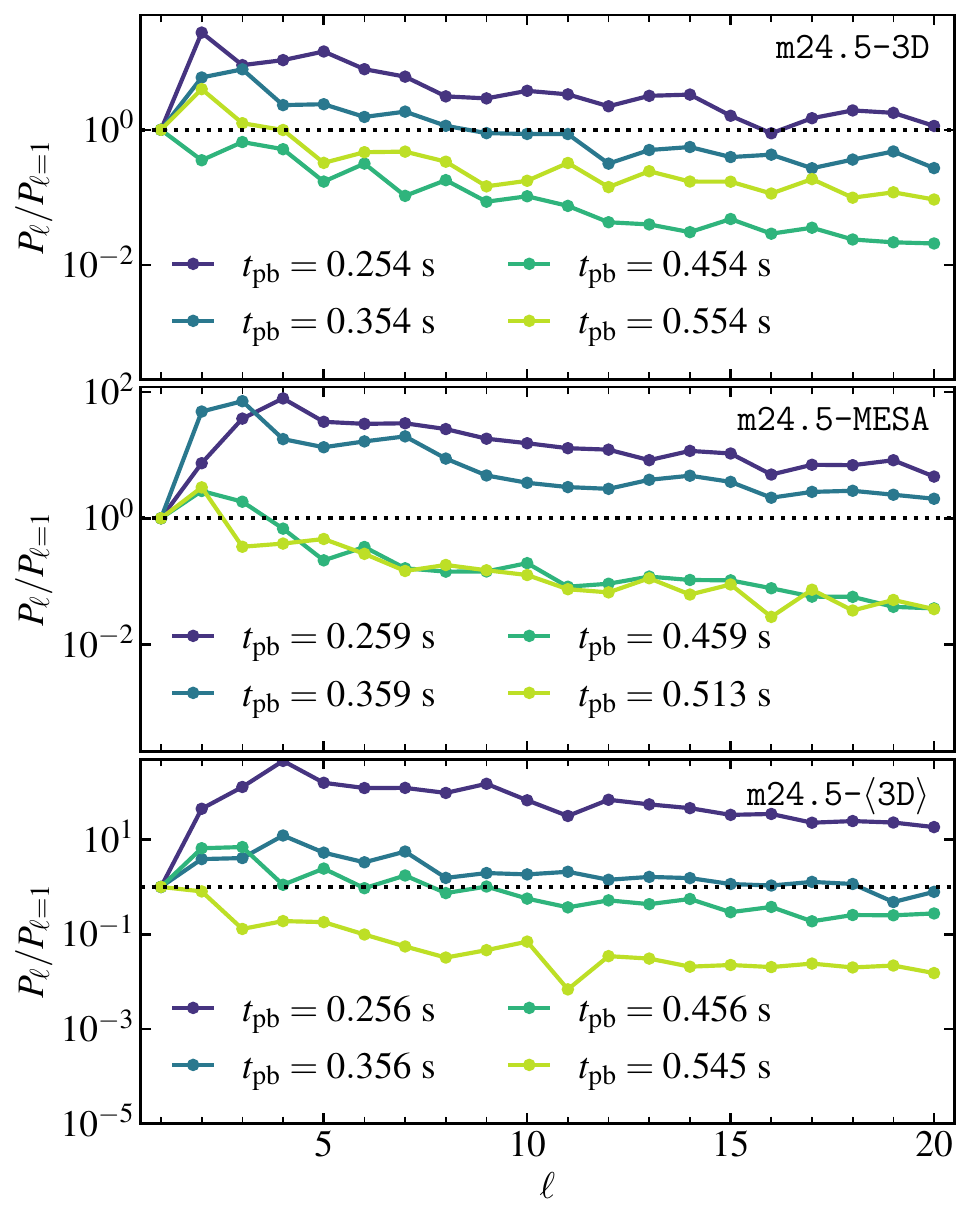}
    \caption{For each model, the angular power spectrum of $v_r^{\prime}$ $50\,$km above the PNS surface at four different times. We chose snapshots close to $0.25$, $0.35$, $0.45$, and $0.55\,$s post-bounce. The power spectrum has been normalised to the power at $\ell=1$. The top, middle, and bottom rows show 
    \texttt{m24.5-3D}, \texttt{m24.5-MESA}, and \texttt{m24.5-$\langle$3D$\rangle$}, respectively.}
    \label{fig:sphericalvr}
\end{figure}
In Fig.~\ref{fig:sphericalvr}, we plot $P_\ell/P_{\ell=1}$ for a radial shell
$50\,$km above the PNS surface at four different times.
For each model, we chose four snapshots corresponding to approximately $0.25$, $0.35$, $0.45$, and $0.55\,$s post-bounce.
We plot the snapshot closest to $0.55\,$s for models \texttt{m24.5-MESA} and \texttt{m24.5-$\langle$3D$\rangle$}, since they did not reach $0.55\,$s post-bounce by the time the simulations were terminated.
The angular power spectra shift towards larger scales as the
simulations progress. In all three models, the snapshots at
$\sim0.45$ and $\sim0.55\,$s carry more power at low $\ell$
relative to high $\ell$ than the earlier snapshots do. This
shift is consistent across the three models, so it is not tied
to the presence of progenitor asymmetries.

The increased GW emission that sets in around $0.4\,$s post-bounce is associated with a corresponding increase in the rate at which kinetic turbulent energy is transported inward to the PNS, while the rate of inward thermal energy transport decreases (see the bottom two panels in Fig.~\ref{fig:trubgw}). The decrease in the thermal component is reflected in the decreased heating rates in the gain layer for all three models (see Fig.~\ref{fig:heating}). At the same time, we find a shift in the structure of the flow towards larger scales. A likely explanation for the increased emission is, therefore, that the reduced heating after shock revival enables the formation of larger-scale fluid patterns which can reach all the way down to the PNS surface. These fluid patterns carry a larger amount of specific turbulent energy and more efficiently perturb the PNS, which in turn leads to an increase in GW emission.
Prior to shock revival, the downflows are broken up to a greater degree as they approach the PNS.

\section{Conclusions}
\label{sec:conclusions}
We have studied the impact of multi-dimensional progenitor structure on
core-collapse supernova dynamics by combining 3D simulations of
late-stage stellar burning with subsequent core-collapse calculations. Starting
from 1D \textsc{MESA} stellar evolution models, we performed
3D \textsc{FLASH} simulations of the final $\sim$10--15 minutes of
stellar evolution for five non-rotating solar-metallicity progenitors with
ZAMS masses of 20, 21.5, 24.5, 26, and 29\,$M_\odot$. For the 24.5\,$M_\odot$
progenitor, we then carried out three core-collapse simulations: one
initialised from the fully 3D \textsc{FLASH} model (\texttt{m24.5-3D}),
one from its angle-averaged counterpart (\texttt{m24.5-}$\langle$\texttt{3D}$\rangle$),
and one directly from the corresponding \textsc{MESA} progenitor (\texttt{m24.5-MESA}).
The angle-averaged comparison case provides a clean
view of the role of progenitor multi-dimensionality, since it removes the
non-radial structure while preserving the spherically averaged thermodynamic
profile of the 3D progenitor.

Our main findings are as follows:
\begin{enumerate}
\item Convection develops in the oxygen-rich layers of all five progenitors
after an initial transient period of $\sim$150\,s, with convective plumes reaching
radial velocities of several hundred km\,s$^{-1}$. The 21.5\,$M_\odot$ model is
an outlier in this respect: the convection in its O-rich layer is roughly an order of magnitude weaker
than in the other four models. Because the convective
turnover time in the O-rich layer is $\tau_\mathrm{conv}\sim 100$--$200\,$s,
our simulations capture only a few turnovers in this region and a fully
developed steady state has likely not been reached.

\item The structure of the inner burning shells differs strongly between
progenitors. Model $24.5\,M_\odot$ develops vigorous convection in the
Si-burning shell, with turnover times of $15$--$30\,$s and tens of turnovers over
the course of the simulation. The 21.5$\,M_\odot$ model develops convection in
the O-burning shell instead, while the 20 and 26$\,M_\odot$ models show
weaker convective activity in the Si-burning shell region. The 29$\,M_\odot$ model
shows no convective activity interior to the O-rich layer.

\item Comparing the \textsc{FLASH} and \textsc{MESA} evolution, we
find systematic differences in the burning-shell composition. The
$^{32}$S mass fraction in the inner burning shells is consistently $20$--$50\,\%$ higher in the \textsc{FLASH} models, predominantly at the expense of $^{40}$Ca and $^{28}$Si.
We attribute the discrepancies mainly to differences in the underlying reaction rates.
Despite these compositional
differences, the iron-core radii and masses agree at the $\sim$10\%
level between the two codes, and the compactness parameter $\xi_{2.5}$
generally agrees to within $\sim$0.04, with \textsc{MESA} models being marginally more
compact.

\item In the core-collapse simulations, the multi-dimensional progenitor
produces clearly enhanced non-radial kinetic energy in the post-shock region
compared to the models based on spherically symmetric progenitors.
The non-radial kinetic energies of \texttt{m24.5-3D} and
\texttt{m24.5-}$\langle$\texttt{3D}$\rangle$ begin to diverge $\sim$50\,ms
after bounce, coincident with the lower edge of the Si shell falling
through the shock. By $0.2\,$s post-bounce, $E_\theta+E_\phi$ is $\sim$35\% 
larger in \texttt{m24.5-3D} than in
\texttt{m24.5-}$\langle$\texttt{3D}$\rangle$, and the difference grows to as
much as $50\,\%$ later in the evolution. The accreted progenitor
perturbations seed turbulent motions behind the shock and enhance neutrino heating in the gain
region \citep{OConnor_18,Muller_17}. As a result, \texttt{m24.5-3D} maintains an average shock radius
5--10\% larger than its angle-averaged counterpart and is the first of
the three models to undergo shock revival.

\item The \texttt{m24.5-MESA} simulation produces a marginally larger, denser,
and hotter PNS than the two \textsc{FLASH}-based models. At $0.1\,$s post-bounce the PNS
radius is $64.5\,$km in \texttt{m24.5-MESA} compared to $63.8\,$km in the others,
the density is 1--8\% higher in the outer regions of the PNS, and the temperature is
1--2\% higher. This translates into 1--2\% larger neutrinosphere
radii and correspondingly higher neutrino luminosities.
The increased neutrino heating partly compensates for the absence of
progenitor asymmetries: \texttt{m24.5-MESA} reaches shock revival earlier than
\texttt{m24.5-}$\langle$\texttt{3D}$\rangle$, but later than \texttt{m24.5-3D}.

\item The GW emission is similar in all three models, but \texttt{m24.5-3D} shows
a weaker prompt-convection signal than the other two models. We observe an increase in the
signal strength in all three models around $0.4\,$s post-bounce, which is after shock revival in all three models. The increase in GW emission after shock revival is not a consequence of progenitor perturbations, but is due to a change in the downflows reaching the PNS surface. We observe an increase in the specific turbulent energy carried by downflows hitting the PNS surface after $\sim 0.4\,$s post-bounce, which leads to an increase in the GW emission from the PNS.
\end{enumerate}

Our results are broadly consistent with the picture established in earlier
work, in which pre-collapse asymmetries strengthen the neutrino-driven
mechanism by seeding turbulence behind the shock
\citep{Couch_13a,Muller_17,OConnor_18,Bollig_21,Vartanyan_22}.
Our finding that the 3D progenitor produces a larger
shock radius and earlier shock revival than its angle-averaged counterpart
agrees qualitatively with the trends reported in similar \textsc{FLASH}
simulations carried out by \cite{OConnor_18}. They found that perturbations
led to larger shock radii and stronger post-shock turbulence, although none
of their models ultimately exploded. We do not find a clear explosion/non-explosion dichotomy in our
simulations, unlike what was reported by \cite{Muller_17}.
Our finding that the multi-dimensional progenitor produces stronger
post-shock turbulence and earlier shock revival is consistent with the
results reported by \cite{Bollig_21} and \cite{Vartanyan_22}, who found that
turbulence generated by progenitor asymmetries aids explosions.
Our results are more difficult to reconcile with \cite{Chen_26}, who
report that progenitor perturbations do not significantly affect the outcome
of their axisymmetric simulations. Importantly, we do not include nuclear burning during collapse,
which \cite{Chen_26} identified as a strong source of seed perturbations for turbulence in their simulations. 
Furthermore, it is not
trivial to compare the impact of turbulent convection between 2D and 3D simulations.

There are two main weaknesses with our current work, which we aim to resolve in future simulations.

The first is the nuclear network. \texttt{approx21} is a minimal
$\alpha$-chain network designed to give reasonable nuclear
energy generation rates at low computational cost
\citep{Timmes2000}, but it provides only a coarse
description of the nucleosynthesis.
Larger networks yield systematically different shell compositions, electron
fractions, and core masses \citep{Farmer_16,Grichener_25}.
Because these quantities feed
directly into both the structure of the iron core at the onset of collapse
and the convective velocities in the burning shells, performing
burning calculations with a larger network is an important next step.
The second is the duration of the 3D burning phase. Our simulations
span only the final $\sim$10--15 minutes prior to collapse, which is
sufficient for the Si-burning shell ($\tau_\mathrm{conv}\sim 15$--$30\,$s)
to reach a quasi-steady state, but captures only a few turnovers in the
O-rich layer ($\tau_\mathrm{conv}\sim 100$--$200\,$s). The convection in the O-rich layer
is, therefore, unlikely to be fully developed at the moment of core collapse. 

Additionally, the issue with the nuclear burning rates in \textsc{MESA} discussed in
Appendix~\ref{apx:rates} means that the composition of the \textsc{MESA} models
used to initialise our \textsc{FLASH} simulations is not what it should be. The
\textsc{FLASH} network is not affected by this issue, but our
simulations started from incorrect initial compositions, and the error therefore
propagates into the final models. We expect, however, that the convection and
the subsequent core collapse would proceed in a similar way if the
simulations were repeated from corrected \textsc{MESA} models, since we did not find a drastic change in the general structure of the \textsc{MESA} models after updating the rates.

\section*{Acknowledgements}
We are grateful to the \textsc{MESA} development team for helpful discussions regarding the nuclear network reaction rates.
The computations were enabled by resources provided by the National Academic Infrastructure for Supercomputing in Sweden (NAISS) at NSC partially funded by the Swedish Research Council through grant agreements no. 2022-06725 and no. 2018-05973. 
We acknowledge NAISS for awarding this project access to the LUMI supercomputer, owned by the EuroHPC Joint Undertaking, hosted by CSC (Finland) and the LUMI consortium.
This work is supported by the Swedish Research Council (Project No. 2020-00452). 
\textbf{{Software}:} \textsc{FLASH} \citep{Fryxell_2000}, \textsc{NuLib}\citep{OConnor_15}, \textsc{MESA} \citep{Paxton2011,Paxton2019},  \textsc{Matplotlib} \citep{Hunter07}, \textsc{NumPy} \citep{numpy}, \textsc{SciPy} \citep{2020SciPy-NMeth}, \textsc{yt} \citep{Turk11}.
\section*{Data Availability}
The \texttt{MESA} inlists, the \texttt{MESA}
models we started our burning simulations from, and the final 
three-dimensional progenitor snapshots are available at \url{https://zenodo.org/records/21994328}.
The gravitational-wave signals, the total neutrino luminosities, and the mean neutrino energies
are available at \url{https://zenodo.org/records/21979612}.
The remaining simulation data can be provided upon reasonable
request.

\bibliographystyle{mnras}
\bibliography{merged}

\begin{thebibliography}{}
\makeatletter
\relax
\def\mn@urlcharsother{\let\do\@makeother \do\$\do\&\do\#\do\^\do\_\do\%\do\~}
\def\mn@doi{\begingroup\mn@urlcharsother \@ifnextchar [ {\mn@doi@}
  {\mn@doi@[]}}
\def\mn@doi@[#1]#2{\def\@tempa{#1}\ifx\@tempa\@empty \href
  {http://dx.doi.org/#2} {doi:#2}\else \href {http://dx.doi.org/#2} {#1}\fi
  \endgroup}
\def\mn@eprint#1#2{\mn@eprint@#1:#2::\@nil}
\def\mn@eprint@arXiv#1{\href {http://arxiv.org/abs/#1} {{\tt arXiv:#1}}}
\def\mn@eprint@dblp#1{\href {http://dblp.uni-trier.de/rec/bibtex/#1.xml}
  {dblp:#1}}
\def\mn@eprint@#1:#2:#3:#4\@nil{\def\@tempa {#1}\def\@tempb {#2}\def\@tempc
  {#3}\ifx \@tempc \@empty \let \@tempc \@tempb \let \@tempb \@tempa \fi \ifx
  \@tempb \@empty \def\@tempb {arXiv}\fi \@ifundefined
  {mn@eprint@\@tempb}{\@tempb:\@tempc}{\expandafter \expandafter \csname
  mn@eprint@\@tempb\endcsname \expandafter{\@tempc}}}

\bibitem[\protect\citeauthoryear{{Abdikamalov}, {Pagliaroli}  \&
  {Radice}}{{Abdikamalov} et~al.}{2022}]{Abdikamalov_22}
{Abdikamalov} E.,  {Pagliaroli} G.,   {Radice} D.,  2022, in {Bambi} C.,
  {Katsanevas} S.,   {Kokkotas} K.~D.,  eds, , Handbook of Gravitational Wave
  Astronomy.
p.~21 (\mn@eprint {arXiv} {2010.04356}),
  \mn@doi{10.1007/978-981-15-4702-7_21-1}

\bibitem[\protect\citeauthoryear{{Alastuey} \& {Jancovici}}{{Alastuey} \&
  {Jancovici}}{1978}]{Alastuey1978}
{Alastuey} A.,  {Jancovici} B.,  1978, \mn@doi [\apj] {10.1086/156681}, \href
  {https://ui.adsabs.harvard.edu/\#abs/1978ApJ...226.1034A} {226, 1034}

\bibitem[\protect\citeauthoryear{{Andresen}, {M{\"u}ller}, {M{\"u}ller}  \&
  {Janka}}{{Andresen} et~al.}{2017}]{Andresen_17}
{Andresen} H.,  {M{\"u}ller} B.,  {M{\"u}ller} E.,   {Janka} H.~T.,  2017,
  \mn@doi [\mnras] {10.1093/mnras/stx618}, \href
  {https://ui.adsabs.harvard.edu/abs/2017MNRAS.468.2032A} {468, 2032}

\bibitem[\protect\citeauthoryear{{Andresen}, {M{\"u}ller}, {Janka}, {Summa},
  {Gill}  \& {Zanolin}}{{Andresen} et~al.}{2019}]{Andresen_19}
{Andresen} H.,  {M{\"u}ller} E.,  {Janka} H.~T.,  {Summa} A.,  {Gill} K.,
  {Zanolin} M.,  2019, \mn@doi [\mnras] {10.1093/mnras/stz990}, \href
  {https://ui.adsabs.harvard.edu/abs/2019MNRAS.486.2238A} {486, 2238}

\bibitem[\protect\citeauthoryear{{Andresen}, {Glas}  \& {Janka}}{{Andresen}
  et~al.}{2021}]{Andresen_21}
{Andresen} H.,  {Glas} R.,   {Janka} H.~T.,  2021, \mn@doi [\mnras]
  {10.1093/mnras/stab675}, \href
  {https://ui.adsabs.harvard.edu/abs/2021MNRAS.503.3552A} {503, 3552}

\bibitem[\protect\citeauthoryear{{Andresen}, {Li}, {Betranhandy}, {O'Connor},
  {Zha}  \& {Couch}}{{Andresen} et~al.}{2026}]{andresen_26}
{Andresen} H.,  {Li} X.,  {Betranhandy} A.,  {O'Connor} E.~P.,  {Zha} S.,
  {Couch} S.~M.,  2026, \mn@doi [arXiv e-prints] {10.48550/arXiv.2603.26408},
  \href {https://ui.adsabs.harvard.edu/abs/2026arXiv260326408A} {p.
  arXiv:2603.26408}

\bibitem[\protect\citeauthoryear{{Angulo} et~al.,}{{Angulo}
  et~al.}{1999}]{Angulo1999}
{Angulo} C.,  et~al., 1999, \mn@doi [\nphysa] {10.1016/S0375-9474(99)00030-5},
  \href {https://ui.adsabs.harvard.edu/abs/1999NuPhA.656....3A} {656, 3}

\bibitem[\protect\citeauthoryear{{Asplund}, {Grevesse}, {Sauval}  \&
  {Scott}}{{Asplund} et~al.}{2009}]{asplund_09}
{Asplund} M.,  {Grevesse} N.,  {Sauval} A.~J.,   {Scott} P.,  2009, \mn@doi
  [\araa] {10.1146/annurev.astro.46.060407.145222}, \href
  {https://ui.adsabs.harvard.edu/abs/2009ARA&A..47..481A} {47, 481}

\bibitem[\protect\citeauthoryear{{Bethe}}{{Bethe}}{1990}]{Bethe_90}
{Bethe} H.~A.,  1990, \mn@doi [Reviews of Modern Physics]
  {10.1103/RevModPhys.62.801}, \href
  {https://ui.adsabs.harvard.edu/abs/1990RvMP...62..801B} {62, 801}

\bibitem[\protect\citeauthoryear{{Bethe} \& {Wilson}}{{Bethe} \&
  {Wilson}}{1985}]{Bethe_85}
{Bethe} H.~A.,  {Wilson} J.~R.,  1985, \mn@doi [\apj] {10.1086/163343}, \href
  {https://ui.adsabs.harvard.edu/abs/1985ApJ...295...14B} {295, 14}

\bibitem[\protect\citeauthoryear{{Blanchet}, {Damour}  \&
  {Sch\"afer}}{{Blanchet} et~al.}{1990}]{blanchet_90}
{Blanchet} L.,  {Damour} T.,   {Sch\"afer} G.,  1990, \mnras, \href
  {http://adsabs.harvard.edu/abs/1990MNRAS.242..289B} {242, 289}

\bibitem[\protect\citeauthoryear{{Blondin}, {Mezzacappa}  \&
  {DeMarino}}{{Blondin} et~al.}{2003}]{Blondin_03}
{Blondin} J.~M.,  {Mezzacappa} A.,   {DeMarino} C.,  2003, \mn@doi [\apj]
  {10.1086/345812}, \href
  {https://ui.adsabs.harvard.edu/abs/2003ApJ...584..971B} {584, 971}

\bibitem[\protect\citeauthoryear{{Blouin}, {Shaffer}, {Saumon}  \&
  {Starrett}}{{Blouin} et~al.}{2020}]{Blouin2020}
{Blouin} S.,  {Shaffer} N.~R.,  {Saumon} D.,   {Starrett} C.~E.,  2020, \mn@doi
  [\apj] {10.3847/1538-4357/ab9e75}, \href
  {https://ui.adsabs.harvard.edu/abs/2020ApJ...899...46B} {899, 46}

\bibitem[\protect\citeauthoryear{{Bollig}, {Yadav}, {Kresse}, {Janka},
  {M{\"u}ller}  \& {Heger}}{{Bollig} et~al.}{2021}]{Bollig_21}
{Bollig} R.,  {Yadav} N.,  {Kresse} D.,  {Janka} H.-T.,  {M{\"u}ller} B.,
  {Heger} A.,  2021, \mn@doi [\apj] {10.3847/1538-4357/abf82e}, \href
  {https://ui.adsabs.harvard.edu/abs/2021ApJ...915...28B} {915, 28}

\bibitem[\protect\citeauthoryear{{Borges}, {Carmona}, {Costa}  \&
  {Don}}{{Borges} et~al.}{2008}]{borges_08}
{Borges} R.,  {Carmona} M.,  {Costa} B.,   {Don} W.~S.,  2008, \mn@doi [jcoph]
  {10.1016/j.jcp.2007.11.038}, \href
  {http://adsabs.harvard.edu/abs/2008JCoPh.227.3191B} {227, 3191}

\bibitem[\protect\citeauthoryear{{Bruenn}}{{Bruenn}}{1985}]{bruenn_85}
{Bruenn} S.~W.,  1985, \mn@doi [\apjs] {10.1086/191056}, \href
  {https://ui.adsabs.harvard.edu/abs/1985ApJS...58..771B} {58, 771}

\bibitem[\protect\citeauthoryear{{Bruenn} et~al.,}{{Bruenn}
  et~al.}{2023}]{Bruenn_23}
{Bruenn} S.~W.,  et~al., 2023, \mn@doi [\apj] {10.3847/1538-4357/acbb65}, \href
  {https://ui.adsabs.harvard.edu/abs/2023ApJ...947...35B} {947, 35}

\bibitem[\protect\citeauthoryear{{Bugli}, {Guilet}, {Foglizzo}  \&
  {Obergaulinger}}{{Bugli} et~al.}{2023}]{bugli_23}
{Bugli} M.,  {Guilet} J.,  {Foglizzo} T.,   {Obergaulinger} M.,  2023, \mn@doi
  [\mnras] {10.1093/mnras/stad496}, \href
  {https://ui.adsabs.harvard.edu/abs/2023MNRAS.520.5622B} {520, 5622}

\bibitem[\protect\citeauthoryear{{Burrows} \& {Vartanyan}}{{Burrows} \&
  {Vartanyan}}{2021}]{Burrows_21}
{Burrows} A.,  {Vartanyan} D.,  2021, \mn@doi [\nat]
  {10.1038/s41586-020-03059-w}, \href
  {https://ui.adsabs.harvard.edu/abs/2021Natur.589...29B} {589, 29}

\bibitem[\protect\citeauthoryear{{Burrows}, {Hayes}  \& {Fryxell}}{{Burrows}
  et~al.}{1995}]{Burrows_95}
{Burrows} A.,  {Hayes} J.,   {Fryxell} B.~A.,  1995, \mn@doi [\apj]
  {10.1086/176188}, \href
  {https://ui.adsabs.harvard.edu/abs/1995ApJ...450..830B} {450, 830}

\bibitem[\protect\citeauthoryear{Burrows, Radice  \& Vartanyan}{Burrows
  et~al.}{2019}]{Burrows_19}
Burrows A.,  Radice D.,   Vartanyan D.,  2019, \mn@doi [Mon. Not. Roy. Astron.
  Soc.] {10.1093/mnras/stz543}, 485, 3153

\bibitem[\protect\citeauthoryear{Burrows, Radice, Vartanyan, Nagakura, Skinner
  \& Dolence}{Burrows et~al.}{2020}]{Burrows_20}
Burrows A.,  Radice D.,  Vartanyan D.,  Nagakura H.,  Skinner M.~A.,   Dolence
  J.,  2020, \mn@doi [Mon. Not. Roy. Astron. Soc.] {10.1093/mnras/stz3223},
  491, 2715

\bibitem[\protect\citeauthoryear{Cardall, Endeve  \& Mezzacappa}{Cardall
  et~al.}{2013}]{Cardall_12}
Cardall C.~Y.,  Endeve E.,   Mezzacappa A.,  2013, \mn@doi [Phys. Rev. D]
  {10.1103/PhysRevD.87.103004}, 87, 103004

\bibitem[\protect\citeauthoryear{{Cassisi}, {Potekhin}, {Pietrinferni},
  {Catelan}  \& {Salaris}}{{Cassisi} et~al.}{2007}]{Cassisi2007}
{Cassisi} S.,  {Potekhin} A.~Y.,  {Pietrinferni} A.,  {Catelan} M.,   {Salaris}
  M.,  2007, \mn@doi [\apj] {10.1086/516819}, \href
  {https://ui.adsabs.harvard.edu/abs/2007ApJ...661.1094C} {661, 1094}

\bibitem[\protect\citeauthoryear{{Chen}, {Lentz}, {Hix}, {Harris}, {Keeling
  Sandoval}  \& {Bruenn}}{{Chen} et~al.}{2026}]{Chen_26}
{Chen} C.-H.,  {Lentz} E.~J.,  {Hix} W.~R.,  {Harris} J.~A.,  {Keeling
  Sandoval} C.,   {Bruenn} S.~W.,  2026, \mn@doi [arXiv e-prints]
  {10.48550/arXiv.2604.09906}, \href
  {https://ui.adsabs.harvard.edu/abs/2026arXiv260409906C} {p. arXiv:2604.09906}

\bibitem[\protect\citeauthoryear{{Chugunov}, {Dewitt}  \&
  {Yakovlev}}{{Chugunov} et~al.}{2007}]{Chugunov2007}
{Chugunov} A.~I.,  {Dewitt} H.~E.,   {Yakovlev} D.~G.,  2007, \mn@doi [\prd]
  {10.1103/PhysRevD.76.025028}, \href
  {https://ui.adsabs.harvard.edu/abs/2007PhRvD..76b5028C} {76, 025028}

\bibitem[\protect\citeauthoryear{Couch}{Couch}{2013}]{Couch_13}
Couch S.~M.,  2013, \mn@doi [The Astrophysical Journal]
  {10.1088/0004-637X/765/1/29}, 765, 29

\bibitem[\protect\citeauthoryear{Couch \& O'Connor}{Couch \&
  O'Connor}{2014}]{Couch_14}
Couch S.~M.,  O'Connor E.~P.,  2014, \mn@doi [The Astrophysical Journal]
  {10.1088/0004-637X/785/2/123}, 785, 123

\bibitem[\protect\citeauthoryear{{Couch} \& {Ott}}{{Couch} \&
  {Ott}}{2013}]{Couch_13a}
{Couch} S.~M.,  {Ott} C.~D.,  2013, \mn@doi [\apjl]
  {10.1088/2041-8205/778/1/L7}, \href
  {https://ui.adsabs.harvard.edu/abs/2013ApJ...778L...7C} {778, L7}

\bibitem[\protect\citeauthoryear{{Couch} \& {Ott}}{{Couch} \&
  {Ott}}{2015}]{Couch_15}
{Couch} S.~M.,  {Ott} C.~D.,  2015, \mn@doi [\apj] {10.1088/0004-637X/799/1/5},
  \href {https://ui.adsabs.harvard.edu/abs/2015ApJ...799....5C} {799, 5}

\bibitem[\protect\citeauthoryear{{Couch}, {Chatzopoulos}, {Arnett}  \&
  {Timmes}}{{Couch} et~al.}{2015}]{Couch_15a}
{Couch} S.~M.,  {Chatzopoulos} E.,  {Arnett} W.~D.,   {Timmes} F.~X.,  2015,
  \mn@doi [\apjl] {10.1088/2041-8205/808/1/L21}, \href
  {https://ui.adsabs.harvard.edu/abs/2015ApJ...808L..21C} {808, L21}

\bibitem[\protect\citeauthoryear{{Couch}, {Warren}  \& {O'Connor}}{{Couch}
  et~al.}{2020}]{Couch_19}
{Couch} S.~M.,  {Warren} M.~L.,   {O'Connor} E.~P.,  2020, \mn@doi [\apj]
  {10.3847/1538-4357/ab609e}, \href
  {https://ui.adsabs.harvard.edu/abs/2020ApJ...890..127C} {890, 127}

\bibitem[\protect\citeauthoryear{Couch, Carlson, Pajkos, O’Shea, Dubey  \&
  Klosterman}{Couch et~al.}{2021}]{couch_21}
Couch S.~M.,  Carlson J.,  Pajkos M.,  O’Shea B.~W.,  Dubey A.,   Klosterman
  T.,  2021, \mn@doi [Parallel Computing] {10.1016/j.parco.2021.102830}, 108,
  102830

\bibitem[\protect\citeauthoryear{{Cyburt} et~al.,}{{Cyburt}
  et~al.}{2010}]{Cyburt2010}
{Cyburt} R.~H.,  et~al., 2010, \mn@doi [\apjs] {10.1088/0067-0049/189/1/240},
  \href {https://ui.adsabs.harvard.edu/abs/2010ApJS..189..240C} {189, 240}

\bibitem[\protect\citeauthoryear{{Dewitt}, {Graboske}  \& {Cooper}}{{Dewitt}
  et~al.}{1973}]{Dewitt1973}
{Dewitt} H.~E.,  {Graboske} H.~C.,   {Cooper} M.~S.,  1973, \mn@doi [\apj]
  {10.1086/152061}, \href
  {https://ui.adsabs.harvard.edu/\#abs/1973ApJ...181..439D} {181, 439}

\bibitem[\protect\citeauthoryear{{Dubey}, {Reid}, {Weide}, {Antypas},
  {Ganapathy}, {Riley}, {Sheeler}  \& {Siegal}}{{Dubey}
  et~al.}{2009}]{Dubey_09}
{Dubey} A.,  {Reid} L.~B.,  {Weide} K.,  {Antypas} K.,  {Ganapathy} M.~K.,
  {Riley} K.,  {Sheeler} D.,   {Siegal} A.,  2009, \mn@doi [arXiv e-prints]
  {10.48550/arXiv.0903.4875}, \href
  {https://ui.adsabs.harvard.edu/abs/2009arXiv0903.4875D} {p. arXiv:0903.4875}

\bibitem[\protect\citeauthoryear{{Eggenberger Andersen}, {Zha}, {da Silva
  Schneider}, {Betranhandy}, {Couch}  \& {O'Connor}}{{Eggenberger Andersen}
  et~al.}{2021}]{Eggenberger-Andersen_21}
{Eggenberger Andersen} O.,  {Zha} S.,  {da Silva Schneider} A.,  {Betranhandy}
  A.,  {Couch} S.~M.,   {O'Connor} E.~P.,  2021, \mn@doi [\apj]
  {10.3847/1538-4357/ac294c}, \href
  {https://ui.adsabs.harvard.edu/abs/2021ApJ...923..201E} {923, 201}

\bibitem[\protect\citeauthoryear{{Eggleton}}{{Eggleton}}{1983}]{Eggleton1983}
{Eggleton} P.~P.,  1983, \mn@doi [\apj] {10.1086/160960}, \href
  {https://ui.adsabs.harvard.edu/abs/1983ApJ...268..368E} {268, 368}

\bibitem[\protect\citeauthoryear{{Einfeldt}}{{Einfeldt}}{1988}]{Einfeldt_88}
{Einfeldt} B.,  1988, in Shock tubes and waves; Proceedings of the Sixteenth
  International Symposium, Aachen, Germany, July 26--31, 1987. VCH Verlag,
  Weinheim, Germany. p.~671

\bibitem[\protect\citeauthoryear{{Farmer}, {Fields}, {Petermann}, {Dessart},
  {Cantiello}, {Paxton}  \& {Timmes}}{{Farmer} et~al.}{2016}]{Farmer_16}
{Farmer} R.,  {Fields} C.~E.,  {Petermann} I.,  {Dessart} L.,  {Cantiello} M.,
  {Paxton} B.,   {Timmes} F.~X.,  2016, \mn@doi [\apjs]
  {10.3847/1538-4365/227/2/22}, \href
  {https://ui.adsabs.harvard.edu/abs/2016ApJS..227...22F} {227, 22}

\bibitem[\protect\citeauthoryear{{Ferguson}, {Alexander}, {Allard}, {Barman},
  {Bodnarik}, {Hauschildt}, {Heffner-Wong}  \& {Tamanai}}{{Ferguson}
  et~al.}{2005}]{Ferguson2005}
{Ferguson} J.~W.,  {Alexander} D.~R.,  {Allard} F.,  {Barman} T.,  {Bodnarik}
  J.~G.,  {Hauschildt} P.~H.,  {Heffner-Wong} A.,   {Tamanai} A.,  2005,
  \mn@doi [\apj] {10.1086/428642}, \href
  {https://ui.adsabs.harvard.edu/abs/2005ApJ...623..585F} {623, 585}

\bibitem[\protect\citeauthoryear{{Fields} \& {Couch}}{{Fields} \&
  {Couch}}{2020}]{Fields_20}
{Fields} C.~E.,  {Couch} S.~M.,  2020, \mn@doi [\apj]
  {10.3847/1538-4357/abada7}, \href
  {https://ui.adsabs.harvard.edu/abs/2020ApJ...901...33F} {901, 33}

\bibitem[\protect\citeauthoryear{{Fields} \& {Couch}}{{Fields} \&
  {Couch}}{2021}]{Fields_21}
{Fields} C.~E.,  {Couch} S.~M.,  2021, \mn@doi [\apj]
  {10.3847/1538-4357/ac24fb}, \href
  {https://ui.adsabs.harvard.edu/abs/2021ApJ...921...28F} {921, 28}

\bibitem[\protect\citeauthoryear{{Finn}}{{Finn}}{1989}]{finn_89}
{Finn} L.~S.,  1989, in {Evans} C.~R.,  {Finn} L.~S.,   {Hobill} D.~W.,  eds,
  Frontiers in Numerical Relativity. Cambridge University Press, Cambridge
  (UK), pp 126--145

\bibitem[\protect\citeauthoryear{{Fromm}, {Mewes}, {Messer}, {Lentz}, {Hix}  \&
  {Harris}}{{Fromm} et~al.}{2026}]{fromm_26}
{Fromm} S.~A.,  {Mewes} V.,  {Messer} O.~E.~B.,  {Lentz} E.~J.,  {Hix} W.~R.,
  {Harris} J.~A.,  2026, \mn@doi [arXiv e-prints] {10.48550/arXiv.2604.20579},
  \href {https://ui.adsabs.harvard.edu/abs/2026arXiv260420579F} {p.
  arXiv:2604.20579}

\bibitem[\protect\citeauthoryear{Fryxell et~al.,}{Fryxell
  et~al.}{2000}]{Fryxell_2000}
Fryxell B.,  et~al., 2000, \mn@doi [The Astrophysical Journal Supplement
  Series] {10.1086/317361}, 131, 273

\bibitem[\protect\citeauthoryear{{Fuller}, {Fowler}  \& {Newman}}{{Fuller}
  et~al.}{1985}]{Fuller1985}
{Fuller} G.~M.,  {Fowler} W.~A.,   {Newman} M.~J.,  1985, \mn@doi [\apj]
  {10.1086/163208}, \href
  {https://ui.adsabs.harvard.edu/abs/1985ApJ...293....1F} {293, 1}

\bibitem[\protect\citeauthoryear{{Georgy} et~al.,}{{Georgy}
  et~al.}{2024}]{Georgy_24}
{Georgy} C.,  et~al., 2024, \mn@doi [\mnras] {10.1093/mnras/stae1381}, \href
  {https://ui.adsabs.harvard.edu/abs/2024MNRAS.531.4293G} {531, 4293}

\bibitem[\protect\citeauthoryear{{Grichener} et~al.,}{{Grichener}
  et~al.}{2025}]{Grichener_25}
{Grichener} A.,  et~al., 2025, \mn@doi [\apjs] {10.3847/1538-4365/ade717},
  \href {https://ui.adsabs.harvard.edu/abs/2025ApJS..279...49G} {279, 49}

\bibitem[\protect\citeauthoryear{{Griffiths}, {Aloy}  \&
  {Obergaulinger}}{{Griffiths} et~al.}{2026a}]{Griffiths_26a}
{Griffiths} A.,  {Aloy} M.-{\'A}.,   {Obergaulinger} M.,  2026a, \mn@doi [arXiv
  e-prints] {10.48550/arXiv.2605.22927}, \href
  {https://ui.adsabs.harvard.edu/abs/2026arXiv260522927G} {p. arXiv:2605.22927}

\bibitem[\protect\citeauthoryear{{Griffiths}, {Aloy}  \&
  {Obergaulinger}}{{Griffiths} et~al.}{2026b}]{Griffiths_26b}
{Griffiths} A.,  {Aloy} M.-{\'A}.,   {Obergaulinger} M.,  2026b, \mn@doi [arXiv
  e-prints] {10.48550/arXiv.2605.22938}, \href
  {https://ui.adsabs.harvard.edu/abs/2026arXiv260522938G} {p. arXiv:2605.22938}

\bibitem[\protect\citeauthoryear{{Hanke}, {Marek}, {M{\"u}ller}  \&
  {Janka}}{{Hanke} et~al.}{2012}]{Hanke_12}
{Hanke} F.,  {Marek} A.,  {M{\"u}ller} B.,   {Janka} H.-T.,  2012, \mn@doi
  [\apj] {10.1088/0004-637X/755/2/138}, \href
  {https://ui.adsabs.harvard.edu/abs/2012ApJ...755..138H} {755, 138}

\bibitem[\protect\citeauthoryear{{Hanke}, {M{\"u}ller}, {Wongwathanarat},
  {Marek}  \& {Janka}}{{Hanke} et~al.}{2013}]{Hanke_13}
{Hanke} F.,  {M{\"u}ller} B.,  {Wongwathanarat} A.,  {Marek} A.,   {Janka}
  H.-T.,  2013, \mn@doi [\apj] {10.1088/0004-637X/770/1/66}, \href
  {https://ui.adsabs.harvard.edu/abs/2013ApJ...770...66H} {770, 66}

\bibitem[\protect\citeauthoryear{Harris et~al.,}{Harris et~al.}{2020}]{numpy}
Harris C.~R.,  et~al., 2020, \mn@doi [Nature] {10.1038/s41586-020-2649-2}, 585,
  357

\bibitem[\protect\citeauthoryear{{Herant}, {Benz}, {Hix}, {Fryer}  \&
  {Colgate}}{{Herant} et~al.}{1994}]{Herant_94}
{Herant} M.,  {Benz} W.,  {Hix} W.~R.,  {Fryer} C.~L.,   {Colgate} S.~A.,
  1994, \mn@doi [\apj] {10.1086/174817}, \href
  {https://ui.adsabs.harvard.edu/abs/1994ApJ...435..339H} {435, 339}

\bibitem[\protect\citeauthoryear{{Horowitz}, {Caballero}, {Lin}, {O'Connor}  \&
  {Schwenk}}{{Horowitz} et~al.}{2017}]{horowitz_17}
{Horowitz} C.~J.,  {Caballero} O.~L.,  {Lin} Z.,  {O'Connor} E.,   {Schwenk}
  A.,  2017, \mn@doi [\prc] {10.1103/PhysRevC.95.025801}, \href
  {https://ui.adsabs.harvard.edu/abs/2017PhRvC..95b5801H} {95, 025801}

\bibitem[\protect\citeauthoryear{Hunter}{Hunter}{2007}]{Hunter07}
Hunter J.~D.,  2007, \mn@doi [Computing in Science \& Engineering]
  {10.1109/MCSE.2007.55}, 9, 90

\bibitem[\protect\citeauthoryear{{Iglesias} \& {Rogers}}{{Iglesias} \&
  {Rogers}}{1993}]{Iglesias1993}
{Iglesias} C.~A.,  {Rogers} F.~J.,  1993, \mn@doi [\apj] {10.1086/172958},
  \href {https://ui.adsabs.harvard.edu/abs/1993ApJ...412..752I} {412, 752}

\bibitem[\protect\citeauthoryear{{Iglesias} \& {Rogers}}{{Iglesias} \&
  {Rogers}}{1996}]{Iglesias1996}
{Iglesias} C.~A.,  {Rogers} F.~J.,  1996, \mn@doi [\apj] {10.1086/177381},
  \href {https://ui.adsabs.harvard.edu/abs/1996ApJ...464..943I} {464, 943}

\bibitem[\protect\citeauthoryear{{Irwin}}{{Irwin}}{2004}]{Irwin2004}
{Irwin} A.~W.,  2004, The FreeEOS Code for Calculating the Equation of State
  for Stellar Interiors, \url {http://freeeos.sourceforge.net/}

\bibitem[\protect\citeauthoryear{{Itoh}, {Totsuji}, {Ichimaru}  \&
  {Dewitt}}{{Itoh} et~al.}{1979}]{Itoh1979}
{Itoh} N.,  {Totsuji} H.,  {Ichimaru} S.,   {Dewitt} H.~E.,  1979, \mn@doi
  [\apj] {10.1086/157590}, \href
  {https://ui.adsabs.harvard.edu/\#abs/1979ApJ...234.1079I} {234, 1079}

\bibitem[\protect\citeauthoryear{{Itoh}, {Hayashi}, {Nishikawa}  \&
  {Kohyama}}{{Itoh} et~al.}{1996}]{Itoh1996}
{Itoh} N.,  {Hayashi} H.,  {Nishikawa} A.,   {Kohyama} Y.,  1996, \mn@doi
  [\apjs] {10.1086/192264}, \href
  {https://ui.adsabs.harvard.edu/abs/1996ApJS..102..411I} {102, 411}

\bibitem[\protect\citeauthoryear{{Jakobus}, {M{\"u}ller}, {Heger}, {Zha},
  {Powell}, {Motornenko}, {Steinheimer}  \& {St{\"o}cker}}{{Jakobus}
  et~al.}{2023}]{Jakobus_23}
{Jakobus} P.,  {M{\"u}ller} B.,  {Heger} A.,  {Zha} S.,  {Powell} J.,
  {Motornenko} A.,  {Steinheimer} J.,   {St{\"o}cker} H.,  2023, \mn@doi [\prl]
  {10.1103/PhysRevLett.131.191201}, \href
  {https://ui.adsabs.harvard.edu/abs/2023PhRvL.131s1201J} {131, 191201}

\bibitem[\protect\citeauthoryear{{Jakobus}, {M{\"u}ller}  \& {Heger}}{{Jakobus}
  et~al.}{2025}]{Jakobus_25}
{Jakobus} P.,  {M{\"u}ller} B.,   {Heger} A.,  2025, \mn@doi [\mnras]
  {10.1093/mnras/staf868}, \href
  {https://ui.adsabs.harvard.edu/abs/2025MNRAS.540.3008J} {540, 3008}

\bibitem[\protect\citeauthoryear{{Janka}}{{Janka}}{2025}]{Janka_25}
{Janka} H.-T.,  2025, \mn@doi [Annual Review of Nuclear and Particle Science]
  {10.1146/annurev-nucl-121423-100945}, \href
  {https://ui.adsabs.harvard.edu/abs/2025ARNPS..75..425J} {75, 425}

\bibitem[\protect\citeauthoryear{{Janka} \& {Mueller}}{{Janka} \&
  {Mueller}}{1995a}]{Janka_95}
{Janka} H.~T.,  {Mueller} E.,  1995a, \mn@doi [\physrep]
  {10.1016/0370-1573(94)00115-J}, \href
  {https://ui.adsabs.harvard.edu/abs/1995PhR...256..135J} {256, 135}

\bibitem[\protect\citeauthoryear{{Janka} \& {Mueller}}{{Janka} \&
  {Mueller}}{1995b}]{Janka_95a}
{Janka} H.-T.,  {Mueller} E.,  1995b, \mn@doi [\apjl] {10.1086/309604}, \href
  {https://ui.adsabs.harvard.edu/abs/1995ApJ...448L.109J} {448, L109}

\bibitem[\protect\citeauthoryear{{Janka} \& {Mueller}}{{Janka} \&
  {Mueller}}{1996}]{Janka_96}
{Janka} H.-T.,  {Mueller} E.,  1996, \aap, \href
  {https://ui.adsabs.harvard.edu/abs/1996A&A...306..167J} {306, 167}

\bibitem[\protect\citeauthoryear{{Jermyn}, {Schwab}, {Bauer}, {Timmes}  \&
  {Potekhin}}{{Jermyn} et~al.}{2021}]{Jermyn2021}
{Jermyn} A.~S.,  {Schwab} J.,  {Bauer} E.,  {Timmes} F.~X.,   {Potekhin} A.~Y.,
   2021, \mn@doi [\apj] {10.3847/1538-4357/abf48e}, \href
  {https://ui.adsabs.harvard.edu/abs/2021ApJ...913...72J} {913, 72}

\bibitem[\protect\citeauthoryear{{Jermyn} et~al.,}{{Jermyn}
  et~al.}{2023}]{Jermyn2023}
{Jermyn} A.~S.,  et~al., 2023, \mn@doi [\apjs] {10.3847/1538-4365/acae8d},
  \href {https://ui.adsabs.harvard.edu/abs/2023ApJS..265...15J} {265, 15}

\bibitem[\protect\citeauthoryear{{Jones}, {Andrassy}, {Sandalski}, {Davis},
  {Woodward}  \& {Herwig}}{{Jones} et~al.}{2017}]{Jones_17}
{Jones} S.,  {Andrassy} R.,  {Sandalski} S.,  {Davis} A.,  {Woodward} P.,
  {Herwig} F.,  2017, \mn@doi [\mnras] {10.1093/mnras/stw2783}, \href
  {https://ui.adsabs.harvard.edu/abs/2017MNRAS.465.2991J} {465, 2991}

\bibitem[\protect\citeauthoryear{{Just}, {Bollig}, {Janka}, {Obergaulinger},
  {Glas}  \& {Nagataki}}{{Just} et~al.}{2018}]{Just_18}
{Just} O.,  {Bollig} R.,  {Janka} H.~T.,  {Obergaulinger} M.,  {Glas} R.,
  {Nagataki} S.,  2018, \mn@doi [\mnras] {10.1093/mnras/sty2578}, \href
  {https://ui.adsabs.harvard.edu/abs/2018MNRAS.481.4786J} {481, 4786}

\bibitem[\protect\citeauthoryear{{Kawahara}, {Kuroda}, {Takiwaki}, {Hayama}  \&
  {Kotake}}{{Kawahara} et~al.}{2018}]{Kawahara_18}
{Kawahara} H.,  {Kuroda} T.,  {Takiwaki} T.,  {Hayama} K.,   {Kotake} K.,
  2018, \mn@doi [\apj] {10.3847/1538-4357/aae57b}, \href
  {https://ui.adsabs.harvard.edu/abs/2018ApJ...867..126K} {867, 126}

\bibitem[\protect\citeauthoryear{{Kitaura}, {Janka}  \&
  {Hillebrandt}}{{Kitaura} et~al.}{2006}]{Kitaura_06}
{Kitaura} F.~S.,  {Janka} H.-T.,   {Hillebrandt} W.,  2006, \mn@doi [\aap]
  {10.1051/0004-6361:20054703}, \href
  {https://ui.adsabs.harvard.edu/abs/2006A&A...450..345K} {450, 345}

\bibitem[\protect\citeauthoryear{{Komm}, {Gu}, {Hill}, {Stark}  \&
  {Fodor}}{{Komm} et~al.}{1999}]{komm_99}
{Komm} R.~W.,  {Gu} Y.,  {Hill} F.,  {Stark} P.~B.,   {Fodor} I.~K.,  1999,
  \mn@doi [\apj] {10.1086/307359}, \href
  {https://ui.adsabs.harvard.edu/abs/1999ApJ...519..407K} {519, 407}

\bibitem[\protect\citeauthoryear{{Kotake} \& {Kuroda}}{{Kotake} \&
  {Kuroda}}{2017}]{Kotake_17}
{Kotake} K.,  {Kuroda} T.,  2017, in {Alsabti} A.~W.,  {Murdin} P.,  eds, ,
  Handbook of Supernovae.
p.~1671, \mn@doi{10.1007/978-3-319-21846-5_9}

\bibitem[\protect\citeauthoryear{{Kuhfuss}}{{Kuhfuss}}{1986}]{kuhfuss_86}
{Kuhfuss} R.,  1986, \aap, \href
  {https://ui.adsabs.harvard.edu/abs/1986A&A...160..116K} {160, 116}

\bibitem[\protect\citeauthoryear{Kuroda, Kotake, Takiwaki  \&
  Thielemann}{Kuroda et~al.}{2018}]{Kuroda_18}
Kuroda T.,  Kotake K.,  Takiwaki T.,   Thielemann F.-K.,  2018, \mn@doi [Mon.
  Not. Roy. Astron. Soc.] {10.1093/mnrasl/sly059}, 477, L80

\bibitem[\protect\citeauthoryear{{Kuroda}, {Arcones}, {Takiwaki}  \&
  {Kotake}}{{Kuroda} et~al.}{2020}]{Kuroda_20}
{Kuroda} T.,  {Arcones} A.,  {Takiwaki} T.,   {Kotake} K.,  2020, \mn@doi
  [\apj] {10.3847/1538-4357/ab9308}, \href
  {https://ui.adsabs.harvard.edu/abs/2020ApJ...896..102K} {896, 102}

\bibitem[\protect\citeauthoryear{{Kuroda}, {Fischer}, {Takiwaki}  \&
  {Kotake}}{{Kuroda} et~al.}{2022}]{Kuroda_22}
{Kuroda} T.,  {Fischer} T.,  {Takiwaki} T.,   {Kotake} K.,  2022, \mn@doi
  [\apj] {10.3847/1538-4357/ac31a8}, \href
  {https://ui.adsabs.harvard.edu/abs/2022ApJ...924...38K} {924, 38}

\bibitem[\protect\citeauthoryear{{Langanke} \&
  {Mart{\'{\i}}nez-Pinedo}}{{Langanke} \&
  {Mart{\'{\i}}nez-Pinedo}}{2000}]{Langanke2000}
{Langanke} K.,  {Mart{\'{\i}}nez-Pinedo} G.,  2000, \mn@doi [Nuclear Physics A]
  {10.1016/S0375-9474(00)00131-7}, \href
  {https://ui.adsabs.harvard.edu/abs/2000NuPhA.673..481L} {673, 481}

\bibitem[\protect\citeauthoryear{Lentz et~al.,}{Lentz et~al.}{2015}]{lentz_15}
Lentz E.~J.,  et~al., 2015, \mn@doi [The Astrophysical Journal Letters]
  {10.1088/2041-8205/807/2/L31}, 807, L31

\bibitem[\protect\citeauthoryear{{Liebend{\"o}rfer}, {Mezzacappa},
  {Thielemann}, {Messer}, {Hix}  \& {Bruenn}}{{Liebend{\"o}rfer}
  et~al.}{2001}]{Liebendorfer_01}
{Liebend{\"o}rfer} M.,  {Mezzacappa} A.,  {Thielemann} F.-K.,  {Messer} O.~E.,
  {Hix} W.~R.,   {Bruenn} S.~W.,  2001, \mn@doi [\prd]
  {10.1103/PhysRevD.63.103004}, \href
  {https://ui.adsabs.harvard.edu/abs/2001PhRvD..63j3004L} {63, 103004}

\bibitem[\protect\citeauthoryear{{Marek}, {Dimmelmeier}, {Janka}, {M{\"u}ller}
  \& {Buras}}{{Marek} et~al.}{2006}]{Marek_06}
{Marek} A.,  {Dimmelmeier} H.,  {Janka} H.~T.,  {M{\"u}ller} E.,   {Buras} R.,
  2006, \mn@doi [\aap] {10.1051/0004-6361:20052840}, \href
  {https://ui.adsabs.harvard.edu/abs/2006A&A...445..273M} {445, 273}

\bibitem[\protect\citeauthoryear{{Marek}, {Janka}  \& {M{\"u}ller}}{{Marek}
  et~al.}{2009}]{Marek_09}
{Marek} A.,  {Janka} H.~T.,   {M{\"u}ller} E.,  2009, \mn@doi [\aap]
  {10.1051/0004-6361/200810883}, \href
  {https://ui.adsabs.harvard.edu/abs/2009A&A...496..475M} {496, 475}

\bibitem[\protect\citeauthoryear{Matsumoto, Asahina, Takiwaki, Kotake  \&
  Takahashi}{Matsumoto et~al.}{2022}]{Matsumoto_22}
Matsumoto J.,  Asahina Y.,  Takiwaki T.,  Kotake K.,   Takahashi H.~R.,  2022,
  \mn@doi [Mon. Not. Roy. Astron. Soc.] {10.1093/mnras/stac2335}, 516, 1752

\bibitem[\protect\citeauthoryear{Melson, Janka  \& Marek}{Melson
  et~al.}{2015a}]{Melson_15}
Melson T.,  Janka H.-T.,   Marek A.,  2015a, \mn@doi [Astrophys. J. Lett.]
  {10.1088/2041-8205/801/2/L24}, 801, L24

\bibitem[\protect\citeauthoryear{Melson, Janka, Bollig, Hanke, Marek  \&
  M\"uller}{Melson et~al.}{2015b}]{Melson_15a}
Melson T.,  Janka H.-T.,  Bollig R.,  Hanke F.,  Marek A.,   M\"uller B.,
  2015b, \mn@doi [Astrophys. J. Lett.] {10.1088/2041-8205/808/2/L42}, 808, L42

\bibitem[\protect\citeauthoryear{{Melson}, {Kresse}  \& {Janka}}{{Melson}
  et~al.}{2020}]{Melson_20}
{Melson} T.,  {Kresse} D.,   {Janka} H.-T.,  2020, \mn@doi [\apj]
  {10.3847/1538-4357/ab72a7}, \href
  {https://ui.adsabs.harvard.edu/abs/2020ApJ...891...27M} {891, 27}

\bibitem[\protect\citeauthoryear{Mezzacappa}{Mezzacappa}{2026}]{Mezzacappa_2026}
Mezzacappa A.,  2026.  (\mn@eprint {arXiv} {2604.24970})

\bibitem[\protect\citeauthoryear{{Mezzacappa} et~al.,}{{Mezzacappa}
  et~al.}{2020}]{Mezzacappa_20}
{Mezzacappa} A.,  et~al., 2020, \mn@doi [\prd] {10.1103/PhysRevD.102.023027},
  \href {https://ui.adsabs.harvard.edu/abs/2020PhRvD.102b3027M} {102, 023027}

\bibitem[\protect\citeauthoryear{{Mezzacappa} et~al.,}{{Mezzacappa}
  et~al.}{2023}]{Mezzacappa_23}
{Mezzacappa} A.,  et~al., 2023, \mn@doi [\prd] {10.1103/PhysRevD.107.043008},
  \href {https://ui.adsabs.harvard.edu/abs/2023PhRvD.107d3008M} {107, 043008}

\bibitem[\protect\citeauthoryear{{Mori}, {Takiwaki}, {Kotake}  \&
  {Horiuchi}}{{Mori} et~al.}{2022}]{Mori_22}
{Mori} K.,  {Takiwaki} T.,  {Kotake} K.,   {Horiuchi} S.,  2022, \mn@doi [\prd]
  {10.1103/PhysRevD.105.063009}, \href
  {https://ui.adsabs.harvard.edu/abs/2022PhRvD.105f3009M} {105, 063009}

\bibitem[\protect\citeauthoryear{{Mori}, {Takiwaki}, {Kotake}  \&
  {Horiuchi}}{{Mori} et~al.}{2025}]{Mori_25}
{Mori} K.,  {Takiwaki} T.,  {Kotake} K.,   {Horiuchi} S.,  2025, \mn@doi
  [\pasj] {10.1093/pasj/psaf007}, \href
  {https://ui.adsabs.harvard.edu/abs/2025PASJ...77L...9M} {77, L9}

\bibitem[\protect\citeauthoryear{{Morozova}, {Radice}, {Burrows}  \&
  {Vartanyan}}{{Morozova} et~al.}{2018}]{Morozova_18}
{Morozova} V.,  {Radice} D.,  {Burrows} A.,   {Vartanyan} D.,  2018, \mn@doi
  [\apj] {10.3847/1538-4357/aac5f1}, \href
  {https://ui.adsabs.harvard.edu/abs/2018ApJ...861...10M} {861, 10}

\bibitem[\protect\citeauthoryear{M\"uller \& Janka}{M\"uller \&
  Janka}{2015}]{Muller_15}
M\"uller B.,  Janka H.~T.,  2015, \mn@doi [Mon. Not. Roy. Astron. Soc.]
  {10.1093/mnras/stv101}, 448, 2141

\bibitem[\protect\citeauthoryear{{M{\"u}ller}, {Janka}  \&
  {Wongwathanarat}}{{M{\"u}ller} et~al.}{2012}]{Muller_12}
{M{\"u}ller} E.,  {Janka} H.~T.,   {Wongwathanarat} A.,  2012, \mn@doi [\aap]
  {10.1051/0004-6361/201117611}, \href
  {https://ui.adsabs.harvard.edu/abs/2012A&A...537A..63M} {537, A63}

\bibitem[\protect\citeauthoryear{{M{\"u}ller}, {Janka}  \&
  {Marek}}{{M{\"u}ller} et~al.}{2013}]{muller_13}
{M{\"u}ller} B.,  {Janka} H.-T.,   {Marek} A.,  2013, \mn@doi [\apj]
  {10.1088/0004-637X/766/1/43}, \href
  {https://ui.adsabs.harvard.edu/abs/2013ApJ...766...43M} {766, 43}

\bibitem[\protect\citeauthoryear{{M{\"u}ller}, {Viallet}, {Heger}  \&
  {Janka}}{{M{\"u}ller} et~al.}{2016}]{muller_16}
{M{\"u}ller} B.,  {Viallet} M.,  {Heger} A.,   {Janka} H.-T.,  2016, \mn@doi
  [\apj] {10.3847/1538-4357/833/1/124}, \href
  {https://ui.adsabs.harvard.edu/abs/2016ApJ...833..124M} {833, 124}

\bibitem[\protect\citeauthoryear{{M{\"u}ller}, {Melson}, {Heger}  \&
  {Janka}}{{M{\"u}ller} et~al.}{2017}]{Muller_17}
{M{\"u}ller} B.,  {Melson} T.,  {Heger} A.,   {Janka} H.-T.,  2017, \mn@doi
  [\mnras] {10.1093/mnras/stx1962}, \href
  {https://ui.adsabs.harvard.edu/abs/2017MNRAS.472..491M} {472, 491}

\bibitem[\protect\citeauthoryear{{M{\"u}ller} et~al.,}{{M{\"u}ller}
  et~al.}{2019}]{Muller_19}
{M{\"u}ller} B.,  et~al., 2019, \mn@doi [\mnras] {10.1093/mnras/stz216}, \href
  {https://ui.adsabs.harvard.edu/abs/2019MNRAS.484.3307M} {484, 3307}

\bibitem[\protect\citeauthoryear{Murphy \& Burrows}{Murphy \&
  Burrows}{2008}]{Murphy_08}
Murphy J.~W.,  Burrows A.,  2008, \mn@doi [Astrophys. J.] {10.1086/592214},
  688, 1159

\bibitem[\protect\citeauthoryear{{Murphy}, {Ott}  \& {Burrows}}{{Murphy}
  et~al.}{2009}]{Murphy_09}
{Murphy} J.~W.,  {Ott} C.~D.,   {Burrows} A.,  2009, \mn@doi [\apj]
  {10.1088/0004-637X/707/2/1173}, \href
  {https://ui.adsabs.harvard.edu/abs/2009ApJ...707.1173M} {707, 1173}

\bibitem[\protect\citeauthoryear{{Murphy}, {Mezzacappa}, {Lentz}  \&
  {Marronetti}}{{Murphy} et~al.}{2025}]{Murphy_25}
{Murphy} R.~D.,  {Mezzacappa} A.,  {Lentz} E.~J.,   {Marronetti} P.,  2025,
  \mn@doi [\prd] {10.1103/5z9g-yr28}, \href
  {https://ui.adsabs.harvard.edu/abs/2025PhRvD.112f3062M} {112, 063062}

\bibitem[\protect\citeauthoryear{Mösta et~al.,}{Mösta
  et~al.}{2014}]{mosta_14}
Mösta P.,  et~al., 2014, \mn@doi [The Astrophysical Journal Letters]
  {10.1088/2041-8205/785/2/L29}, 785, L29

\bibitem[\protect\citeauthoryear{{Nakamura}, {Takiwaki}  \&
  {Kotake}}{{Nakamura} et~al.}{2022}]{Nakamura_22}
{Nakamura} K.,  {Takiwaki} T.,   {Kotake} K.,  2022, \mn@doi [\mnras]
  {10.1093/mnras/stac1586}, \href
  {https://ui.adsabs.harvard.edu/abs/2022MNRAS.514.3941N} {514, 3941}

\bibitem[\protect\citeauthoryear{{Nakamura}, {Takiwaki}, {Matsumoto}  \&
  {Kotake}}{{Nakamura} et~al.}{2025}]{Nakamura_25}
{Nakamura} K.,  {Takiwaki} T.,  {Matsumoto} J.,   {Kotake} K.,  2025, \mn@doi
  [\mnras] {10.1093/mnras/stae2611}, \href
  {https://ui.adsabs.harvard.edu/abs/2025MNRAS.536..280N} {536, 280}

\bibitem[\protect\citeauthoryear{{Nordlund}, {Stein}  \& {Asplund}}{{Nordlund}
  et~al.}{2009}]{Nordlund_09}
{Nordlund} {\r{A}}.,  {Stein} R.~F.,   {Asplund} M.,  2009, \mn@doi [Living
  Reviews in Solar Physics] {10.12942/lrsp-2009-2}, \href
  {https://ui.adsabs.harvard.edu/abs/2009LRSP....6....2N} {6, 2}

\bibitem[\protect\citeauthoryear{O'Connor}{O'Connor}{2015}]{OConnor_15}
O'Connor E.,  2015, \mn@doi [Astrophys. J. Suppl.]
  {10.1088/0067-0049/219/2/24}, 219, 24

\bibitem[\protect\citeauthoryear{{O'Connor} \& {Couch}}{{O'Connor} \&
  {Couch}}{2018a}]{oconnor_18b}
{O'Connor} E.~P.,  {Couch} S.~M.,  2018a, \mn@doi [\apj]
  {10.3847/1538-4357/aaa893}, \href
  {https://ui.adsabs.harvard.edu/abs/2018ApJ...854...63O} {854, 63}

\bibitem[\protect\citeauthoryear{{O'Connor} \& {Couch}}{{O'Connor} \&
  {Couch}}{2018b}]{OConnor_18}
{O'Connor} E.~P.,  {Couch} S.~M.,  2018b, \mn@doi [\apj]
  {10.3847/1538-4357/aadcf7}, \href
  {https://ui.adsabs.harvard.edu/abs/2018ApJ...865...81O} {865, 81}

\bibitem[\protect\citeauthoryear{{O'Connor} \& {Ott}}{{O'Connor} \&
  {Ott}}{2011}]{OConnor_11}
{O'Connor} E.,  {Ott} C.~D.,  2011, \mn@doi [\apj]
  {10.1088/0004-637X/730/2/70}, \href
  {https://ui.adsabs.harvard.edu/abs/2011ApJ...730...70O} {730, 70}

\bibitem[\protect\citeauthoryear{Obergaulinger \& Aloy}{Obergaulinger \&
  Aloy}{2020}]{obergaulinger_20}
Obergaulinger M.,  Aloy M.~A.,  2020, \mn@doi [Monthly Notices of the Royal
  Astronomical Society] {10.1093/mnras/staa096}, 492, 4613

\bibitem[\protect\citeauthoryear{Obergaulinger \& Aloy}{Obergaulinger \&
  Aloy}{2021}]{Obergaulinger_21}
Obergaulinger M.,  Aloy M.-A.,  2021, \mn@doi [Mon. Not. Roy. Astron. Soc.]
  {10.1093/mnras/stab295}, 503, 4942

\bibitem[\protect\citeauthoryear{Obergaulinger \& Aloy}{Obergaulinger \&
  Aloy}{2022}]{obergaulinger_22}
Obergaulinger M.,  Aloy M.~A.,  2022, \mn@doi [Monthly Notices of the Royal
  Astronomical Society] {10.1093/mnras/stac613}, 512, 2489

\bibitem[\protect\citeauthoryear{{Oda}, {Hino}, {Muto}, {Takahara}  \&
  {Sato}}{{Oda} et~al.}{1994}]{Oda1994}
{Oda} T.,  {Hino} M.,  {Muto} K.,  {Takahara} M.,   {Sato} K.,  1994, \mn@doi
  [Atomic Data and Nuclear Data Tables] {10.1006/adnd.1994.1007}, \href
  {https://ui.adsabs.harvard.edu/abs/1994ADNDT..56..231O} {56, 231}

\bibitem[\protect\citeauthoryear{Oohara, Nakamura  \& Shibata}{Oohara
  et~al.}{1997}]{oohara_97}
Oohara K.-i.,  Nakamura T.,   Shibata M.,  1997, \mn@doi [Progress of
  Theoretical Physics Supplement] {10.1143/PTPS.128.183}, 128, 183

\bibitem[\protect\citeauthoryear{Park, Lindberg  \& Vernon~III}{Park
  et~al.}{1987}]{park_87}
Park J.,  Lindberg C.~R.,   Vernon~III F.~L.,  1987, \mn@doi [Journal of
  Geophysical Research: Solid Earth] {https://doi.org/10.1029/JB092iB12p12675},
  92, 12675

\bibitem[\protect\citeauthoryear{{Paxton}, {Bildsten}, {Dotter}, {Herwig},
  {Lesaffre}  \& {Timmes}}{{Paxton} et~al.}{2011}]{Paxton2011}
{Paxton} B.,  {Bildsten} L.,  {Dotter} A.,  {Herwig} F.,  {Lesaffre} P.,
  {Timmes} F.,  2011, \mn@doi [\apjs] {10.1088/0067-0049/192/1/3}, \href
  {https://ui.adsabs.harvard.edu/abs/2011ApJS..192....3P} {192, 3}

\bibitem[\protect\citeauthoryear{{Paxton} et~al.,}{{Paxton}
  et~al.}{2013}]{Paxton2013}
{Paxton} B.,  et~al., 2013, \mn@doi [\apjs] {10.1088/0067-0049/208/1/4}, \href
  {https://ui.adsabs.harvard.edu/abs/2013ApJS..208....4P} {208, 4}

\bibitem[\protect\citeauthoryear{{Paxton} et~al.,}{{Paxton}
  et~al.}{2015}]{Paxton2015}
{Paxton} B.,  et~al., 2015, \mn@doi [\apjs] {10.1088/0067-0049/220/1/15}, \href
  {https://ui.adsabs.harvard.edu/abs/2015ApJS..220...15P} {220, 15}

\bibitem[\protect\citeauthoryear{{Paxton} et~al.,}{{Paxton}
  et~al.}{2018}]{Paxton2018}
{Paxton} B.,  et~al., 2018, \mn@doi [\apjs] {10.3847/1538-4365/aaa5a8}, \href
  {https://ui.adsabs.harvard.edu/abs/2018ApJS..234...34P} {234, 34}

\bibitem[\protect\citeauthoryear{{Paxton} et~al.,}{{Paxton}
  et~al.}{2019}]{Paxton2019}
{Paxton} B.,  et~al., 2019, \mn@doi [\apjs] {10.3847/1538-4365/ab2241}, \href
  {https://ui.adsabs.harvard.edu/abs/2019ApJS..243...10P} {243, 10}

\bibitem[\protect\citeauthoryear{{Potekhin} \& {Chabrier}}{{Potekhin} \&
  {Chabrier}}{2010}]{Potekhin2010}
{Potekhin} A.~Y.,  {Chabrier} G.,  2010, \mn@doi [Contributions to Plasma
  Physics] {10.1002/ctpp.201010017}, \href
  {https://ui.adsabs.harvard.edu/abs/2010CoPP...50...82P} {50, 82}

\bibitem[\protect\citeauthoryear{{Poutanen}}{{Poutanen}}{2017}]{Poutanen2017}
{Poutanen} J.,  2017, \mn@doi [\apj] {10.3847/1538-4357/835/2/119}, \href
  {https://ui.adsabs.harvard.edu/abs/2017ApJ...835..119P} {835, 119}

\bibitem[\protect\citeauthoryear{{Powell} \& {M{\"u}ller}}{{Powell} \&
  {M{\"u}ller}}{2019}]{Powell_19}
{Powell} J.,  {M{\"u}ller} B.,  2019, \mn@doi [\mnras] {10.1093/mnras/stz1304},
  \href {https://ui.adsabs.harvard.edu/abs/2019MNRAS.487.1178P} {487, 1178}

\bibitem[\protect\citeauthoryear{{Powell} \& {M{\"u}ller}}{{Powell} \&
  {M{\"u}ller}}{2020}]{Powell_20}
{Powell} J.,  {M{\"u}ller} B.,  2020, \mn@doi [\mnras]
  {10.1093/mnras/staa1048}, \href
  {https://ui.adsabs.harvard.edu/abs/2020MNRAS.494.4665P} {494, 4665}

\bibitem[\protect\citeauthoryear{{Powell}, {M{\"u}ller}  \& {Heger}}{{Powell}
  et~al.}{2021}]{Powell_21}
{Powell} J.,  {M{\"u}ller} B.,   {Heger} A.,  2021, \mn@doi [\mnras]
  {10.1093/mnras/stab614}, \href
  {https://ui.adsabs.harvard.edu/abs/2021MNRAS.503.2108P} {503, 2108}

\bibitem[\protect\citeauthoryear{{Powell}, {M{\"u}ller}, {Aguilera-Dena}  \&
  {Langer}}{{Powell} et~al.}{2023}]{Powell_23}
{Powell} J.,  {M{\"u}ller} B.,  {Aguilera-Dena} D.~R.,   {Langer} N.,  2023,
  \mn@doi [\mnras] {10.1093/mnras/stad1292}, \href
  {https://ui.adsabs.harvard.edu/abs/2023MNRAS.522.6070P} {522, 6070}

\bibitem[\protect\citeauthoryear{{Radice}, {Ott}, {Abdikamalov}, {Couch},
  {Haas}  \& {Schnetter}}{{Radice} et~al.}{2016}]{Radice_16}
{Radice} D.,  {Ott} C.~D.,  {Abdikamalov} E.,  {Couch} S.~M.,  {Haas} R.,
  {Schnetter} E.,  2016, \mn@doi [\apj] {10.3847/0004-637X/820/1/76}, \href
  {https://ui.adsabs.harvard.edu/abs/2016ApJ...820...76R} {820, 76}

\bibitem[\protect\citeauthoryear{{Radice}, {Morozova}, {Burrows}, {Vartanyan}
  \& {Nagakura}}{{Radice} et~al.}{2019}]{Radice_19}
{Radice} D.,  {Morozova} V.,  {Burrows} A.,  {Vartanyan} D.,   {Nagakura} H.,
  2019, \mn@doi [\apjl] {10.3847/2041-8213/ab191a}, \href
  {https://ui.adsabs.harvard.edu/abs/2019ApJ...876L...9R} {876, L9}

\bibitem[\protect\citeauthoryear{{Rampp} \& {Janka}}{{Rampp} \&
  {Janka}}{2000}]{Rampp_00}
{Rampp} M.,  {Janka} H.-T.,  2000, \mn@doi [\apjl] {10.1086/312837}, \href
  {https://ui.adsabs.harvard.edu/abs/2000ApJ...539L..33R} {539, L33}

\bibitem[\protect\citeauthoryear{Reynolds}{Reynolds}{1895}]{Reynolds_95}
Reynolds O.,  1895, \mn@doi [Philosophical Transactions of the Royal Society of
  London, Series A: Containing Papers of a Mathematical or Physical Character]
  {10.1098/rsta.1895.0004}, pp 123--164

\bibitem[\protect\citeauthoryear{{Ritter}}{{Ritter}}{1988}]{Ritter1988}
{Ritter} H.,  1988, \aap, \href
  {https://ui.adsabs.harvard.edu/abs/1988A%26A...202...93R} {202, 93}

\bibitem[\protect\citeauthoryear{{Rogers} \& {Nayfonov}}{{Rogers} \&
  {Nayfonov}}{2002}]{Rogers2002}
{Rogers} F.~J.,  {Nayfonov} A.,  2002, \mn@doi [\apj] {10.1086/341894}, \href
  {https://ui.adsabs.harvard.edu/abs/2002ApJ...576.1064R} {576, 1064}

\bibitem[\protect\citeauthoryear{{Salpeter}}{{Salpeter}}{1954}]{Salpeter1954}
{Salpeter} E.~E.,  1954, \mn@doi [Australian Journal of Physics]
  {10.1071/PH540373}, \href
  {https://ui.adsabs.harvard.edu/\#abs/1954AuJPh...7..373S} {7, 373}

\bibitem[\protect\citeauthoryear{{Saumon}, {Chabrier}  \& {van Horn}}{{Saumon}
  et~al.}{1995}]{Saumon1995}
{Saumon} D.,  {Chabrier} G.,   {van Horn} H.~M.,  1995, \mn@doi [\apjs]
  {10.1086/192204}, \href
  {https://ui.adsabs.harvard.edu/abs/1995ApJS...99..713S} {99, 713}

\bibitem[\protect\citeauthoryear{{Shibagaki}, {Kuroda}, {Kotake}, {Takiwaki}
  \& {Fischer}}{{Shibagaki} et~al.}{2024}]{Shibagaki_24}
{Shibagaki} S.,  {Kuroda} T.,  {Kotake} K.,  {Takiwaki} T.,   {Fischer} T.,
  2024, \mn@doi [\mnras] {10.1093/mnras/stae1361}, \href
  {https://ui.adsabs.harvard.edu/abs/2024MNRAS.531.3732S} {531, 3732}

\bibitem[\protect\citeauthoryear{Shibata, Kiuchi, Sekiguchi  \& Suwa}{Shibata
  et~al.}{2011}]{Shibata_11}
Shibata M.,  Kiuchi K.,  Sekiguchi Y.-i.,   Suwa Y.,  2011, \mn@doi [Prog.
  Theor. Phys.] {10.1143/PTP.125.1255}, 125, 1255

\bibitem[\protect\citeauthoryear{{Steiner}, {Hempel}  \& {Fischer}}{{Steiner}
  et~al.}{2013}]{steiner_13}
{Steiner} A.~W.,  {Hempel} M.,   {Fischer} T.,  2013, \mn@doi [\apj]
  {10.1088/0004-637X/774/1/17}, 774, 17

\bibitem[\protect\citeauthoryear{{Summa}, {Janka}, {Melson}  \&
  {Marek}}{{Summa} et~al.}{2018}]{Summa_18}
{Summa} A.,  {Janka} H.-T.,  {Melson} T.,   {Marek} A.,  2018, \mn@doi [\apj]
  {10.3847/1538-4357/aa9ce8}, \href
  {https://ui.adsabs.harvard.edu/abs/2018ApJ...852...28S} {852, 28}

\bibitem[\protect\citeauthoryear{{Takata}, {Mori}, {Nakamura}  \&
  {Kotake}}{{Takata} et~al.}{2025}]{Takata_25}
{Takata} T.,  {Mori} K.,  {Nakamura} K.,   {Kotake} K.,  2025, \mn@doi [\prd]
  {10.1103/PhysRevD.111.103028}, \href
  {https://ui.adsabs.harvard.edu/abs/2025PhRvD.111j3028T} {111, 103028}

\bibitem[\protect\citeauthoryear{Takiwaki, Kotake  \& Suwa}{Takiwaki
  et~al.}{2014}]{Takiwaki_14}
Takiwaki T.,  Kotake K.,   Suwa Y.,  2014, \mn@doi [Astrophys. J.]
  {10.1088/0004-637X/786/2/83}, 786, 83

\bibitem[\protect\citeauthoryear{Thomson}{Thomson}{1982}]{thomson_82}
Thomson D.,  1982, \mn@doi [Proceedings of the IEEE] {10.1109/PROC.1982.12433},
  70, 1055

\bibitem[\protect\citeauthoryear{{Timmes} \& {Swesty}}{{Timmes} \&
  {Swesty}}{2000}]{Timmes2000}
{Timmes} F.~X.,  {Swesty} F.~D.,  2000, \mn@doi [\apjs] {10.1086/313304}, \href
  {https://ui.adsabs.harvard.edu/abs/2000ApJS..126..501T} {126, 501}

\bibitem[\protect\citeauthoryear{{Turk}, {Smith}, {Oishi}, {Skory}, {Skillman},
  {Abel}  \& {Norman}}{{Turk} et~al.}{2011}]{Turk11}
{Turk} M.~J.,  {Smith} B.~D.,  {Oishi} J.~S.,  {Skory} S.,  {Skillman} S.~W.,
  {Abel} T.,   {Norman} M.~L.,  2011, \mn@doi [\apjs]
  {10.1088/0067-0049/192/1/9}, \href
  {https://ui.adsabs.harvard.edu/abs/2011ApJS..192....9T} {192, 9}

\bibitem[\protect\citeauthoryear{{Varma} \& {M{\"u}ller}}{{Varma} \&
  {M{\"u}ller}}{2021}]{Varma_21}
{Varma} V.,  {M{\"u}ller} B.,  2021, \mn@doi [\mnras] {10.1093/mnras/stab883},
  \href {https://ui.adsabs.harvard.edu/abs/2021MNRAS.504..636V} {504, 636}

\bibitem[\protect\citeauthoryear{{Varma} et~al.,}{{Varma}
  et~al.}{2026}]{Varma_26}
{Varma} V.,  et~al., 2026, \mn@doi [arXiv e-prints]
  {10.48550/arXiv.2608.04954}, \href
  {https://ui.adsabs.harvard.edu/abs/2026arXiv260804954V} {p. arXiv:2608.04954}

\bibitem[\protect\citeauthoryear{Vartanyan, Burrows, Radice, Skinner  \&
  Dolence}{Vartanyan et~al.}{2019}]{Vartanyan_19a}
Vartanyan D.,  Burrows A.,  Radice D.,  Skinner A.~M.,   Dolence J.,  2019,
  \mn@doi [Mon. Not. Roy. Astron. Soc.] {10.1093/mnras/sty2585}, 482, 351

\bibitem[\protect\citeauthoryear{{Vartanyan}, {Coleman}  \&
  {Burrows}}{{Vartanyan} et~al.}{2022}]{Vartanyan_22}
{Vartanyan} D.,  {Coleman} M. S.~B.,   {Burrows} A.,  2022, \mn@doi [\mnras]
  {10.1093/mnras/stab3702}, \href
  {https://ui.adsabs.harvard.edu/abs/2022MNRAS.510.4689V} {510, 4689}

\bibitem[\protect\citeauthoryear{{Vartanyan}, {Burrows}, {Wang}, {Coleman}  \&
  {White}}{{Vartanyan} et~al.}{2023}]{Vartanyan_23}
{Vartanyan} D.,  {Burrows} A.,  {Wang} T.,  {Coleman} M. S.~B.,   {White}
  C.~J.,  2023, \mn@doi [\prd] {10.1103/PhysRevD.107.103015}, \href
  {https://ui.adsabs.harvard.edu/abs/2023PhRvD.107j3015V} {107, 103015}

\bibitem[\protect\citeauthoryear{{Vartanyan}, {Tsang}, {Kasen}, {Burrows},
  {Wang}  \& {Teryoshin}}{{Vartanyan} et~al.}{2025}]{Vartanyan_25}
{Vartanyan} D.,  {Tsang} B. T.-H.,  {Kasen} D.,  {Burrows} A.,  {Wang} T.,
  {Teryoshin} L.,  2025, \mn@doi [\apj] {10.3847/1538-4357/adb1e4}, \href
  {https://ui.adsabs.harvard.edu/abs/2025ApJ...982....9V} {982, 9}

\bibitem[\protect\citeauthoryear{{Viallet}, {Meakin}, {Arnett}  \&
  {Moc{\'a}k}}{{Viallet} et~al.}{2013}]{Viallet_13}
{Viallet} M.,  {Meakin} C.,  {Arnett} D.,   {Moc{\'a}k} M.,  2013, \mn@doi
  [\apj] {10.1088/0004-637X/769/1/1}, \href
  {https://ui.adsabs.harvard.edu/abs/2013ApJ...769....1V} {769, 1}

\bibitem[\protect\citeauthoryear{Virtanen et~al.,}{Virtanen
  et~al.}{2020a}]{scipy}
Virtanen P.,  et~al., 2020a, \mn@doi [Nature Methods]
  {10.1038/s41592-019-0686-2}, \href {https://rdcu.be/b08Wh} {17, 261}

\bibitem[\protect\citeauthoryear{Virtanen et~al.,}{Virtanen
  et~al.}{2020b}]{2020SciPy-NMeth}
Virtanen P.,  et~al., 2020b, \mn@doi [Nature Methods]
  {10.1038/s41592-019-0686-2}, \href {https://rdcu.be/b08Wh} {17, 261}

\bibitem[\protect\citeauthoryear{{Wang}, {Vartanyan}, {Burrows}  \&
  {Coleman}}{{Wang} et~al.}{2022}]{wang_22}
{Wang} T.,  {Vartanyan} D.,  {Burrows} A.,   {Coleman} M. S.~B.,  2022, \mn@doi
  [\mnras] {10.1093/mnras/stac2691}, \href
  {https://ui.adsabs.harvard.edu/abs/2022MNRAS.517..543W} {517, 543}

\bibitem[\protect\citeauthoryear{{Woosley} \& {Heger}}{{Woosley} \&
  {Heger}}{2007}]{woosley_07}
{Woosley} S.~E.,  {Heger} A.,  2007, \mn@doi [\physrep]
  {10.1016/j.physrep.2007.02.009}, \href
  {http://adsabs.harvard.edu/abs/2007PhR...442..269W} {442, 269}

\bibitem[\protect\citeauthoryear{{Woosley}, {Heger}  \& {Weaver}}{{Woosley}
  et~al.}{2002}]{woosley_02}
{Woosley} S.~E.,  {Heger} A.,   {Weaver} T.~A.,  2002, \mn@doi [Reviews of
  Modern Physics] {10.1103/RevModPhys.74.1015}, \href
  {https://ui.adsabs.harvard.edu/abs/2002RvMP...74.1015W} {74, 1015}

\bibitem[\protect\citeauthoryear{{Yadav}, {M{\"u}ller}, {Janka}, {Melson}  \&
  {Heger}}{{Yadav} et~al.}{2020}]{Yadav_20}
{Yadav} N.,  {M{\"u}ller} B.,  {Janka} H.~T.,  {Melson} T.,   {Heger} A.,
  2020, \mn@doi [\apj] {10.3847/1538-4357/ab66bb}, \href
  {https://ui.adsabs.harvard.edu/abs/2020ApJ...890...94Y} {890, 94}

\bibitem[\protect\citeauthoryear{{Yamada} et~al.,}{{Yamada}
  et~al.}{2024}]{Yamada_24}
{Yamada} S.,  et~al., 2024, \mn@doi [Proceedings of the Japan Academy, Series
  B] {10.2183/pjab.100.015}, \href
  {https://ui.adsabs.harvard.edu/abs/2024PJAB..100..190Y} {100, 190}

\bibitem[\protect\citeauthoryear{{Yoshida}, {Takiwaki}, {Kotake}, {Takahashi},
  {Nakamura}  \& {Umeda}}{{Yoshida} et~al.}{2019}]{yoshida_19}
{Yoshida} T.,  {Takiwaki} T.,  {Kotake} K.,  {Takahashi} K.,  {Nakamura} K.,
  {Umeda} H.,  2019, \mn@doi [\apj] {10.3847/1538-4357/ab2b9d}, \href
  {https://ui.adsabs.harvard.edu/abs/2019ApJ...881...16Y} {881, 16}

\bibitem[\protect\citeauthoryear{{Yoshida}, {Takiwaki}, {Kotake}, {Takahashi},
  {Nakamura}  \& {Umeda}}{{Yoshida} et~al.}{2021}]{yoshida_21}
{Yoshida} T.,  {Takiwaki} T.,  {Kotake} K.,  {Takahashi} K.,  {Nakamura} K.,
  {Umeda} H.,  2021, \mn@doi [\apj] {10.3847/1538-4357/abd3a3}, \href
  {https://ui.adsabs.harvard.edu/abs/2021ApJ...908...44Y} {908, 44}

\bibitem[\protect\citeauthoryear{{de Jager}, {Nieuwenhuijzen}  \& {van der
  Hucht}}{{de Jager} et~al.}{1988}]{dejager_88}
{de Jager} C.,  {Nieuwenhuijzen} H.,   {van der Hucht} K.~A.,  1988, \aaps,
  \href {https://ui.adsabs.harvard.edu/abs/1988A&AS...72..259D} {72, 259}

\makeatother
\end{thebibliography}

\appendix

\section{Rate Differences} \label{apx:rates}
When we examined the reaction rates in more detail, particularly around 
$T \sim 10^9\,\mathrm{K}$, 
we observed systematic offsets between the rates that were independent of temperature.
Temperatures of $T \sim 10^9\,\mathrm{K}$ are typical for the O- and Si-burning shells 
where we see the largest differences between \textsc{FLASH} and \textsc{MESA}.
While we found differences in several rates, we examined two specific reactions in more detail:
\begin{equation}
{}^{36}\mathrm{Ar} + \alpha \longrightarrow {}^{39}\mathrm{K} + p.
\end{equation}
The observed ratio between the \textsc{MESA} and \textsc{FLASH} rates was
$\approx 7.1$ and, notably, roughly constant in temperature over $T = (2$--$4)\times10^9\,$K.
The same behaviour is found for
\begin{equation}
{}^{28}\mathrm{Si} + \alpha \longrightarrow {}^{31}\mathrm{P} + p,
\end{equation}
where we find an offset of $\approx 6.9$, again independent of temperature.
Numerically, these two rates are computed as the reverse rates of
\begin{equation}
{}^{39}\mathrm{K} + p \longrightarrow {}^{36}\mathrm{Ar} + \alpha,
\end{equation}
and
\begin{equation}
{}^{31}\mathrm{P} + p \longrightarrow {}^{28}\mathrm{Si} + \alpha.
\end{equation}
In equilibrium, reverse rates are obtained from detailed balance. For the present discussion, it suffices to note that
\begin{equation} \label{eq:revrates}
\frac{\lambda_{\rm rev}}{\lambda_{\rm fwd}} \propto \Big (\frac{A_{\rm in}}{A_{\rm out}}\Big)^{3/2} \exp\Big(-\frac{Q}{k_b T}\Big),
\end{equation}
where $A_{\rm in}$ and $A_{\rm out}$ denote the products of the mass numbers of the reactants and products, respectively, and $Q$ is the forward energy release. $\lambda_{\rm rev}$ denotes the reverse rate and $\lambda_{\rm fwd}$ the forward rate. 
For the reactions above we have
\begin{equation}
\Big (\frac{A_{\rm in}}{A_{\rm out}}\Big)^{3/2} = \Big(\frac{144}{39}\Big)^{3/2} \approx 7.10,
\end{equation}
and
\begin{equation}
\Big (\frac{A_{\rm in}}{A_{\rm out}}\Big)^{3/2} = \Big(\frac{112}{31}\Big)^{3/2} \approx 6.87, \end{equation}
which are in close agreement with the offsets between the two codes.

In \textsc{MESA}, we found that the exponent of the term
$({A_{\rm in}}/{A_{\rm out}})$
in Eq.~\eqref{eq:revrates} carried an additional factor
$n = |N_{\rm out} - N_{\rm in}|$, where $N_{\rm in}$ and
$N_{\rm out}$ are the numbers of reactants and products, so that
the exponent read $3n/2$ rather than $3/2$. This explains the
differences between the rates in \textsc{MESA} and
\textsc{FLASH}, since $n=0$ for these reactions. 
We reported this bug to the developers of 
\textsc{MESA} and it was fixed in a recent release (\url{https://github.com/MESAHub/mesa/pull/975}). The factor
$\exp(-{Q}/{k_b T})$ 
in Eq.~\eqref{eq:revrates} suppresses reverse reactions at low $T$, and the issue only becomes relevant at $T \sim 10^9\,\mathrm{K}$,
in other words during advanced burning stages.
Additionally, we found smaller differences in forward reaction rates 
used by the \texttt{approx21} implementations in \textsc{FLASH} and
\textsc{MESA}. Importantly, despite these differences
the evolution of quantities such as the temperature and density is similar
across the two codes.

\begin{figure*}
    \centering
    \includegraphics[width=0.49\linewidth]{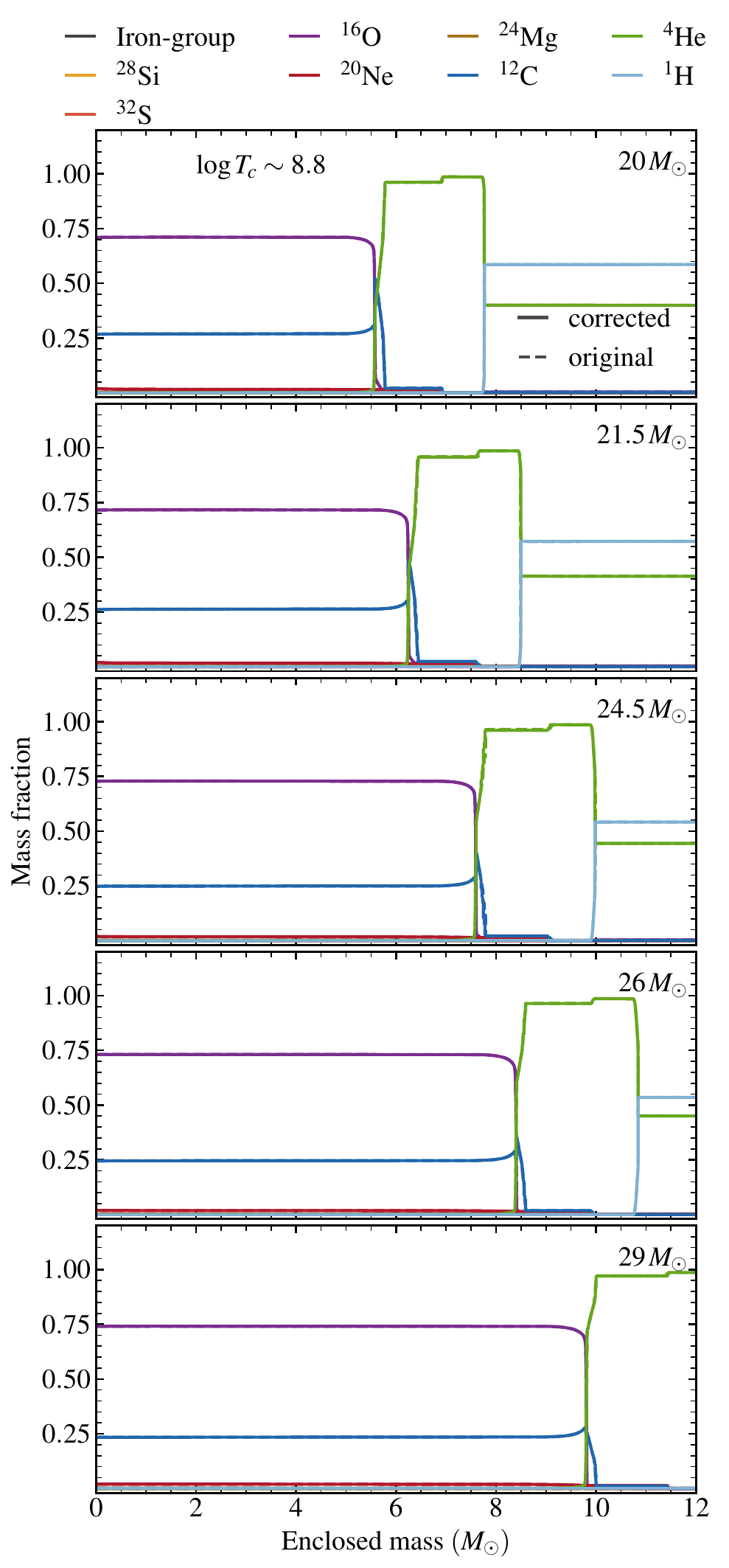}
    \includegraphics[width=0.49\linewidth]{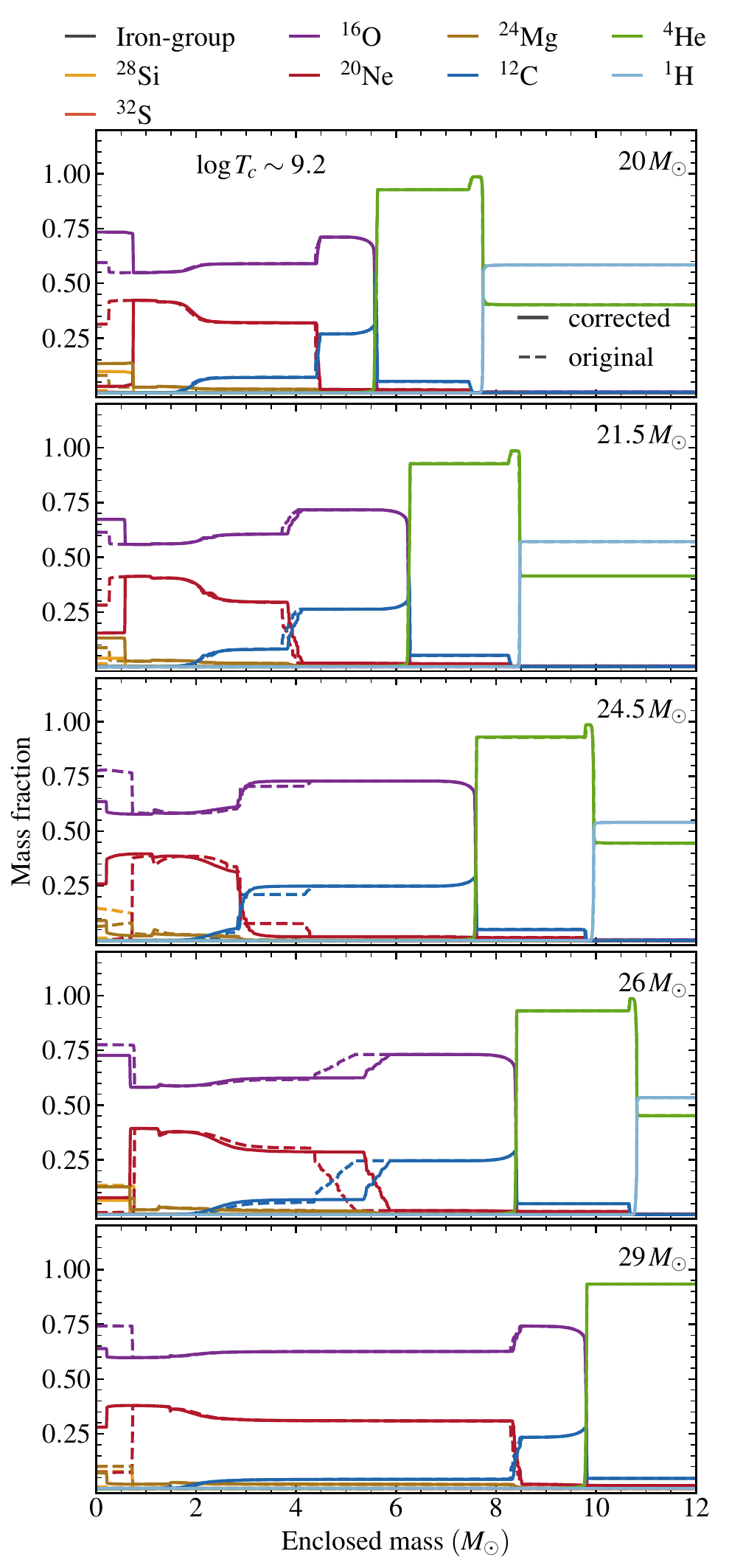}
    \caption{Comparison of composition profiles from \textsc{MESA} with the original rates and with the fixed rates. The left column shows the mass fractions during central carbon burning ($\log T_c = 8.8$) and the right column shows the mass fractions during central oxygen burning ($\log T_c = 9.2$).  Solid lines correspond to the corrected \textsc{MESA} runs and dashed lines to the original runs.}
    \label{fig:rates_burning}
\end{figure*}

\begin{figure*}
    \centering
    \includegraphics[width=0.49\linewidth]{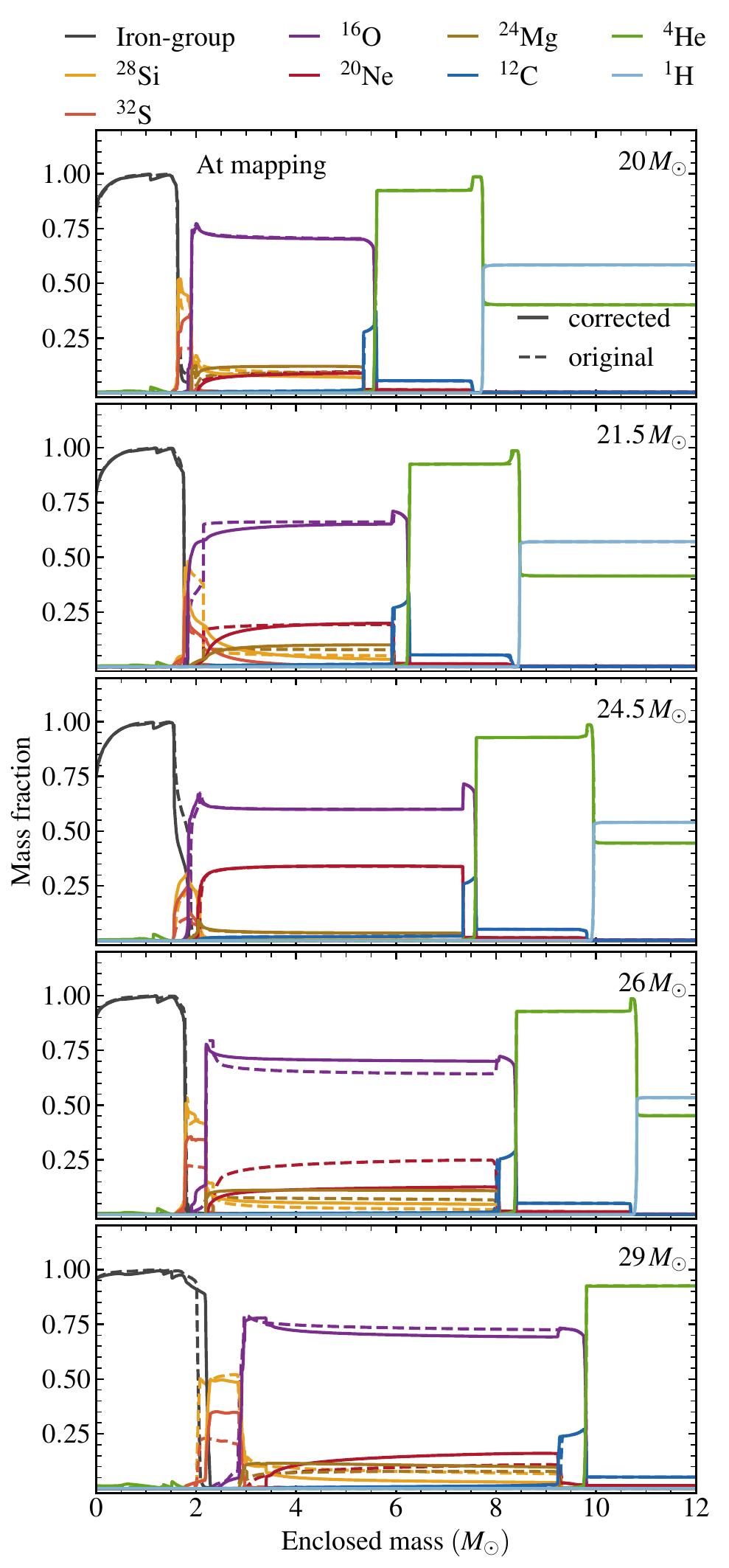}
    \includegraphics[width=0.49\linewidth]{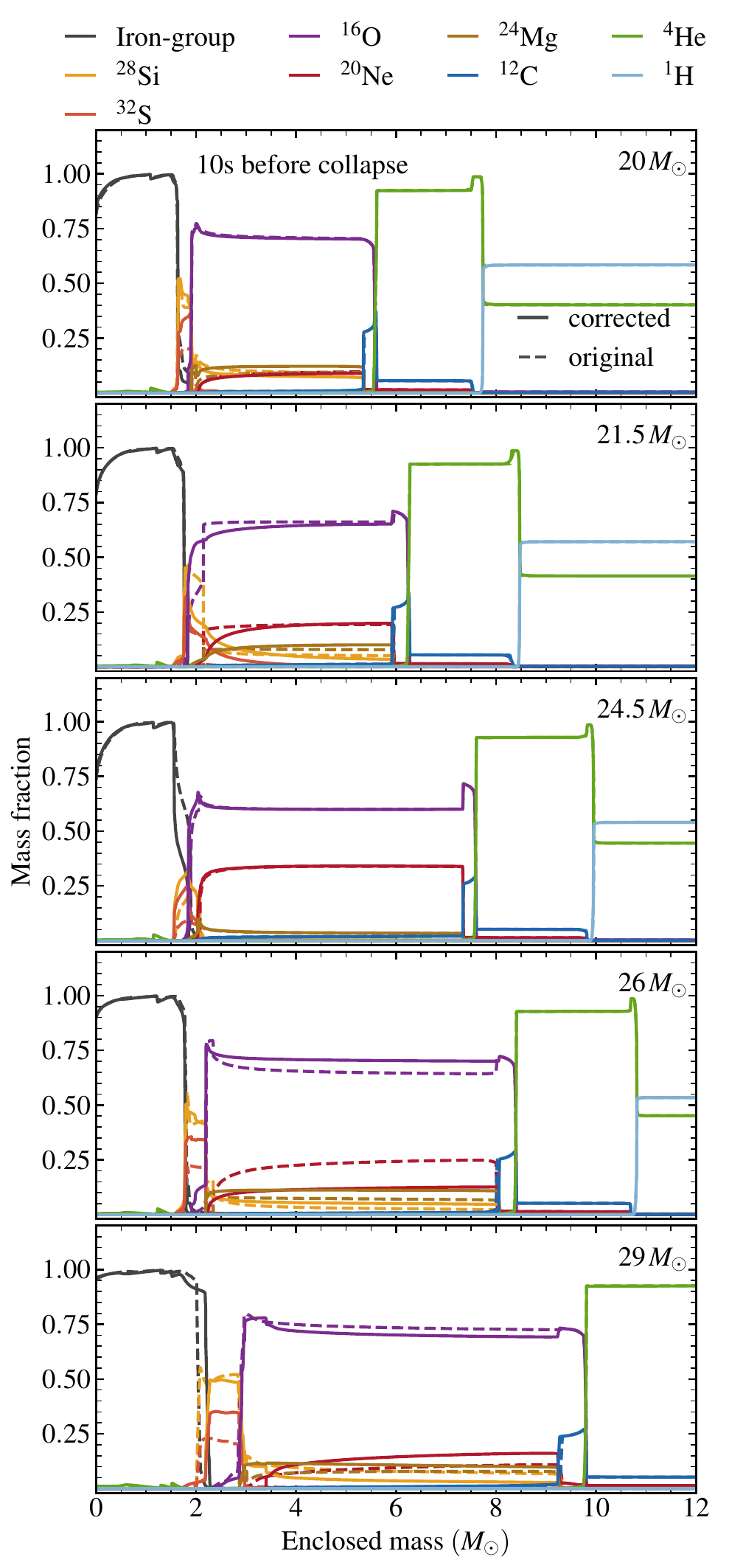}
    \caption{Comparison of composition profiles from \textsc{MESA} with the original rates and with the fixed rates. The left column shows the mass fractions at the time of mapping from \textsc{MESA} to \textsc{FLASH}  and the right column shows the mass fractions $\sim10\,$s prior to core collapse. Solid lines correspond to the corrected \textsc{MESA} runs and dashed lines to the original runs.}
    \label{fig:rates_collapse}
\end{figure*}
To investigate the impact of the updated reaction rates on the initial conditions for our
\textsc{FLASH} simulations, we compare the mass fractions of our original \textsc{MESA} runs
and new runs performed after fixing the issue. We will refer to the new runs as the corrected runs and our original runs as the original runs.
In Fig.~\ref{fig:rates_burning} and Fig.~\ref{fig:rates_collapse}
we show the mass fraction at four informative epochs in the evolution of each of our five
progenitors. The two times shown in Fig.~\ref{fig:rates_burning} are
selected to be during core carbon burning and core oxygen burning.
The former is taken to be the first time the central temperature ($T_c$) crosses
$\log T_c = 8.8$ ($T_c \simeq 6.3 \times 10^{8}\,\mathrm{K}$), and central
oxygen burning is taken as the first crossing of $\log T_c = 9.2$ ($T_c \simeq 1.6 \times
10^{9}\,\mathrm{K}$). These values correspond to burning temperatures
reported in \cite{woosley_02}. The two times shown by Fig.~\ref{fig:rates_collapse} are 
the time at which we map from \textsc{MESA} to \textsc{FLASH} and $10\,$s before
core collapse.

From Fig.~\ref{fig:rates_burning}, we see that the compositions agree well up to and including the carbon burning stage. The solid and dashed lines in the left panel of  
Fig.~\ref{fig:rates_burning} are indistinguishable. However, there are clear differences between the two simulation sets once the central temperature
exceeds $\sim10^{9}\,$K and oxygen burning starts. Interestingly, the differences are not
monotonic across our models. For example, the original $20$ and $21.5\,M_{\odot}$ models show a deficit of $^{16}$O in the inner regions, when compared to the new runs.
On the other hand, the situation is reversed for the other three models. Both the $^{12}$C
and $^{20}$Ne burning are strongly affected by the incorrect rates in model $26\,M_{\odot}$,
but this difference is less pronounced for the other models.

\begin{figure}
    \centering
    \includegraphics[width=\linewidth]{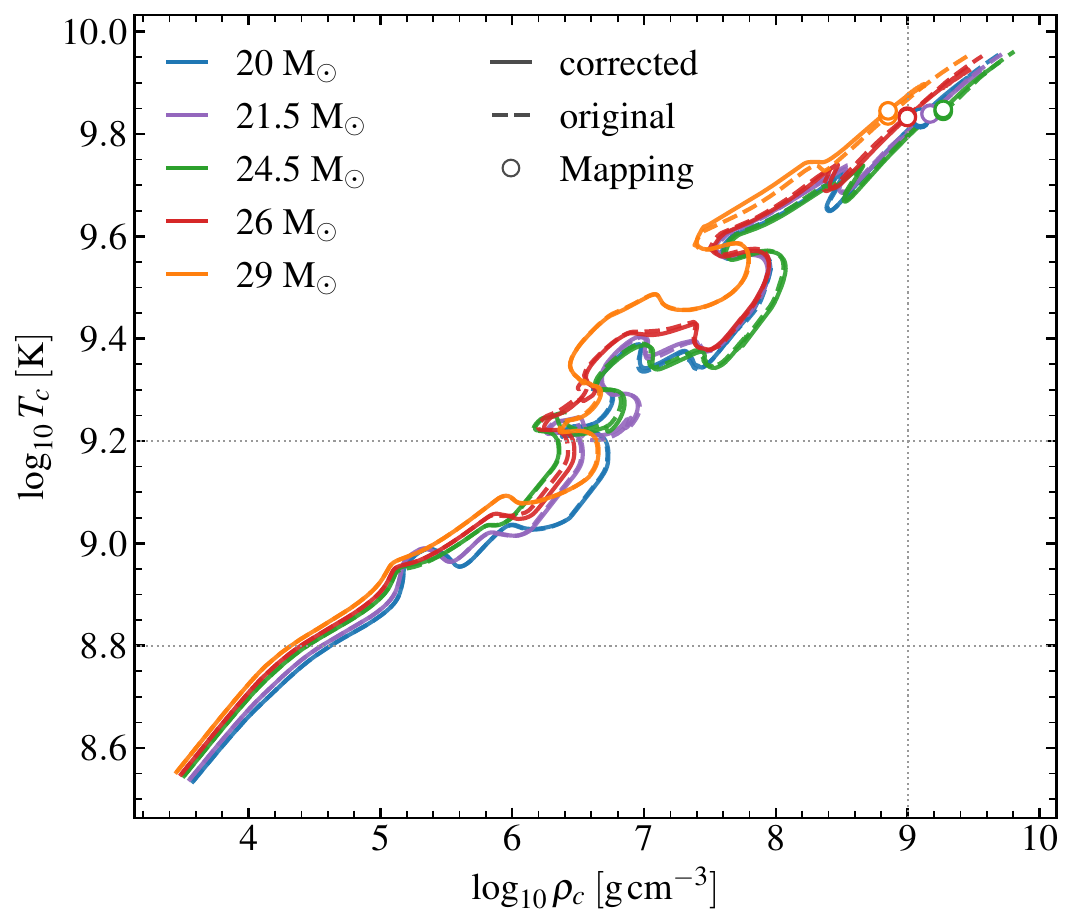} \\
    \includegraphics[width=\linewidth]{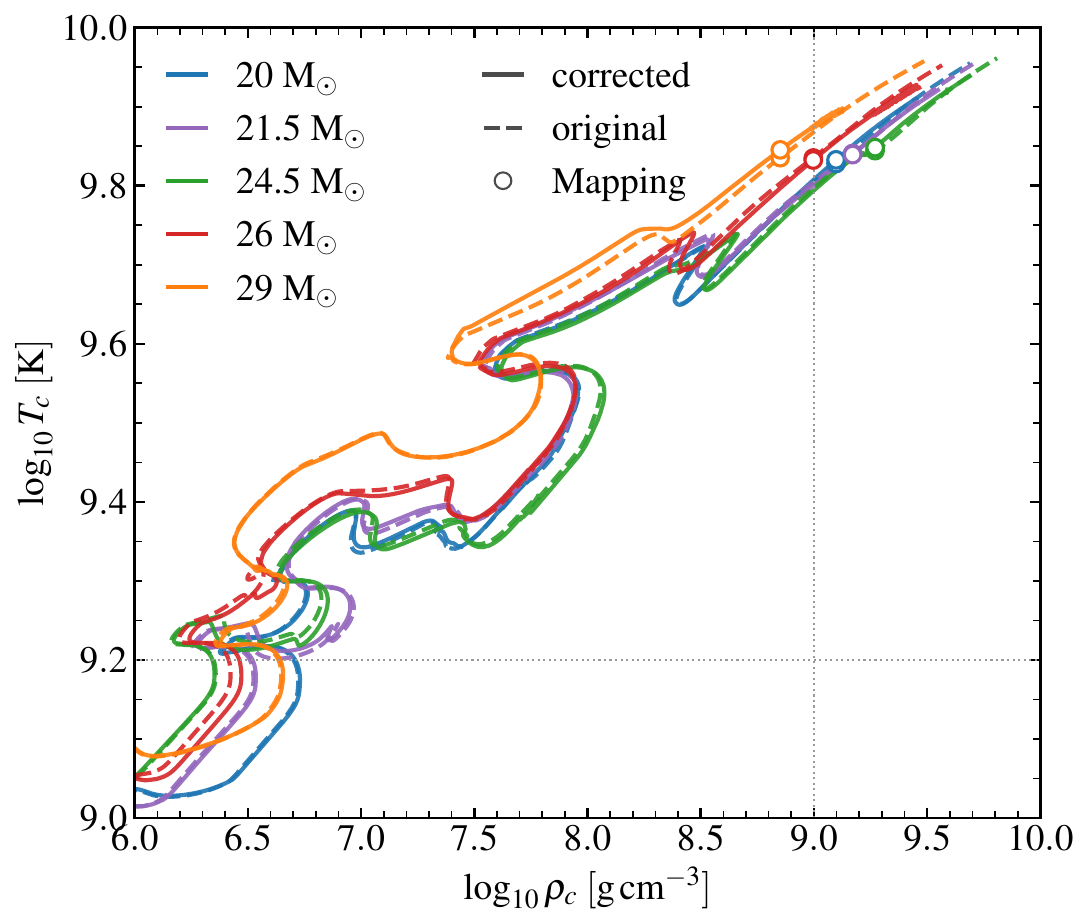} 
    \caption{Evolutionary tracks for the five stellar models, central temperature ($T_c$) versus central density ($\rho_c$).  
    The top panel shows the  
    full evolution from helium depletion to core collapse, the bottom 
    panel zooms in on the phase following carbon ignition. Solid lines
    show the corrected \textsc{MESA} runs and dashed lines the original runs. 
    Circles mark the central  density at which we mapped the original models into
    \textsc{FLASH}. The two horizontal dotted lines mark $\log T_c = 8.8$ and  
    $\log T_c = 9.2$, which are markers for carbon and oxygen burning, 
    respectively. The vertical dotted lines mark $\rho_c = 10^{9}\,
    \mathrm{g\,cm^{-3}}$.}
    \label{fig:mesa_tracks}
\end{figure}
We observe relatively large differences in the composition profiles of the original and corrected runs at the time of mapping (see Fig.~\ref{fig:rates_collapse}). As was the case during oxygen burning, the trends are not monotonic across our model set. The original runs have some models that show an enhancement of $^{16}$O in the oxygen-rich layer, compared to the new corrected runs, while the situation is reversed for other models. Comparing the right panels of Fig.~\ref{fig:rates_collapse} with the same panels in
Fig.~\ref{fig:composition_inner} we see that the differences between the corrected and original runs are similar to the differences we first observed between our \textsc{FLASH} and
\textsc{MESA} runs. Mapping the models to \textsc{FLASH} does not undo the composition differences, but the correct rates in our \textsc{FLASH} runs evolved the mass fractions in the right direction.

We show evolutionary tracks for our five stellar models, both the original set and the corrected set, in Fig.~\ref{fig:mesa_tracks}. We plot the central temperature versus the central density ($\rho_c$). The upper panel of Fig.~\ref{fig:mesa_tracks} shows the full evolution from helium depletion to collapse and the bottom panel shows the
final stages of the evolution (from carbon burning until collapse).
In general, the models show reasonably good agreement all the way until collapse, but clear differences
are visible once central oxygen burning sets in. The shapes of the tracks are similar between the original and corrected runs. However, as is clearly visible for $29\,M_{\odot}$, the curves are shifted
between the two runs. Once $\log_{10} T_c$ exceeds 9.2, we observe differences between the two simulation sets for all models during the advanced burning stages.
Additionally, the corrected runs do not evolve as far as the original runs and towards the end we see spurious oscillations in the tracks. We found similar oscillations in the
energy generation and compositions in the core of our corrected runs. The oscillations are
likely due to the operator split nuclear burning scheme used in our \textsc{MESA} simulations close
to core collapse.

For all of the models, there are differences in the tracks throughout the evolution, but the differences are smaller than what we observe during the final burning stages. The early time differences are most easily seen in the two simulations of the $26\,M_{\odot}$ progenitor, where we see clear separation of the two tracks prior to central oxygen burning.

We did not re-run the \textsc{FLASH} simulations from the
corrected set. The overall structure of the stars at the time of
mapping agrees reasonably well between the two realisations,
even if the compositions and the exact locations of the shell
boundaries differ. Repeating the simulations would likely alter
the details of how convection develops in individual models, but
turbulent burning would still be established and the subsequent
core collapse would proceed in a similar manner. We therefore do
not expect the main conclusions of this work to be affected.

\bsp	
\label{lastpage}
\end{document}